\documentclass{jfm}

\usepackage{subfigure}
\usepackage{amsmath,graphicx,amssymb}
\usepackage{multirow,color}
\usepackage{url}
\usepackage{verbatim}

\usepackage{bm,caption}

\renewcommand{\vec}[1]{\bm{#1}}
\newcommand{\fvec}[1]{\hat{\vec{#1}}}
\newcommand{\beq}{\begin{equation}}
\newcommand{\eeq}{\end{equation}}

\newcommand{\vx}{\vec{x}}

\newcommand{\vk}{\vec{k}}
\newcommand{\vp}{\vec{p}}
\newcommand{\vq}{\vec{q}}

\newcommand{\dt}{\partial_t}

\newcommand{\bk}{b_{s_k}}
\newcommand{\bp}{b_{s_p}}
\newcommand{\bq}{b_{s_q}}
\newcommand{\uk}{u_{s_k}}
\newcommand{\up}{u_{s_p}}
\newcommand{\uq}{u_{s_q}}
\newcommand{\Bp}{B_{s_p}}
\newcommand{\Up}{U_{s_p}}

\newcommand{\g}{g_{kpq\:}}

\begin{document}
\shorttitle{Helical mode interactions in MHD turbulence} 
\shortauthor{M. Linkmann, A. Berera, M. McKay and J. J\"{a}ger} 

\title{Helical mode interactions and spectral transfer processes in magnetohydrodynamic turbulence}


\author
 {
 Moritz Linkmann
  \corresp{\email{m.linkmann@ed.ac.uk}},
  Arjun Berera
  \corresp{\email{ab@ph.ed.ac.uk}},
  Mairi McKay
  \and
  Julia J\"{a}ger
  }

\affiliation
{
School of Physics and Astronomy, University of Edinburgh, Edinburgh, EH9 3FD, UK
}

\date{\today}

\maketitle

\begin{abstract}
Spectral transfer processes in magnetohydrodynamic (MHD) turbulence are 
investigated analytically by decomposition of the velocity and magnetic fields in 
Fourier space into helical modes.  
Steady solutions of the dynamical system 
which governs the evolution of the helical modes are determined, and a stability 
analysis of these solutions is carried out. 
The interpretation of the analysis is that
unstable solutions lead to energy transfer between the
interacting modes while stable solutions do not. From this, a dependence of possible 
interscale energy and helicity transfers on the helicities of the interacting modes
is derived. 
As expected from the inverse cascade of magnetic helicity in 3D MHD turbulence, 
mode interactions with like helicities lead to transfer of energy and magnetic helicity to 
smaller wavenumbers. However, some interactions of modes with unlike helicities also contribute
to an inverse energy transfer. As such, an inverse energy cascade for nonhelical magnetic 
fields is shown to be possible.   
Furthermore, it is found that high values of the cross-helicity may have an 
asymmetric effect on forward and reverse transfer of energy, where forward
transfer is more quenched in regions of high cross-helicity than reverse transfer.
This conforms with recent observations of solar wind turbulence.
For specific helical interactions the relation to dynamo action is established. 
The present analysis provides new theoretical insights into physical
processes where inverse cascade and dynamo action are involved, such as the evolution of
cosmological and astrophysical magnetic fields and laboratory plasmas.
\end{abstract}


\section{Introduction}

Since the contributions by Richardson and Kolmogorov it is well established
that the average transfer of kinetic energy occurring in 
isotropic non-conducting turbulent fluids in three dimensions proceeds from the large scales
to the small scales, or, in the Fourier representation, from small to large wavenumbers 
(see e.~g.~\cite{Frisch95,McComb14a}). 
However, many turbulent flows occurring in nature and/or in 
industrial applications deviate from this behaviour,
showing some backwards energy transfer, such as rotating flows \citep{Mininni09a},
two-dimensional flows e.~g.~\citep{Kraichnan67,Boffetta10,Mininni13} as well as flows
doted with polymeric additives \citep{Dubief13}. 

Electrically conducting turbulent flows also show a variety of 
phenomena resulting in a transfer of energy from the small scales to the 
large scales \citep{Biskamp93}. 
One of these is the inverse cascade of magnetic helicity, 
first proposed by \citet{Frisch75} after the derivation of absolute equilibrium
spectra for magnetic and kinetic energies as well as cross and magnetic
helicities and subsequently confirmed numerically 
\citep{Pouquet76,Pouquet78}. By virtue of the realisability condition, which
states that magnetic energy at a given wavenumber $k$ is bounded from below
by $k/2$ times the magnitude of the magnetic helicity
(see e.~g.~\cite{Moffatt78,Biskamp93}), it also drives a transfer of magnetic 
energy from the small to the large scales.
Another process, which is of particular importance in astrophysical fluid dynamics 
due to the generation of magnetic fields of astrophysical bodies, is the large-scale dynamo,
by which a magnetic field on scales {\em larger} than the typical scale of the 
largest eddies is generated from a magnetic seed field. 
The $\alpha$-effect of mean-field electrodynamics is a classic example of a 
large-scale dynamo, and it relies on the presence of kinetic helicity \citep{Krause80,Brandenburg01}. 
Thus in both these cases energy is transferred in spectral space from 
large to small wavenumbers, and a lack of mirror symmetry 
(i.e.~the presence of kinetic and/or magnetic helicity)
facilitates these types of energy transfer.
Moreover, results from recent numerical simulations 
\citep{Brandenburg15,Zrake14,Berera14a} show that some kind of 
inverse spectral transfer also occurs in conducting flows with
vanishing magnetic and kinetic helicities. 
Recently, \citet{Stepanov15} calculated energy and helicity transfer fluxes numerically from 
a helical shell model, where helicity and energy input was separated in order to test
the influence of magnetic helicity on the turbulent dynamics. They found that the inverse 
cascade of magnetic helicity is connected to a transfer of purely 
magnetic energy to lower wavenumbers. By alternating the sign of helicity injected into 
the system it was shown that even if the average injection rate of magnetic helicity
vanishes, the reverse transfer of magnetic energy induced by the instantaneous injection
of magnetic helicity remains active.

The effect of helicity on energy transfer and evolution in non-conducting 
turbulent fluids has received considerable attention 
\citep{Moffatt69,Andre77,Pelz86,Polifke89,Polifke91,Waleffe92,Chen03a,
Chen03b,Biferale12,Biferale13a,Biferale13} 
and has been studied in a variety of ways,
e.~g.~using analytical methods, closure calculations, conventional direct 
numerical simulations (DNSs) and novel approaches in DNS. 
\citet{Waleffe92} decomposed the Fourier 
transform of the velocity field into eigenfunctions of the curl operator 
and derived evolution equations for these eigenfunctions 
by substitution of the decomposed field into the Navier-Stokes equation 
for incompressible flow. Since the nonzero eigenvalues of the corresponding 
eigenfunctions are related to the helicity of given velocity field mode,  
the evolution equations were further analysed in order to derive the dependence 
of the possible energy transfers on the helicities of the interacting modes. 
If the largest two wavenumbers of a given wavevector triad had 
helicities of opposite sign energy was transferred forward in wavenumber 
space, while a reverse transfer of energy became possible 
if the helicities were of the same sign.

\citet{Biferale12} investigated numerically whether 
this reverse spectral transfer caused by interactions of helical modes of the
 same sign occurs. By defining a projection operation on the nonlinear term 
the authors altered the Navier-Stokes equation so to ensure that only modes 
of, say, positive helicity were present in the system. The altered 
Navier-Stokes equation was subsequently solved numerically using the standard
pseudospectral method in conjunction with 
small-scale forcing, and, as predicted by Waleffe's analysis, it was found
that kinetic energy was indeed transferred downwards in wavenumber space. This
was the first observation of an inverse energy cascade in three-dimensional
isotropic turbulence.
 
In a subsequent paper \citep{Biferale13a}, the same authors forced the system
at the large scales in order to study the predicted forward cascade of kinetic
helicity, which was indeed observed in the simulations. Since the subset of 
positively helical modes does not transfer energy to the small scales, 
it was expected that the resulting dynamical system would not show finite
dissipation in the limit of infinite Reynolds number. Hence the projected Navier-Stokes
equation which governs the evolution should be globally regular, 
which was subsequently proven by \citet{Biferale13}. 

Thus, in summary, the decomposition of the Fourier transform of 
the turbulent velocity field fluctuations into helical modes has been proven 
to be very useful in terms of understanding some fundamental features of turbulent
flows which go beyond the established Kolmogorov-Richardson (direct) cascade of 
kinetic energy. 
In view of the effects of kinetic and magnetic helicities 
on the direction of energy transfer in MHD turbulence  
and inspired by the successes of the helical decomposition used
in hydrodynamics, in this paper we propose to use the decomposition 
of both the magnetic and velocity fields into helical modes in order 
to perhaps shed some more light on why magnetohydrodynamic (MHD) 
turbulence shows much more transfer from the small scales to the 
large scales than turbulence in non-conducting fluids. 
This analysis provides deeper insight into fundamental transfer processes in MHD turbulence. 
It will help in the further theoretical understanding of various physical processes 
involving inverse cascade and dynamo action, such as the evolution and generation of 
cosmological and galactic magnetic fields.



Before outlining the structure of this paper we pause briefly to discuss 
the terminology used.
As the precise meaning of the term varies in the literature, 
it is not always evident what is meant by an inverse cascade.
In the astrophysical literature, 
transfer of energy and helicity from higher to lower wavenumbers is often described as an 
inverse cascade 
\citep{Son99,Christensson01,Cho11}, 
while the fluid dynamics literature requires any 
cascade process to possess a wavenumber-independent flux  
\citep{McComb14a,Biferale12,Biferale13,Mueller12,Alexakis06,Brandenburg15}. 
It is thus of interest to not only classify the different types of reverse 
transfer that can occur in MHD turbulence, but also to perhaps clarify the 
terminology. Therefore we propose in this paper the general term reverse (or inverse) spectral
transfer, which encompasses all the phenomena described above as subcategories 
according to their properties.
We define it as any process that produces an increase in a spectral quantity 
(total energy, magnetic helicity, etc.) at low wavenumbers due to transfer of that quantity away from higher 
wavenumbers into smaller wavenumbers. In this framework an inverse cascade is a 
reverse spectral transfer showing constant flux of the cascading quantity over a 
certain wavenumber range. We point out that in MHD concerns have been raised
over the use of the term `cascade' \citep{Mueller12}, as it may be understood to imply energy 
(or magnetic helicity) transfer mainly due to local interactions, 
which might not be the case in MHD turbulence 
\citep{Alexakis05a,Debliquy05,Cho10,Mueller12}. 
We will come back to this point in the discussion section of this paper.    
 
This paper is organised as follows. First we explain the helical 
decomposition of the velocity and magnetic fields in 
sec.~\ref{sec:decomposition} and use this decomposition to outline the 
derivation of the evolution equations for the helical coefficients from the MHD
 equations in sec.~\ref{sec:MHD_helical} following the work by 
\citet{Lessinnes09}. This leads to a system of 
coupled ordinary differential equations (ODEs) describing the interaction of modes corresponding to a single 
triad of wavenumbers.
In sections \ref{sec:steady_state} and \ref{sec:instability} the linear stability of 
steady solutions of this system is examined.
Section \ref{sec:instab_assump} explains the assumptions used to interpret results from 
the stability analysis followed by a presentation of first results on energy transfers, 
which are used in sec.~\ref{sec:fluxes} to calculate the contribution to inertial range energy 
and magnetic helicity cascades. 
In sec.~\ref{sec:dynamo} we analyse specific cases where the MHD 
equations can be decoupled and relate our 
analysis to large- and small-scale kinematic dynamo results such as the 
$\alpha$-effect \citep{Moffatt78, Parker79, Krause80}. 
Our results are summarised and discussed in sec.~\ref{sec:discussion} 
in the context of numerical simulations of MHD turbulence and observations of turbulence 
in the solar wind, at this point we also provide suggestions for further 
work. 

\section{Problem statement and basic equations}
\label{sec:decomposition}
In this paper we will be studying the interscale energy and helicity transfers
that govern the dynamics of MHD turbulence in the absence of a background magnetic
field. We consider the turbulent flow to be incompressible and make no explicit assumptions
about the value of the magnetic Prandtl number. The MHD equations for incompressible
flow are
\begin{align}
\label{eq:momentum}
\partial_t \vec{u}&= - \frac{1}{\rho}\nabla P -(\vec{u}\cdot \nabla)\vec{u}
 + \frac{1}{\rho}(\nabla \times \vec{b}) \times \vec{b} + \nu \Delta \vec{u}  \ , \\
\label{eq:induction}
\partial_t \vec{b}&= (\vec{b}\cdot \nabla)\vec{u}-(\vec{u}\cdot \nabla)\vec{b} + \eta \Delta \vec{b}\ , \\
\label{eq:incompr}
&\nabla \cdot \vec{u} = 0 \ \ \mbox{and} \ \  \nabla \cdot \vec{b} = 0 \ ,  
\end{align}
where $\vec{u}$ denotes the velocity field, $\vec{b}$ the magnetic induction
expressed in Alfv\'{e}n units, $\nu$ the kinematic viscosity, $\eta$ the
resistivity, $P$ the pressure and $\rho$ the density, which is set to unity for convenience.

For simplicity at first we consider periodic boundary conditions 
on a domain $\Omega = [0,L]^3 \subset \mathbb{R}^3$, thus working with the discrete 
Fourier transformed MHD equations 
\begin{align}
\label{eq:Fmomentum}
(\partial_t+\nu k^2) \fvec{u}(\vec{k}) = & \ -FT \left[\nabla \left( P+\frac{|\vec{u}|^2}{2}\right)\right] \nonumber \\
& +\sum_{\vec{k}+\vec{p}+\vec{q}=0} \left[ -(i\vec{p}\times \fvec{u}(\vec{p}))\times \fvec{u}(\vec{q})+(i\vec{p}\times \fvec{b}(\vec{p}))\times \fvec{b}(\vec{q})\right] \ , \\
\label{eq:Finduction}
(\partial_t+\nu k^2) \fvec{b}(\vec{k}) = & \  i\vec{k} \times \sum_{0=\vec{k}+\vec{p}+\vec{q}} \fvec{u}(\vec{p}) \times \fvec{b}(\vec{q}) \ , 
\end{align}
where $FT$ denotes the three-dimensional Fourier transform as a linear operator
acting on $L^2(\Omega)$ functions, $\fvec{u}$ and $\fvec{b}$ denote the 
Fourier transforms of the velocity and magnetic fields respectively and the inertial
term $(\vec{u}\cdot \nabla)\vec{u}$ in the momentum equation \eqref{eq:momentum} has been written in rotational
form 
$(\vec{u}\cdot \nabla)\vec{u} = (\nabla \times \vec{u}) \times \vec{u} + 0.5\nabla |\vec{u}|^2$.
In order to determine the contribution of specific interactions to 
the fluxes of magnetic helicity and magnetic energy, eventually we will formally take 
the limit $L \to \infty$ in sec.~\ref{sec:fluxes}, necessarily assuming 
that the relevant functions are then well-behaved at infinity to ensure 
the convergence of the respective Fourier integrals 
\footnote{A discussion of this point can be found in the book by \citet{Titchmarsh39}}.

\subsection{Helical decomposition}
The decomposition of the Fourier transform of a solenoidal 
vector field in circularly polarised waves as  
proposed by \citet{Constantin88} has been
used in several investigations of hydrodynamic turbulence
\citep{Waleffe92,Biferale12,Biferale13} in order to establish
the properties of energy transfer depending on the kinetic
helicity. For conciseness we only 
review the fundamental properties of the helical decomposition
and refer to the relevant literature for details and derivations.

The action of the curl operator on square integrable real vector field 
$\vec{v}(\vx)$ can be viewed in spectral space 
as the action of a linear operator acting on the Fourier transform $\fvec{v}(\vk)$
of $\vec{v}(\vx)$,
\begin{align}
I_k: \mathbb{C}^3 & \longrightarrow \mathbb{C}^3 \nonumber \\
       \fvec{v}(\vec{k}) & \longrightarrow i\vec{k} \times \fvec{v}(\vec{k})  \ .\nonumber
\end{align} 
As such the linear operator $I_k(\cdot) = i\vec{k} \times (\cdot)$ 
has a set of linearly independent eigenvectors defining a  
basis of $\mathbb{C}^3$, thus $\fvec{v}(\vec{k})$ can be expanded in this 
basis. That is, it can be expressed as a linear combination of eigenvectors 
$i\vec{k}$, $\vec{h}_{+}(\vec{k})$ and $\vec{h}_{-}(\vec{k})$ of the curl operator $I_k$, where  
\begin{equation}
\label{eq:eigenvec}
i\vec{k}\times \vec{h}_{s_k}=s_k k \vec{h}_{s_k} \ ,
\end{equation} 
\begin{equation}
-i\vec{k}\times \vec{h}_{s_k}^{*}=s_k k \vec{h}_{s_k}^* \ ,
\end{equation}
$s_k = \pm 1$ and $s_k k = \pm k$ are the nonzero eigenvalues of the curl 
operator in spectral space 
\footnote{Note that the curl operator can have eigenvectors with nonzero 
eigenvalues, as it involves the cross product of two {\em complex} vectors. This 
is not necessarily orthogonal to the plane spanned by the two complex vectors, instead 
it is orthogonal to the plane spanned by the {\em complex conjugates} of the two vectors.}, 
and $*$ denotes the complex conjugate. 
The complex eigenvectors are orthogonal to each other 
and are fully helical. They are normalised to unit vectors
for the remainder of this paper. 

Since $\fvec{u}(\vec{k})$ and $\fvec{b}(\vec{k})$ 
are solenoidal, they can be expressed in terms of $\vec{h}_-$, $\vec{h}_+$ 
only
\beq
\fvec{u}(\vec{k},t)=u_-(\vec{k},t) \vec{h}_-(\vec{k}) + u_+(\vec{k},t) \vec{h}_+(\vec{k})
=\sum_{s_k} u_{s_k}(\vec{k},t) \vec{h}_{s_k}(\vec{k}) \ ,  
\label{eq:basis_u}
\eeq
\beq
\fvec{b}(\vec{k},t)=b_-(\vec{k},t) \vec{h}_-(\vec{k}) + b_+(\vec{k},t) \vec{h}_+(\vec{k})
=\sum_{s_k} b_{s_k}(\vec{k},t) \vec{h}_{s_k}(\vec{k}) \ , 
\label{eq:basis_b}
\eeq
where $u_{s_k}$ and $b_{s_k}$ are complex coefficients. These coefficients 
can be calculated by taking the inner product of the basis vectors with the appropriate fields
\begin{equation}
u_{s_k}(\vec{k},t)=\frac{\vec{h}^*_{s_k}(\vec{k})\cdot \vec{u}(\vec{k},t)}{\vec{h}_{s_k}(\vec{k})\cdot \vec{h}_{s_k}^*(\vec{k})} \ ,
\end{equation}
and
\begin{equation}
b_{s_k}(\vec{k},t)=\frac{\vec{h}^*_{s_k}(\vec{k})\cdot \vec{b}(\vec{k},t)}{\vec{h}_{s_k}(\vec{k})\cdot \vec{h}_{s_k}^*(\vec{k})} \ .
\end{equation}
In order to keep the derivations concise we suppress the dependence of 
the coefficients on time and wavevector from now on in the notation. 

The helical decompositon of a solenoidal vector field was first applied to 
incompressible MHD flows by \citet{Lessinnes09}, 
who derived a dynamical system in Fourier space describing helical triadic 
interactions in MHD. 
This system was subsequently used to construct a helical shell model of MHD turbulence. 
In the following section we briefly review the derivation 
carried out by \citet{Lessinnes09}.
  
\section{The evolution of the helical modes}
\label{sec:MHD_helical}
The equations describing the evolution of the helical coefficients 
$\uk$ and $\bk$ are derived by substituting the decompositions 
\eqref{eq:basis_u} and \eqref{eq:basis_b} into the MHD equations 
for incompressible flow and 
then taking the inner product with $\vec{h}_{s_k}$ on both sides of
the respective equations.
The resulting evolution equation for the helical coefficient 
$u_{s_k}$ is
\begin{align}
\label{eq:a_equation}
(\partial_t+\nu k^2) u_{s_k}= & \
\frac{\vec{h}_{s_k}^*}{2} {\bm \cdot}
\left( -FT \left[\nabla \left( P+\frac{|\vec{u}|^2}{2}\right)\right] \right) \nonumber \\
&+\frac{\vec{h}_{s_k}^*}{2} {\bm \cdot} \sum_{\vec{k}+\vec{p}+\vec{q}=0} \left[ -(i\vec{p}\times \fvec{u}(\vec{p}))\times \fvec{u}(\vec{q})+(i\vec{p}\times 
\fvec{b}(\vec{p}))\times \fvec{b}(\vec{q})\right]  \nonumber \\ 
= & \ -\frac{1}{2}\sum_{s_p, s_q} \sum_{0=\vec{k}+\vec{p}+\vec{q}} (s_p  p-s_q  q)\left[\vec{h}_{s_p}^* 
\times \vec{h}_{s_q}^* \cdot \vec{h}_{s_k}^* \right] (u_{s_p}^* u_{s_q}^* - b_{s_p}^* b_{s_q}^*)  \ , 
\end{align}
where the dummy variables $\vec{p}$ and $\vec{q}$ were exchanged in order to 
symmetrise the momentum equation with respect to $\vec{p}$ and 
$\vec{q}$ and thus to obtain the factor $(s_p  p-s_q  q)/2$. 
Following an analogous procedure \citep{Lessinnes09} for the helical 
coefficient $b_{s_k}$ of the magnetic field 
leads to
\begin{align}
\label{eq:c_equation}
(\partial_t+\eta k^2) b_{s_k} &=  \frac{\vec{h}^*_{s_k}}{2}\left[i\vec{k} 
\times \sum_{0=\vec{k}+\vec{p}+\vec{q}} \fvec{u}(\vec{p}) \times \fvec{b}(\vec{q})\right] \nonumber \\ 
&=  \frac{s_k k}{2} \sum_{s_p, s_q} \sum_{0=\vec{k}+\vec{p}+\vec{q}} \left[ \vec{h}_{s_p}^* 
\times \vec{h}_{s_q}^* \cdot \vec{h}_{s_k}^* \right](u_{s_p}^{*} b_{s_q}^{*}-b_{s_p}^{*} u_{s_q}^{*}) \ .
\end{align}

In order to study the interaction of helical modes, that is the evolution of 
the helical coefficients due to the mode coupling only, the diffusivities are 
from now on omitted. 
For a given triad $\vec{k},\vec{p},\vec{q}$ of wavevectors, 
expressions for the first time-derivatives of each helical coefficient are obtained 
from \eqref{eq:a_equation} and \eqref{eq:c_equation} and 
from the corresponding equations
for $\bp$, $\bq$, $\up$ and $\uq$. This leads to the following system of coupled ODEs
describing the evolution of the helical coefficients in a single triad interaction
\begin{align}
\dt \uk &= (s_pp-s_qq)\: \g (\up^* \uq^* - \bp^* \bq^*) \ , \nonumber \\
\dt \up &= (s_qq-s_kk)\: \g (\uq^* \uk^* - \bq^* \bk^*) \ , \nonumber \\
\dt \uq &= (s_kk-s_pp)\: \g (\uk^* \up^* - \bk^* \bp^*) \ , 
\label{eq:a_eqs_coup} \\
\dt \bk &= -s_kk\:\g (\up^* \bq^* - \bp^* \uq^*) \ , \nonumber \\
\dt \bp &= -s_pp\:\g (\uq^* \bk^* - \bq^* \uk^*) \ , \nonumber \\
\dt \bq &= -s_qq\:\g (\uk^* \bp^* - \bk^* \up^*) \ ,
\label{eq:c_eqs_coup}
\end{align}
where the geometric factor 
\beq
\label{eq:geom_factor}
g_{kpq}=-\frac{1}{2}\vec{h}_{s_p}^* \times \vec{h}_{s_q}^* \cdot \vec{h}_{s_k}^* \ ,
\eeq
is introduced for conciseness, following \citet{Waleffe92} and \citet{Lessinnes09}. 
It can also be written as
\beq
\label{eq:geo_fac}
g_{kpq}=\frac{s_k s_p s_q}{2} e^{i\alpha(k,p,q)}\frac{N}{2kpq}(s_k k + s_p p + s_q q) \ ,
\eeq
where $\alpha$ is a wavenumber-dependent real number determined by the 
orientation of the triad and $N$ a factor depending on the shape of the triad. 
Further details and a derivation of \eqref{eq:geo_fac} can be found in
the paper by \citet{Waleffe92}.

The three ideal invariants total energy, magnetic helicity and cross-helicity are defined respectively as
\begin{align}
E_{tot} =& \frac{1}{2}\sum_{\vec{k}} \langle |\fvec{u}(\vec{k})|^2 + |\fvec{b}(\vec{k})|^2 \rangle =\frac{1}{2}\sum_{\vec{k}, s_k} \left(|\uk|^2+|\bk|^2 \right) \ , \\
H_{mag} =& \sum_{\vec{k}} \langle \fvec{a}(\vec{k})\fvec{b}(-\vec{k}) =\sum_{\vec{k}, s_k} \frac{s_k}{k} |\bk|^2 \ , \\ 
H_{c} =& \sum_{\vec{k}} \langle \fvec{u}(\vec{k})\fvec{b}(-\vec{k})\rangle =\sum_{\vec{k}, s_k} \mbox{Re}\left(\uk\bk^*\right) \ , 
\end{align}
where $\vec{a}$ denotes the vector potential of the magnetic field, Re the real part of a 
complex number and angle brackets an ensemble average. They are conserved in single triad 
interactions \citep{Lessinnes09}. 

\section{Stability of steady solutions}
\label{sec:steady_state}
Examining the linear stability of steady solutions of the system 
\eqref{eq:a_eqs_coup}-\eqref{eq:c_eqs_coup} can reveal the influence which
the helicities of the interacting modes have on the interscale
transfer of a given quantity of interest. 

The system \eqref{eq:a_eqs_coup} without a 
magnetic field (that is for $b_{s}=0$) was analysed by \citet{Waleffe92}
with respect to the linear stability of its steady solutions. 
Linearly unstable solutions were found  
depending on the helicities of the interacting modes.
This result was then interpreted following the {\em instability assumption}
inspired by the formal analogy to rigid-body rotation,
where rotation around the axis of middle inertia is unstable.
The existence of a linearly unstable solution involving a velocity 
field mode $\fvec{u}$ is interpreted as
the $\fvec{u}$-mode losing energy to the other two modes it interacts with.
An equivalent assumption had already been used by \citet{Kraichnan67} 
for two-dimensional hydrodynamic turbulence.
In the remainder of this paper we take a similar approach and investigate 
the linear stability of steady solutions of the system 
\eqref{eq:a_eqs_coup}-\eqref{eq:c_eqs_coup}
in view of possible applications to spectral transfer processes in MHD and 
in particular for the inverse transfers of total energy and magnetic helicity. 
In principle, a similar analysis could be carried out for the remaining
ideal invariant, the cross-helicity. 

\subsection{The steady solutions}
The system \eqref{eq:a_eqs_coup}-\eqref{eq:c_eqs_coup} 
of six coupled ODEs has several steady solutions one can linearise about.
To simplify the notation, a (formal) solution of the system 
\eqref{eq:a_eqs_coup}-\eqref{eq:c_eqs_coup} consisting of
helical $\fvec{u}$- and $\fvec{b}$-field 
modes interacting in a given triad $\vk,\vp,\vq$ is written as:
\begin{equation} (u_{s_k},u_{s_p},u_{s_q};b_{s_k},b_{s_p},b_{s_q}) \ . 
\label{eq:notation}
\end{equation} 

In order to find the steady solutions of the system 
\eqref{eq:a_eqs_coup}-\eqref{eq:c_eqs_coup}, we assume (without loss of generality) 
that the middle components $b_{s_p}=B_{s_p}$ 
and $u_{s_p}=U_{s_p}$ are constant in time. Then \eqref{eq:a_eqs_coup} and 
\eqref{eq:c_eqs_coup}
require the other four components to vanish by the following argument. 
A steady solution requires $\partial_t{u}_{s_k}=0$, and 
the only way that this can happen nontrivially is if both products 
$u_{s_p}^*u_{s_q}^*$ and $b_{s_p}^*b_{s_q}^*$ vanish 
\footnote{This requires assuming that no cancellations occur. However, 
the occurrence of cancellations would require the system to be in a specific state, 
which is unlikely to happen frequently in a chaotic system.}. 
Since $u_{s_p}=U_{s_p}$ is constant in time, 
$u_{s_q}=0$ and similarly $b_{s_q}=0$. This leaves us with
\begin{equation} 
(u_{s_k},U_{s_p},0;b_{s_k},B_{s_p},0) \ .\nonumber
\end{equation} 
Applying the same argument to $\partial_t{u}_{s_q}$, it follows that 
$u_{s_k}$ and $b_{s_k}$ must also vanish. Therefore a steady solution
of the system \eqref{eq:a_eqs_coup}-\eqref{eq:c_eqs_coup} has the form
\begin{equation} 
(0,U_{s_p},0;0,B_{s_p},0) \ . \nonumber
\end{equation} 
It can now be checked for consistency that $\partial_t{b}_s=0$ for 
$k$, $p$ and $q$ as well. Therefore the solution is steady for the magnetic 
field and for the velocity field alike. 
Asides from the just explained example, steady solutions of the form 
$(U_{s_k},0,0;B_{s_k},0,0)$ and $(0,0,U_{s_q};0,0,B_{s_q})$
are obtained in the same way.

Thus the steady solutions of \eqref{eq:a_eqs_coup}-\eqref{eq:c_eqs_coup} 
are of the same form as for the hydrodynamic case \citep{Waleffe92}, where 
at least two of the three interacting modes vanish. However, there are two special cases: 
one where the magnetic field component $B_s$ also vanishes, while $U_s \neq 0$ 
and the other, where the velocity field component $U_s$ vanishes, 
while $B_s \neq 0$. The former case may perhaps be connected to a dynamo process. 
At this point we note that for the kinematic dynamo, where the back-reaction 
of the magnetic field on the velocity field can be neglected, the (linear) 
stability of the velocity field coefficients $u_s$ is only determined by 
hydrodynamic interactions. We will come back to this 
point in sec.~\ref{sec:dynamo}. 

\subsection{Linear stability analysis}
\label{sec:stability}
In order to assess whether a given steady solution is linearly
stable in our particular setting, we assume without loss of generality that the 
coefficients $u_{s_p}$ and $b_{s_p}$ corresponding to wavevector $\vec{p}$ are 
nonzero and constant in time, that is, we study the linear stability 
of the solution $(0,U_{s_p},0; 0, B_{s_p},0)$ 
with respect to infinitesimal perturbations of the four modes that
had been set to zero.
As the first-order equations involve the coupling of all three modes
of a given triad, little information can be obtained from them at first
sight. Therefore we pass to second-order time-derivatives of the evolution 
equations for the perturbations $\uk$, $\bk$, $\uq$ and $\bq$. 
Taking time-derivatives on both sides of 
\eqref{eq:a_eqs_coup}-\eqref{eq:c_eqs_coup} and subsequently
substituting any occurrence of a first-order time-derivative on the right-hand side
by the appropriate evolution equation, we obtain  
\begin{align}
\label{eq:a_second_der}
\dt^2\uk&=|\g|^2(s_pp-s_qq)\left[\big((s_kk-s_pp)|\Up|^2+s_qq\:|\Bp|^2\big)\uk \right] \nonumber \\
        & - |\g|^2(s_pp-s_qq)\left[\big((s_kk-s_pp)\Up^* \Bp + s_qq\:\Up\Bp^*\big)\bk\right] \ , \\
\label{eq:c_second_der}
\dt^2\bk&=|\g|^2s_kk\left[\big(s_qq\:\Up^*\Bp+(s_kk-s_pp)\Up\Bp^*\big)\uk \right] \nonumber \\
& -|\g|^2s_kk\left[\big(s_qq\:|\Up|^2 + (s_kk-s_pp)|\Bp|^2\big)\bk\right] \ ,
\end{align}
where terms of second order in small quantities (such as e.g. $u_s^2$) 
have been neglected. Note that these equation do not depend on modes at wavenumber $q$. 
The evolution equations of the helical coefficients $\uq$ and $\bq$ can be obtained 
similarly and show no dependence on $k$, therefore we restrict our attention to 
the evolution of $\uk$ and $\bk$. 

The system \eqref{eq:a_second_der} and \eqref{eq:c_second_der} can be written as a matrix ODE
\begin{equation}
\ddot{\vec{x}}
= 
  \begin{pmatrix}
    \alpha & \beta \\
    \gamma & \delta
  \end{pmatrix}
\vec{x} \ ,
\label{eq:matrix_ode}
\end{equation}
where $\vec{x} \equiv (u_{s_k},b_{s_k})$ and the matrix elements are
\begin{align}
  \alpha &= |\g|^2(s_pp-s_qq)\big[(s_kk-s_pp)|\Up|^2+s_qq\:|\Bp|^2\big] \ , \\
  \beta &= -|\g|^2(s_pp-s_qq)\big[(s_kk-s_pp)\Up^* \Bp + s_qq\:\Up\Bp^*\big] \ , \\
  \gamma &= |\g|^2s_kk \: \big[s_qq\:\Up^*\Bp+(s_kk-s_pp)\Up\Bp^*\big] \ , \\
  \delta &= -|\g|^2s_kk \:\big[s_qq\:|\Up|^2 + (s_kk-s_pp)|\Bp|^2\big] \ .
\label{eq:entries}
\end{align}
The linear stability of this system can be determined from the 
eigenvalues $\lambda_1$ and $\lambda_2$ of the matrix in \eqref{eq:matrix_ode}. 
These eigenvalues depend not only on the helicities of the interacting modes and on the 
magnitudes of $U_{s_p}$ and $B_{s_p}$ relative to each other, but also on the 
alignment between the magnetic and velocity field modes at wavevector 
$\vec{p}$, that is, on the cross-helicity.
For a given steady solution to be unstable the perturbations have 
to be exponentially growing, and so at least one of the eigenvalues 
$\sqrt{\lambda_i}$ (for $i=1,2$) must have a positive real part. We will 
now assess under which conditions this is possible. 

The eigenvalues $\lambda_i$ ($i=1,2$) are given by
\begin{equation}
  \lambda_{1,2} = \frac{\alpha + \delta}{2} \pm \sqrt{\frac{(\alpha + \delta)^2}{4} - \alpha \delta + \beta \gamma} \ .
\end{equation}
For convenience define 
\beq
x \equiv  \frac{\alpha + \delta}{2} \ \mbox{ and } \ Q\equiv  \alpha \delta - \beta \gamma \ ,
\eeq
such that 
\begin{align}
x =& \ -\frac{|g_{kpq}|^2}{2}|U_{s_p}|^2[s_kk s_qq + (s_k k -s_p p)(s_q q -s_p p)] \nonumber \\
& \ -\frac{|g_{kpq}|^2}{2}|B_{s_p}|^2[s_k k(s_k k -s_p p) + s_q q(s_q q -s_p p)] \ ,
\label{eq:aplusd}
\end{align}
and 
\beq
Q=|g_{kpq}|^4 s_kk s_qq (s_k k -s_p p)(s_q q -s_p p)\left (|U_{s_p}|^4+|B_{s_p}|^4+2|\Up|^2|\Bp|^2-4H_c(p)^2) \right) \ ,
\label{eq:Q}
\eeq 
hence the cross-helicity $H_c(p)$ enters the dynamics through the parameter $Q$.
The derivation of \eqref{eq:Q} can be found in appendix \ref{app:crosshel}. 
Note that the term $|U_{s_p}|^4+|B_{s_p}|^4+2|\Up|^2|\Bp|^2-4H_c(p)^2$ is always positive,
regardless of the value of $H_c$ since $|H_c(p)| \leqslant |\Up||\Bp|$, thus the sign of $Q$
is determined by the helicities of the interacting modes and the wavenumber ordering.

The eigenvalues $\lambda_i$ can now be written more concisely as
\beq
\lambda_{1,2} = x \pm \sqrt{x^2-Q} \ ,
\eeq 
therefore the possibility of finding exponential solutions of the system 
\eqref{eq:matrix_ode} depends on the values of $x$ and $Q$. 
Apart from the trivial case, where $x=0$ and $Q=0$, there is only one case for 
which no linear instability occurs: this is if $x<0$ and 
$|x| > |\sqrt{x^2-Q}|$, since then  
$\sqrt{\lambda_1}$ and $\sqrt{\lambda_2}$ are imaginary numbers 
allowing only oscillatory solutions of the matrix ODE \eqref{eq:matrix_ode}. 
All other cases lead to exponentially growing as well as exponentially 
decaying solutions.

Cases in which $x>0$ and $Q<0$ result in the largest eigenvalues and thus in 
the fastest growing exponential solution. These cases are therefore of special 
interest, as within the framework of the instability assumption they may lead to 
the largest energy transfer and thus can yield 
information about which combination of parameters facilitates most of the 
energy and helicity transfers. We will consider this point in further detail in 
sec.~\ref{sec:fluxes}.

As can be seen from the structure of the terms $x$ and $Q$, the relative magnitudes
and the ordering of the wavenumbers in a given triad will influence the stability of
steady solutions. 
In view of the continuous interest in nonlocality of interactions in MHD turbulence 
\citep{Brandenburg01,Alexakis05a,Debliquy05,Cho10,Mueller12}, 
we point towards specific results for local and nonlocal interactions
where appropriate. 
Following \citet{Waleffe92}, for wavenumbers ordered
$k < p < q$, the nonlocal limit is defined as $k<<p \simeq q$, while local interactions
are characterised by $k \simeq p \simeq q$.  

\section{Instability and helical interactions}
\label{sec:instability}
Since $s=\pm 1$, interactions between helical modes which all 
have helicities of opposite signs are not possible, and at least 
two modes will always have helicities of the same sign. 
Therefore we have four classes of possible helicity combinations  
\beq
s_k = s_p \neq s_q \ , \ \  s_k=s_q \neq s_p \ , \ \ s_k \neq s_q = s_p \ \ \mbox{ and } \ s_k=s_q = s_p \ , \nonumber 
\eeq
each of which occur twice as $s$ can take the values $\pm 1$.
These four possible (classes of) combinations are now studied on a case-by-case approach
in order to determine when a certain combination of helicities leads to exponentially growing
solutions of the system \eqref{eq:matrix_ode}.

\subsection{The case $s_k = s_q \neq s_p$}
\label{sec:no_trans}
Since the expressions in square brackets of \eqref{eq:aplusd} become
\beq
kq + (k + p)(q + p) > 0 \ \text{ and } k(k + p) + q(q + p) > 0 \ ,
\eeq
one obtains $x=(\alpha + \delta)/2 < 0$. For an unstable solution 
$|x| < |\sqrt{x^2-Q}|$, however, we obtain $Q>0$ since
\beq
Q \sim s_k s_q kq (s_k k -s_p p)(s_q q -s_p p) \ ,
\eeq
which is positive for $s_k = s_q \neq s_p$. Furthermore we obtain $Q < x^2$ (see appendix
\ref{app:Q_calc}) and thus 
$|x| > |\sqrt{x^2-Q}|$, which results in imaginary eigenvalues of the matrix in 
\eqref{eq:matrix_ode}. 
Therefore we do not obtain unstable solutions for the case
$s_k = s_q \neq s_p$, and this is independent of the ordering of the
wavenumbers $k, p$ and $q$. Note that this implies that exponentially growing
solutions of \eqref{eq:matrix_ode} are impossible if the perturbations 
$u_{s_k}$, $\uq$, $\bk$ and $\bq$
have helicities opposite to the helicities of the modes $\Up$ and $\Bp$ 
constituting the steady solution. 


For the remaining helicity combinations, which do result in unstable solutions,
the ordering of wavenumbers matters. 
The arguments used to decide whether or not an exponentially growing solution 
becomes possible are similar to the procedure employed for the case 
$s_k = s_q \neq s_p$ described above. 

\subsection{The case $s_k \neq s_p = s_q$}
In this case we obtain 
\beq
Q \sim kq (k+p)(q-p) 
\eeq
and
\beq
x=-\frac{|g_{kpq}|^2}{2}|U_{s_p}|^2[-kq - (k + p)(q - p)] -\frac{|g_{kpq}|^2}{2}|B_{s_p}|^2[k(k + p) + q(q - p)] \ .
\eeq
The stability of a steady solution depends on the signs of these terms which 
in turn depend on wavenumber ordering, cross-helicity and the ratio $|\Up|/|\Bp|$.

\begin{itemize}
\item For $k<p<q$ we obtain unstable solutions if $|U_{s_p}| > |B_{s_p}|$, since
then $x>0$. For $|B_{s_p}| > |U_{s_p}|$ unstable solutions are still 
possible, provided $H_c(p)$ is small and $|B_{s_p}|$ not much larger than $|U_{s_p}|$. 
Thus in regions of large cross-helicity unstable solutions only occur for
weak magnetic fields.  
The method by which these results are obtained is explained 
in appendix \ref{app:graphs}. 

For nonlocal interactions ($k << p \simeq q$) 
we obtain $Q=0$ and the sign of $x$ determines whether unstable 
solutions occur. The term $x$ is now of the form
\beq
x \simeq \frac{|g_{kpq}|^2}{2}kq (|U_{s_p}|^2 -|B_{s_p}|^2)\ ,
\eeq
hence nonlocal interactions lead to unstable solutions if
$|B_{s_p}| < |U_{s_p}|$.

\item For $k<q<p$, 
$Q$ will become negative, leading to unstable
solutions regardless of the ratio $|U_{s_p}|/|B_{s_p}|$ and the value of $H_c(p)$.

\item For $p<k<q$ again we obtain unstable solutions if $|U_{s_p}| > |B_{s_p}|$, since
then $x>0$. For $|B_{s_p}| > |U_{s_p}|$ unstable solutions are still
possible, provided $H_c(p)$ is small and $|B_{s_p}|/|U_{s_p}|$ not $>> 1$ 
(see appendix \ref{app:graphs}).
Nonlocal interactions ($p<<k\simeq q$) lead to unstable solutions if $|\Up| > |\Bp|$, because
then
\beq
x \simeq |g_{kpq}|^2 k^2(|U_{s_p}|^2 -|B_{s_p}|^2) > 0 \ .
\eeq
\end{itemize}
In summary, a given steady solution in this case is more 
likely to be stable if the nonzero mode is at medium or low 
wavenumbers in regions of high cross-helicity.

\subsection{The case $s_k = s_p = s_q$}
In this case we obtain
\beq
Q \sim kq (k-p)(q-p)  
\eeq
and
\beq
x = -\frac{|g_{kpq}|^2}{2}|U_{s_p}|^2[kq + (k - p)(q - p)] 
-\frac{|g_{kpq}|^2}{2}|B_{s_p}|^2[k(k - p) + q(q - p)]  \ .
\eeq
\begin{itemize}
\item For $k<p<q$ we obtain $Q<0$ and thus $x+\sqrt{x^2-Q} \geqslant 0$,
leading to exponentially growing solutions independent of $H_c(p)$ and the ratio 
$|U_{s_p}|/|B_{s_p}|$. 
We note that both velocity and magnetic field modes have positive and negative 
contributions to the sign of $x$. This is of 
interest since if $x>0$ the resulting eigenvalue would be larger and thus the 
solution would grow faster. However, in this case this cannot
be determined from the ratio $|U_{s_p}|/|B_{s_p}|$ and thus there is 
little information about what contributes to a faster growing exponential 
and thus to a more unstable solution. 
 
For both local ($k \simeq p \simeq q$) and nonlocal ($k << p \simeq q$)
interactions we obtain $Q=0$ and the sign of $x$ determines whether unstable 
solutions occur. For the nonlocal case only the magnetic field term is positive, 
and $x$ has the form
\beq
x \simeq \frac{|g_{kpq}|^2}{2}kq (|B_{s_p}|^2 -|U_{s_p}|^2)\ .
\eeq
leading to unstable solutions if 
$|B_{s_p}|>|U_{s_p}|$, while for local interactions 
no instability occurs as the only term in $x$ that does not vanish is 
$-|g_{kpq}|^2|U_{s_p}|^2kq < 0$.  

\item For $k<q<p$, the possibility of exponentially
growing solutions depends on the ratio $|U_{s_p}|/|B_{s_p}|$ and on the 
relative magnitudes of the wavenumbers $k$,$p$ and $q$, as now $Q>0$. 
Since the magnetic field term in $x$ is now positive, instabilities
occur for $|U_{s_p}|/|B_{s_p}|<1$. If $|U_{s_p}|/|B_{s_p}|>1$ it depends 
also on the cross-helicity whether instabilities occur.  
For maximal $H_c(p)$ we obtain $x^2 - Q >0$, hence the solutions will be stable.
If $H_c(p) = 0$ and $|U_{s_p}|/|B_{s_p}|$ is not too small, instabilities will
occur, depending also on the shape of the triad (see appendix \ref{app:graphs} 
for further details).
In general, the smaller $|U_{s_p}|/|B_{s_p}|$ the more unstable is the solution.

\item For $p<k<q$ we obtain $x<0$ and $Q>0$, furthermore 
$x^2-Q > 0$ independent of $|U_{s_p}|/|B_{s_p}|$ and $H_c(p)$ 
(see appendix \ref{app:graphs}), thus no unstable solutions occur.
Nonlocal interactions ($p<<k\simeq q$) do not lead
to unstable solutions, since 
\beq
x \simeq -|g_{kpq}|^2 [k^2-kp](|U_{s_p}|^2 +|B_{s_p}|^2) < 0 \ .
\eeq

\end{itemize}

\subsection{The case $s_k = s_p \neq s_q$}
The terms determining the stability in this case are
\beq
Q \sim kq(k-p)(q+p) 
\eeq
and
\beq
x = -\frac{|g_{kpq}|^2}{2}|U_{s_p}|^2[-kq - (k - p)(q + p)]  
-\frac{|g_{kpq}|^2}{2}|B_{s_p}|^2[k(k - p) + q(q + p)]  \ .
\eeq
\begin{itemize}
\item For  $k<p<q$ unstable solutions occur
independent of the ratio $|U_{s_p}|/|B_{s_p}|$, and since both magnetic
and velocity field terms have positive and negative contributions 
to the sign of $x$, we are in 
a similar situation to the previous case. However, in the present case 
$Q\simeq 0$ only for local ($k \simeq p \simeq q$) interactions. 
It is now the velocity field term $|g_{kpq}|^2|U_{s_p}|^2kq > 0$
which ensures that exponentially growing solutions exist for local interactions
provided $|\Up|>2|\Bp|$.

\item For $k<q<p$ the result is the same, since reversing the relative ordering
of $p$ and $q$ does not change the sign of $Q$. That is, 
exponentially growing solutions occur. 
\item For $p<k<q$ the term $Q$ is positive and the term proportional to $|U_{s_p}|^2$ 
is positive while the term proportional to $|B_{s_p}|^2$ is negative. 
Thus instabilities occur if $|U_{s_p}|/|B_{s_p}|>1$. For $|U_{s_p}|/|B_{s_p}|<1$,
the occurrence of instabilities depends on the value of $H_c(p)$. If $H_c(p)$ is 
maximal and magnetic and velocity field are fully aligned, then the solutions are stable.
For zero cross-helicity and $|B_{s_p}|$ being not much larger than $|U_{s_p}|$, solutions are 
unstable (see appendix \ref{app:graphs}). 

This type of helicity combination is another possibility for
nonlocal interactions of the type $p<<k\simeq q$ leading to
exponentially growing solutions if $|\Up| > |\Bp|$, since then
\beq
x \simeq |g_{kpq}|^2 k^2(|U_{s_p}|^2 -|B_{s_p}|^2) > 0 \ .
\eeq
\end{itemize}

The results of the dependence of the occurrence of unstable solutions 
on combinations of helicities, wavenumber ordering, relative magnitudes 
of the $\vec{u}$ and $\vec{b}$ modes and cross-helicities at wavenumber 
$p$ are summarised in tables \ref{tbl:split}-\ref{tbl:forward}. 

\begin{table}
\begin{center}
\begin{tabular}{l|c|c|c}
Helicities & $H_c$ & constraint & stability  \\
\hline
$s_k\neq s_q =s_p$ & n/a & $|U_{s_p}| > |B_{s_p}|$ & unstable  \\
 & max & $|B_{s_p}| > |U_{s_p}|$  & stable  \\
 & 0 & $|B_{s_p}|/|U_{s_p}|$ not $>>1$ & unstable  \\
\hline
$s_k= s_p \neq s_q$ & n/a & n/a & unstable  \\
\hline
$s_k = s_q =s_p$ & n/a & n/a & unstable \\
\end{tabular}
\caption{Summary of possible unstable solutions for the middle wavenumber modes $k<p<q$.} 
\label{tbl:split}
\end{center}
\end{table}

\begin{table}
\begin{center}
\begin{tabular}{l|c|c|c}
Helicities & $H_c$ & constraint & stability  \\
\hline
$s_k\neq s_q =s_p$ & n/a & n/a & unstable  \\
\hline
$s_k= s_p \neq s_q$ & n/a & n/a & unstable  \\
\hline
$s_k = s_q =s_p$ & n/a & $|B_{s_p}| > |U_{s_p}|$ & unstable \\
 & max & $|U_{s_p}| > |B_{s_p}|$ & stable \\
 & 0 & $|U_{s_p}|/|B_{s_p}|$ not $>> 1$ & unstable \\
\end{tabular}
\caption{Summary of possible unstable solutions for the largest wavenumber modes $k<q<p$.} 
\label{tbl:reverse}
\end{center}
\end{table}

\begin{table}
\begin{center}
\begin{tabular}{l|c|c|c}
Helicities & $H_c$ & constraint & stability  \\
\hline
$s_k\neq s_q =s_p$ & n/a & $|U_{s_p}| > |B_{s_p}|$ & unstable  \\
 & max & $|B_{s_p}| > |U_{s_p}|$  & stable  \\
 & 0 & $|B_{s_p}|/|U_{s_p}|$ not $>>1$ & unstable  \\
\hline
$s_k= s_p \neq s_q$ & n/a & $|U_{s_p}| > |B_{s_p}|$ & unstable  \\
 & max & $|B_{s_p}| > |U_{s_p}|$  & stable  \\
 & 0 & $|B_{s_p}|/|U_{s_p}|$ not $>> 1$  & unstable  \\
\hline
$s_k = s_q =s_p$ & n/a & n/a & stable \\
\end{tabular}
\caption{Summary of possible unstable solutions for the smallest wavenumber modes $p<k<q$.} 
\label{tbl:forward}
\end{center}
\end{table}

\section{Energy transfers and the instability assumption}
\label{sec:instab_assump}
In order to use the results of the previous section to derive results for the 
transfers of the ideal invariants total energy $E_{tot}$ and magnetic 
helicity $H_{mag}$, we invoke the {\em instability 
assumption} \citep{Waleffe92}. Generalised to MHD turbulence, this assumption asserts that
energy is transferred away from modes whose evolution equation for the
helicity coefficient is linearly unstable, into the other two modes it is 
coupled to by a triad interaction given through the system 
\eqref{eq:a_eqs_coup}-\eqref{eq:c_eqs_coup}.

Therefore the results of the stability analysis determine
whether a given helicity combination mainly contributes to forward or reverse
transfer of energy. That is, if a steady solution at wavenumber $p$ is unstable
and energy is transferred away from $B_{s_p}$ and $U_{s_p}$ into the modes they 
interact with (note that $B_{s_p}$ and $U_{s_p}$ do not interact with 
each other directly), then the wavenumber ordering $k<q<p$ results in reverse
transfer of energy, while $p<k<q$ results in forward transfer and $k<p<q$
in a split transfer with contributions to forward and reverse directions of
energy transfer. 

Several immediate results can be deduced from the summary of the stability 
analysis for the different helicity combinations presented in tables 
\ref{tbl:split}-\ref{tbl:forward}. 
First, unlike in non-conducting fluids modes corresponding to the largest 
wavenumber in a given triad can be unstable,
leading to more possibilites for reverse spectral energy transfer in MHD
compared to hydrodynamics. Second, all three helicities influence the
direction of energy transfers, and reverse transfers are also possible 
for cases of unlike helicities. Third, forward transfers appear to be more 
quenched in regions of high cross-helicity than reverse transfers. 
Fourth, very nonlocal triads contribute mainly to 
reverse transfers in magnetically dominated systems through interactions
of modes with like helicity. They
only contribute to forward transfers through interactions of modes 
with unlike helicity and mostly if the kinetic energy is larger 
than the magnetic energy.  

Therefore we obtain that reverse spectral transfer becomes much
more likely in MHD turbulence than in turbulence of non-conducting fluids, 
which reflects the predictions from absolute equilibrium spectra 
\citep{Frisch75,Zhu14} and the well-established numerical results on inverse
cascades, and more generally reverse transfer, in MHD turbulence 
\citep{Pouquet76,Pouquet78,Balsara99,Alexakis06,Brandenburg01,Mueller12,
Berera14a,Brandenburg15}. 

We note that the transfer directions deduced so far may or may not 
contribute to forward and inverse cascades of energy and magnetic
helicity, as no information on the constancy, or otherwise, of the 
fluxes of these quantities through a given wavenumber is available
at this point.
The aim of the next section is to determine the contribution of the individual
transfers to energy and magnetic helicity {\em cascades}. 

\section{Transfer and cascades of total energy and magnetic helicity}
\label{sec:fluxes}
In order to determine the contribution of a given interaction 
of helical modes to energy and magnetic helicity cascades, the fluxes 
of these quantities need to be calculated and studied in the respective
inertial ranges where they are wavenumber-independent. 
However, several technical details need to be discussed 
before we can proceed to this calculation. 

In the discrete Fourier representation the evolution equations 
of the kinetic and magnetic energy spectra 
$E_{kin}(k)$ and $E_{mag}(k)$ are obtained by 
multiplying the relevant equations in the system \eqref{eq:a_eqs_coup} 
by $u^*_{s_k}$ and $b^*_{s_k}$, respectively, then summing over all triads and
helicity combinations and finally carrying out shell- and ensemble averages. For the 
kinetic energy spectrum this leads to
\beq
\partial_t E_{kin}(k) = \frac{1}{2} \sum_{p,q}^{\Delta} \sum_{i=1}^{8} (t^{(i)}_{HD}(k,p,q)+ t^{(i)}_{LF}(k,p,q)) \ ,
\eeq
where $\sum_{p,q}^{\Delta}$ denotes a sum over all wavenumbers $p$ and $q$ whose 
wavevectors $\vp$ and $\vq$ form a triad with $\vk$ such that 
$\vec{k} + \vec{p} + \vec{q} =0$ and the superscript $(i)$
labels the eight possible helicity combinations. 
The transfer terms in this equation are given by  
\beq
t^{(i)}_{HD}(k,p,q) = 
(s_p p - s_q q) \sum_{S(k,p,q)}g_{kpq} \langle u_{s_k}U_{s_p}u_{s_q} \rangle + \mbox{c.c.} \ , 
\eeq
and 
\beq
t^{(i)}_{LF}(k,p,q) = 
-(s_p p - s_q q) \sum_{S(k,p,q)}g_{kpq} \langle u_{s_k}B_{s_p}b_{s_q} \rangle + \mbox{c.c.} \ , 
\eeq
where $S(k,p,q)$ indicates a summation over all wavevectors 
in shells of radius $k,p$ and $q$ and $c.c.$ denotes the complex conjugate.
For the magnetic energy spectrum one obtains
\beq
\partial_t E_{mag}(k) = \frac{1}{2} \sum_{p,q}^{\Delta} \sum_{i=1}^{8} t^{(i)}_{mag}(k,p,q) \ ,
\eeq
where
\beq
t^{(i)}_{mag}(k,p,q) = 
-s_k k \sum_{S(k,p,q)}g_{kpq} \langle b_{s_k}B_{s_p}u_{s_q}- b_{s_k}U_{s_p}b_{s_q} \rangle + \mbox{c.c.} \ . 
\eeq

The evolution equation for the total energy spectrum $E(k) = E_{kin}(k)+E_{mag}(k)$ is given 
by the sum of the respective evolution equations for $E_{kin}(k)$ and $E_{mag}(k)$ 
\beq
\partial_t E(k) = \frac{1}{2} \sum_{p,q}^{\Delta} \sum_{i=1}^{8} t^{(i)}(k,p,q) \ ,
\eeq
and total energy transfer term $t^{(i)}(k,p,q)$ therefore consists of three types of transfers
\beq
t^{(i)}(k,p,q)=t^{(i)}_{HD}(k,p,q)+ t^{(i)}_{LF}(k,p,q)
               + t^{(i)}_{mag}(k,p,q)  \ .
\eeq 
The term $t^{(i)}_{HD}(k,p,q)$ denotes purely hydrodynamic transfer due to the 
coupling of the velocity field to itself, $t^{(i)}_{LF}(k,p,q)$ the contribution
due to the Lorentz force acting on the fluid and $t^{(i)}_{mag}(k,p,q)$ the contributions
due to advection of the magnetic field by the flow and conversion of kinetic to 
magnetic energy, that is, due to dynamo action. In real space the nonlinear term 
$\nabla \times (\vec{u} \times \vec{b})$ corresponding to the magnetic transfer 
term can be split into an advective term $(\vec{u} \cdot \nabla)\vec{b}$ and 
a dynamo term $(\vec{b} \cdot \nabla)\vec{u}$, however this splitting is  
obscured in Fourier space.

These terms are still written in the discrete Fourier representation of the magnetic and velocity fields.
However, the calculation of the energy and magnetic helicity fluxes requires a continuous Fourier
representation. 
The continuous transfer terms are given in terms of Fourier integrals and 
can formally be obtained by taking the 
period $L$ 
to infinity, 
assuming that the respective integrals are well-defined.
The sums then become integrals and the continuous counterpart of e.g.~the 
hydrodynamic transfer term $t^{(i)}_{HD}$ becomes
\begin{align}
T^{(i)}_{HD}(k,p,q) & dk \ dp \ dq = \lim_{L\to \infty} t^{(i)}_{HD}(k,p,q) \nonumber \\
&= (s_p p - s_q q) \int_{|\vec{k}|=k} d\vec{k} \ \int_{|\vec{p}|=p} d\vec{p} \ \int_{|\vec{q}|=q}d\vec{q} \ 
g_{kpq}\langle u_{s_k}U_{s_p}u_{s_q} \rangle + \mbox{c.c.} \ .
\end{align} 
The transfer terms $T^{(i)}_{LF}$ and $T^{(i)}_{mag}$  are defined analogously.

\subsection{Total energy transfer}
\label{sec:totalE}
In the absence of dissipation the total energy is conserved and the transfer term $T(k,p,q)$ in the 
spectral evolution equation of the total energy redistributes energy between the Fourier modes and
vanishes if integrated over all space.  
Therefore the flux of total energy through wavenumber $k$ due to a given interaction $(i)$, 
\begin{equation}  
\Pi^{(i)}(k)=-\int_0^k dk^{\prime } \int_k^{\infty} \int_k^\infty 
T^{(i)}(k^{\prime },p,q) dp dq \ ,
\end{equation}
can be written as the sum of two contributions: the flux of total energy into all modes
at wavenumber
$k'$ due to triads with $p,q<k<k'$ minus the flux of total energy into all modes at $k'$ due
to triads with $k'<k<p,q$
\begin{equation}  
\Pi^{(i)}(k)=
\frac{1}{2}\int_k^\infty  dk^{\prime } \int_0^{k} \int_0^k T^{(i)}(k^{\prime },p,q) dp \ dq 
-\frac{1}{2}\int_0^k dk^{\prime } \int_k^{\infty} \int_k^\infty T^{(i)}(k^{\prime },p,q) dp \ dq \ .
\label{eq:Etransfer} 
\end{equation}

We now follow the procedure introduced by \citet{Waleffe92} in order to
render the two integrals in \eqref{eq:Etransfer} independent of $k$. 
This is achieved using a scaling argument, where the two integrals are 
treated separately. For conciseness we only outline the procedure briefly 
for the first integral on the RHS of \eqref{eq:Etransfer} and refer to the
original work of \citet{Waleffe92} for the full derivation.  
The aim is to express the transfer function in the first integral on the RHS of 
\eqref{eq:Etransfer} in terms of new variables 
\begin{equation} 
v=\frac{q}{p}, \ \ w=\frac{k'}{p}, \ \ u=\frac{k}{p} \ ,
\end{equation}
in order to remove $k$ from the integration limits.  
Since $T^{(i)}_{HD}(k',p,q)$ may scale differently compared to 
$T^{(i)}_{LF}(k',p,q)$ and $T^{(i)}_{mag}(k',p,q)$, 
the term $T^{(i)}(k',p,q)$ in \eqref{eq:Etransfer} 
must be replaced by the individual transfer terms. 
The transfer terms are now expressed individually 
in terms of the new variables $u$, $v$ and $w$
\beq
T^{(i)}_{HD}(k',p,q)=p^{-\beta}T^{(i)}_{HD}(w,1,v) = \left(\frac{k}{u}\right)^{-\beta} T^{(i)}_{HD}(w,1,v) \ ,
\eeq
\beq
T^{(i)}_{LF}(k',p,q)=p^{-\beta'}T^{(i)}_{LF}(w,1,v) = \left(\frac{k}{u}\right)^{-\beta'} T^{(i)}_{LF}(w,1,v) \ ,
\eeq
and
\beq
T^{(i)}_{mag}(k',p,q)=p^{-\beta'}T^{(i)}_{mag}(w,1,v) = \left(\frac{k}{u}\right)^{-\beta'} T^{(i)}_{mag}(w,1,v) \ ,
\eeq
where $\beta$ is related to the exponent of the kinetic
energy spectrum provided it has a power-law dependence on $k$, while 
the exponent $\beta'$ is related to the exponents of the kinetic {\em and} 
magnetic energy spectra as explained in further detail in appendix \ref{app:similarity}.  
The first term on the RHS of \eqref{eq:Etransfer} then becomes
\begin{align}  
\frac{1}{2}\int_k^\infty  &dk^{\prime } \int_0^{k} \int_0^k T^{(i)}(k^{\prime },p,q) \ dp \ dq \nonumber \\ 
=\ &k^{3-\beta}\int_0^1 dv \ \int_1^{1+v} \  dw \int_1^w du \ \left(\frac{1}{u}\right)^{4-\beta} T^{(i)}_{HD}(w,1,v) \nonumber \\ 
&+k^{3-\beta'}\int_0^1 dv \ \int_1^{1+v} \  dw \int_1^w du \ \left(\frac{1}{u}\right)^{4-\beta'} \left [T^{(i)}_{LF}(w,1,v)+ T^{(i)}_{mag}(w,1,v)\right ] \ . 
\end{align}
The second term on the RHS of \eqref{eq:Etransfer} can be treated similarly  
\citep{Waleffe92}, and we obtain
\begin{align}  
\frac{1}{2}\int_0^k  &dk^{\prime } \int_k^{\infty} \int_k^{\infty} T^{(i)}(k^{\prime },p,q) \ dp \ dq \nonumber \\ 
= \ &k^{3-\beta}\int_0^1 dv \ \int_1^{1+v} \  dw \int_v^1 du \ \left(\frac{1}{u}\right)^{4-\beta} T^{(i)}_{HD}(v,1,w) \nonumber \\ 
&+k^{3-\beta'}\int_0^1 dv \ \int_1^{1+v} \  dw \int_v^1 du \ \left(\frac{1}{u}\right)^{4-\beta'} \left [T^{(i)}_{LF}(v,1,w)+ T^{(i)}_{mag}(v,1,w)\right ] \ .
\end{align}
Combining the two results and integrating over $u$ leads to 
the following expression for the total energy transfer flux  
\begin{align}
\label{eq:fluxE}
\Pi^{(i)}(k)&= 
k^{3-\beta} \int_0^1 dv \ \int_1^{1+v} \ dw  
\left(
T^{(i)}_{HD}(w,1,v)\left[ \frac{w^{\beta-3}-1}{\beta-3}\right]  
+  T^{(i)}_{HD}(v,1,w) \left[ \frac{v^{\beta-3}-1}{\beta-3}\right]
\right) \nonumber \\
&+k^{3-\beta'} \int_0^1 dv \ \int_1^{1+v} \ dw  
\left(
T^{(i)}_{LF}(w,1,v) \left[ \frac{w^{\beta'-3}-1}{\beta'-3}\right]  
+  T^{(i)}_{LF}(v,1,w) \left[ \frac{v^{\beta'-3}-1}{\beta'-3}\right]
\right ) \nonumber \\
&+k^{3-\beta'} \int_0^1 dv \ \int_1^{1+v} \ dw  
\left(
 T^{(i)}_{mag}(w,1,v) \left[ \frac{w^{\beta'-3}-1}{\beta'-3}\right]  
+   T^{(i)}_{mag}(v,1,w) \left[ \frac{v^{\beta'-3}-1}{\beta'-3}\right]
\right ) \ ,
\end{align}
where $0\leqslant v \leqslant 1 \leqslant w \leqslant 1+v$ due to the triad geometry.
This now enables us to study the contribution to the total energy transfer 
from a given interaction $(i)$, 
where the scaling of the
magnetic and kinetic energy spectra will influence the transfer through 
the exponents $\beta$ and $\beta'$. 
In the inertial range of total energy the energy transfer flux through a 
given wavenumber $k$ does not depend on that wavenumber, which leads to
the characteristic values of the scaling exponents $\beta'=\beta = 3$, 
making the split of the total energy transfer term into its individual components
redundant in this wavenumber range. 
In sec.~\ref{sec:Ecascade} we concentrate on the contributions
of the different interactions to transfers in the 
inertial range of total energy and set $\beta = 3$, thus taking into account 
only the region in wavenumber space where this scaling is established.
Since the values of $\beta$ and $\beta'$ may 
influence the direction of energy transfer, a similar approach may be useful 
to calculate energy and helicity transfer at the very low wavenumbers. However, 
this awaits consensus on the low-wavenumber scaling of the magnetic and 
kinetic energy spectra. Furthermore, the integrals must be cut off 
at some wavenumber such that a single scaling exponent for the wavenumber 
range of interest can be studied. As the extent of the inertial range 
will grow with increasing Reynolds number, contributions
from the production and dissipation ranges can safely be neglected, 
as they will become very small compared
to the extent of the inertial range. However, in the low wavenumber region, this 
argument is not applicable and further work is necessary in order to establish
if very nonlocal interactions contribute significantly to the transfers of 
magnetic energy and helicity in the low wavenumber range or not.

\subsection{Magnetic helicity transfer}
\label{sec:helicity}
Using the decomposition into helical modes, the transfer term in the 
evolution equation of the magnetic helicity can be expressed through the 
transfer term in the evolution equation of the magnetic energy, that is
\beq
T^{(i)}_H(k,p,q)= \frac{s_k}{k}T^{(i)}_{mag}(k,p,q) \ ,
\eeq
and only the transfer term which originates from the induction equation is present, since
$H_{mag}$ is a purely magnetic quantity and as such only implicitly depends on 
the evolution of the velocity field.

Since the magnetic helicity is an ideal invariant, the transfer term in the 
spectral evolution equation of the magnetic helicity
vanishes if integrated over all space, therefore  
similar to the flux of total energy, the flux of magnetic helicity through wavenumber 
$k$ due to a given interaction $(i)$, 
\begin{equation}  
\Pi_H^{(i)}(k)=-\int_0^k \frac{s_{k^{\prime }}}{k^{\prime }} dk^{\prime } \int_k^{\infty} \int_k^\infty 
T^{(i)}_{mag}(k^{\prime },p,q) dp dq \ ,
\end{equation}
can be written as the sum of two contributions
\begin{align}  
\Pi_H^{(i)}(k)=&
\frac{1}{2}\int_k^\infty \frac{s_{k^{\prime }}}{k^{\prime }} dk^{\prime } \int_0^{k} \int_0^k T^{(i)}_{mag}(k^{\prime },p,q) dp \ dq \nonumber \\ 
&\ -\frac{1}{2}\int_0^k \frac{s_{k^{\prime }}}{k^{\prime }} dk^{\prime } \int_k^{\infty} \int_k^\infty T^{(i)}_{mag}(k^{\prime },p,q) dp \ dq \ .
\end{align}
Following the approach explained in sec.~\ref{sec:totalE} the integral becomes independent of $k$
and one obtains the following expression
for the flux of magnetic helicity through $k$
\begin{align}
\label{eq:Htransfer} 
\Pi_H(k) =& k^{2-\beta'} \int_0^1 dv \ \int_1^{1+v} \ \frac{dw}{w} \nonumber \\ 
& \ \times \left(
s_wT^{(i)}_{mag}(w,1,v)\left[ \frac{w^{\beta'-2}-1}{\beta'-2}\right]  
+ s_v T^{(i)}_{mag}(v,1,w) \left[ \frac{v^{\beta'-2}-1}{\beta'-2}\right]
\right) \ .
\end{align}

\subsection{Cascades and wavenumber-dependent transfers of total energy and magnetic helicity}
From the expressions \eqref{eq:fluxE} and 
\eqref{eq:Htransfer} for the fluxes of total energy and magnetic helicity, respectively,
it is now possible determine the sign of the fluxes and hence the direction of
energy and magnetic helicity transfers using the results from the stability analysis.
If the total energy flux is positive, energy is
transferred from smaller to larger wavenumbers and if it is negative, 
energy is transferred from larger to smaller wavenumbers. As the magnetic helicity is not
positive definite, the situation is slightly different. For positive magnetic helicity a positive 
flux indicates forward transfer just as for the total energy. For negative magnetic helicity
a negative flux indicates forward transfer while a positive flux indicates inverse transfer. 
However, as this situation is symmetric we assume positive helicity throughout the analysis.

In sec.~\ref{sec:instab_assump} unstable solutions of 
\eqref{eq:a_eqs_coup} and \eqref{eq:c_eqs_coup} were interpreted as
leading to energy transfer out of the unstable mode into the 
two modes it interacts with for a given helical mode interaction $(i)$. 
If $\Up$ and $\Bp$ are the unstable modes, 
this interpretation leads to 
\beq
\label{eq:tmagsign} 
\partial_t |B_{s_p}|^2 = T^{(i)}_{mag}(p,k,q) < 0 \ ,  
\eeq
and 
\beq 
\label{eq:thydrosign} 
\partial_t |U_{s_p}|^2 = T^{(i)}_{HD}(p,k,q) + T^{(i)}_{LF}(p,k,q) < 0 \ .  
\eeq
The instability assumption therefore 
attributes signs to the transfer terms, which will determine their respective
contributions to the overall energy (and magnetic helicity) transfer.
Note that $\partial_t |U_{s_p}|^2$ and $\partial_t |B_{s_p}|^2$ cannot
have different signs, as both signs are determined 
from the existence of exponentially growing solutions of the system 
\eqref{eq:matrix_ode}.
 
We now treat the three helicity combinations which lead to unstable solutions 
separately 
assuming without loss of generality that $s_p=1$.
Having determined the signs of the transfer terms within our 
framework, we now use these results to calculate the contributions
of the individual transfer terms to the fluxes of total energy and magnetic helicity 
though a given wavenumber. 

\subsubsection{Total energy cascades}
\label{sec:Ecascade}
For the (inertial range) energy cascade the flux is wavenumber-independent
leading to $\beta=3$ in \eqref{eq:fluxE}. Hence the integrand
in \eqref{eq:fluxE}, which determines the sign of the total energy flux, 
becomes
\begin{equation}
I_E = T^{(i)}(w,1,v)\ln{w} +  T^{(i)}(v,1,w) \ln{v} \ ,
\end{equation}
where we remind the reader of the wavenumber ordering $v \leqslant 1 \leqslant w$.
That is, the term $T^{(i)}(w,1,v)$ describes energy transfer in and out of the
largest wavenumber modes while $T^{(i)}(v,1,w)$ describes energy transfer in and out 
of the smallest wavenumber modes. 

Using the signs of the transfer terms determined for the three 
helicity combinations depending on wavenumber ordering, 
we can now deduce which helicity combinations 
contribute to forward or inverse cascades of total energy. 
\begin{itemize}
\item $s_v=s_1=s_w$ \\ 
For this case we can deduce from the results of the stability analysis
summarised in tables \ref{tbl:split}-\ref{tbl:forward} that 
$T^{(i)}(1,v,w) < 0$, as modes corresponding to the middle wavenumber are unstable,
while  
$T^{(i)}(v,1,w) > 0$, as modes corresponding to the smallest wavenumber are stable and hence
these modes can only receive energy from the modes at higher
wavenumbers. The sign of $T^{(i)}(w,1,v)$ depends on the 
values of cross-helicity and the ratio of magnetic to kinetic energy. 
For a magnetically dominated system $T^{(i)}(w,1,v) <0$ and we obtain in this case 
an inverse cascade of total energy, as 
\begin{equation}
I_E = T^{(i)}(w,1,v)\ln{w} +  T^{(i)}(v,1,w) \ln{v} < 0 \ .
\end{equation}
If the kinetic energy is much larger than the magnetic energy, 
cancellations between the two terms in $I_E$ occur. The term $T^{(i)}(w,1,v)\ln{w}$
is now positive, since the modes at the largest wavenumber can only
receive energy, thus contributing to a forward cascade. 
For intermediate cases the value of the cross-helicity becomes decisive 
as high cross-helicity quenches the inverse transfer in this case.  
In summary, we expect inverse cascade contributions from this combination
of helicities if the magnetic energy dominates, while for larger kinetic
energy high values of cross-helicity quench the inverse transfer contribution 
to some extent. 

\item $s_v=s_1\neq s_w$ \\ 
From tables \ref{tbl:split}-\ref{tbl:forward} the instability assumption
imposes $T^{(i)}(w,1,v)<0$ and $T^{(i)}(1,v,w)<0$ as modes corresponding to the middle and 
largest wavenumbers are unstable. The stability of the lowest wavenumber modes
now depends on several parameters. However, the detailed conservation 
property
\beq
T^{(i)}(w,1,v) + T^{(i)}(1,v,w) + T^{(i)}(v,1,w) = 0 \,
\eeq
asserts that if two of the transfer terms are of like sign, then the third one 
must be of opposite sign. Since there are no constraints on the instability
of the modes corresponding to the two largest wavenumbers, 
we take $T^{(i)}(v,1,w)$ to be positive and 
conclude that this combination of helicities leads to an inverse energy
cascade as $I_E < 0$, and we note that this case behaves very differently 
from its hydrodynamic analogue, where it led to a forward cascade of kinetic energy 
\citep{Waleffe92}. We also note that this inverse cascade should always be present, 
as it is not subject to constraints from $H_c(p)$ and $|U_{s_p}|/|B_{s_p}|$.  

\item $s_v\neq s_1=s_w$ \\
Analogously, we obtain 
$T^{(i)}(1,v,w)<0$, since the modes corresponding to the middle
wavenumber are unstable. As the stability of the two other transfer terms  
depends on several constraints, no clear assessment is possible. 
If we assume them to be of like 
sign and thus positive, as not all transfer term can have the
same sign due to the detailed conservation property, we obtain
a contribution towards an inverse cascade. However, if we assume them being of opposite
sign, contributions to inverse and direct cascades are possible. We note
that the instability leading to forward transfer in this case 
is damped by high values of $H_c(p)$. 
\end{itemize}

\subsubsection{Magnetic helicity transfer in the inertial range of total energy}
For $\beta'=3$, the integrand $I_H$ in \eqref{eq:Htransfer} becomes
\begin{equation}
I_H = T^{(i)}_{mag}(w,1,v)s_w(w-1) + T^{(i)}_{mag}(v,1,w) s_v(v-1) \ .
\label{eq:IH}
\end{equation}
Using the signs of the transfer terms determined for the three
helicity combinations, we can now deduce which helicity combinations
contribute to a forward or inverse cascade of magnetic helicity.
As can be seen in \eqref{eq:IH}, there is an explicit dependence 
of the magnetic helicity flux on the helicities of the interacting modes. In the following we 
assume $s_1=1$.
\begin{itemize}
\item $s_v=s_1=s_w$ \\
The integrand $I_H$ becomes
\begin{equation}
I_H = T^{(i)}_{mag}(w,1,v)(w-1) + T^{(i)}_{mag}(v,1,w)(v-1) \ .
\end{equation}
As the signs of the magnetic energy transfer term deduced from 
the stability analysis are the same as for the total energy and 
$\ln{w}$ and $w-1$ are both positive while $\ln{v}$ and $v-1$ are both negative,
the result for the helicity transfer
reflects the results for the total energy cascade, thus for this helicity combination
total energy and magnetic helicity will be transferred in the same 
direction, which can be both forward and inverse in this case.  

\item $s_v=s_1\neq s_w$ \\
The integrand $I_H$ becomes
\begin{equation}
I_H = -T^{(i)}_{mag}(w,1,v)(w-1) + T^{(i)}_{mag}(v,1,w)(v-1) \ ,
\end{equation}
where the contributions from the largest wavenumber modes now enter with the
opposite sign.
Compared to the total energy flux, which was purely inverse in this case, 
we obtain the possibility of simultaneously a forward helicity flux and an inverse 
energy flux.

\item $s_v\neq s_1=s_w$ \\
The integrand $I_H$ becomes
\begin{equation}
I_H = T^{(i)}_{mag}(w,1,v)(w-1) - T^{(i)}_{mag}(v,1,w)(v-1) \ ,
\end{equation}
where the contributions from the smallest wavenumber modes now enter with the
opposite sign.
Compared to the total energy cascade, again we find that it is possible to 
have a transfer of magnetic helicity in the opposite direction to the transfer 
of total energy. 
\end{itemize}

In this subsection we determined the direction of the magnetic helicity
transfer in the inertial range of total energy for different combinations of 
helicities and compared the results to those for the total energy cascade. We found that a
cascade of total energy is possible in one direction 
while the transfer of magnetic helicity may proceed 
in the opposite direction. 

\subsubsection{Magnetic helicity cascades}
In the inertial range of magnetic helicity the flux of magnetic helicity is 
wavenumber-independent resulting in $\beta' = 2$ in \eqref{eq:Htransfer}. Therefore 
the integrand $I_H$ in \eqref{eq:Htransfer} becomes
\begin{equation}
I_H = T^{(i)}_{mag}(w,1,v)s_w\ln{w} +  T^{(i)}_{mag}(v,1,w) s_v\ln{v} \ .
\label{eq:IH_inert}
\end{equation}
For the three different helicity combinations this leads to
\begin{itemize}
\item $s_v=s_1=s_w$ \\
The integrand in this case is of the same form as the integrand $I_E$ for
the total energy cascade (that is, if $\beta=3$ in $I_E$)
\begin{equation}
I_H = T^{(i)}_{mag}(w,1,v)\ln{w} +  T^{(i)}_{mag}(v,1,w) \ln{v} \ ,
\end{equation}
hence the results for the cascades of magnetic helicity are the same 
as for the cascasdes of total energy.

\item $s_v=s_1\neq s_w$ \\
The integrand in this case has a different form compared to the integrand $I_E$ for
the total energy
\begin{equation}
I_H = -T^{(i)}_{mag}(w,1,v)\ln{w} +  T^{(i)}_{mag}(v,1,w) \ln{v} \ ,
\end{equation}
hence the results for the cascades of magnetic helicity differ from the 
total energy cascades. In particular, this case may lead to a nonhelical reverse
energy transfer while the helicity cascade may be forwards, due to the 
contribution from $T^{(i)}_{mag}(w,1,v)$ now having the opposite sign in $I_H$ 
compared to $I_E$.

\item $s_v\neq s_1=s_w$ \\
Again, the integrand in this case has a different form compared to the integrand $I_E$ for
the total energy
\begin{equation}
I_H = T^{(i)}_{mag}(w,1,v)\ln{w} - T^{(i)}_{mag}(v,1,w) \ln{v} \ ,
\end{equation}
hence the results for the cascades of magnetic helicity are different from the 
total energy cascades. In particular, this case may lead to a nonhelical reverse
energy transfer while the helicity cascade may be forwards, due to the
contribution from $T^{(i)}_{mag}(v,1,w)$ now having the opposite sign in $I_H$ 
compared to $I_E$.
\end{itemize}

\subsubsection{Magnetic energy transfer in the inertial range of magnetic helicity}
For $\beta'=2$, 
the contributions to the integrand $I_E$ due to magnetic energy transfer are
\beq
I_{E_{mag}} = 
-T^{(i)}_{mag}(w,1,v)\left(\frac{1}{w}-1 \right)-  T^{(i)}_{mag}(v,1,w) \left(\frac{1}{v}-1 \right) \ .
\eeq
The signs of $T_{mag}$ and $T$ are the same by 
eqs.~\eqref{eq:tmagsign}-\eqref{eq:thydrosign}, and 
$\ln{w}$ and $w-1$ are both positive while $\ln{v}$ and $v-1$ are both negative. 
Hence, the result for the contributions of these terms to the total energy transfer 
in the inertial range of magnetic helicity is the same as in the inertial range of 
total energy for all helicity combinations. That is, magnetic energy transfer and conversion 
in the inertial ranges of total energy and magnetic helicity proceed in the 
same direction. 

This assessment of contributions to forward and inverse transfers and cascades is based 
on an analysis of the nonlinear terms in the MHD equations only, thus neglecting
the symmetry-breaking effect of dissipation creating an energy sink at the small
scales. Accounting for this effect, it is plausible that the contributions from transfer terms 
leading to forward transfer are higher weighted than contributions
leading to inverse transfer. This is particularly relevant
in interactions where forwards and reverse contributions are present and the 
overall transfer depends on cancellations between the two terms. It
would perhaps be safest to attribute these cases to forwards rather than inverse energy cascades.   

Although it is not possible to exactly determine which helical interactions 
are higher weighted than others, some information can be obtained from the 
magnitude of the geometric factor $g_{kpq}$ defined in 
eq.~\eqref{eq:geom_factor}.
The magnitude of $g_{kpq}$ depends on the 
helicity combinations since it involves the helicity-dependent factor 
$I=s_k k+ s_p p + s_q q$. Therefore it parametrises the 
strength of a given helical interaction, and the case of 
all helicities being of the same sign gives the largest value of 
$|I|$, since in this case $|I|=|k+p+q|$. 

For the reverse transfers, that is, for $k<p,q$, the factor $|I|$ 
takes the smallest value for the case $s_k=s_p \neq s_q$, 
since $|I|=|k+(p-q)|$. Note that in this case $I$ becomes small for 
small $k$ even in the 
nonlocal limit $k<<p\simeq q$, suggesting that the nonhelical 
reverse transfer found in this case is 
less efficient in increasing spectral power at the very low wavenumbers. 
The remaining class of helical interactions $s_k \neq s_p = s_q$ 
leads to $|I| = |k-(p+q)|$. In this case
$|I|$ does not necessarily become small for small $k$ which is due to the 
contribution of nonlocal interactions, where $p$ and $q$ are large compared to $k$. 
According to the results from the stability analysis, in the 
nonlocal limit unstable solutions occur for the case $s_k=s_p = s_q$  
only if $|\Bp|>|\Up|$ and for the case $s_k\neq s_p = s_q$ if 
$|\Up|>|\Bp|$.     

It is therefore possible to deduce within the framework
of the instability assumption that most of the increase in 
energy at the very largest scales (in a magnetically dominated system) 
is mainly due to a breaking of mirror-symmetry,
which had been established before by \citet{Frisch75} using a 
different approach.
That is, it is due to the presence of kinetic and magnetic helicity, since 
interactions of the type $s_k=s_p = s_q$, which account for most 
of the inverse transfer, can only occur in significant numbers for 
fields consisting of many modes with the same helicity.
Recent numerical results in hydrodynamics
showed that there is an overall reverse flux of energy only when the
system mainly contains helical modes of the same sign.
As soon as a small amount of oppositely-polarised modes are introduced,
the usual direct cascade is recovered \citep{Sahoo15}.

In summary, in this section we determined the {\em direction} of total energy and magnetic
helicity transfers in their respective inertial ranges. Not surprisingly, we found that
fully helical magnetic fields lead to inverse cascades of magnetic helicity and 
magnetic energy, but the analysis also showed that an inverse energy cascade is 
possible for nonhelical magnetic fields, which is a new theoretical result. 
However, due to the coupling of the momentum and induction equations,  
within this framework it is not possible to determine the {\em nature} 
of the energy transfers resulting from an instability of 
a given steady solution, since the same eigenvalue controls the growth
of the exponential solution of \eqref{eq:matrix_ode} for both the 
magnetic and the velocity field. Nevertheless, for some special cases the evolution 
equations \eqref{eq:a_eqs_coup}-\eqref{eq:c_eqs_coup} 
decouple and more detailed information becomes available. 
These cases are treated in the following section.  

\section{Special solutions and the (kinematic) dynamo}
\label{sec:dynamo}
Having established the general case, we now draw attention to special cases 
where the analysis becomes much simpler and which are relevant to specific
problems in MHD such as the kinematic dynamo. In sec.~\ref{sec:stability} we analysed
the stability of general steady solutions of the dynamical system 
\eqref{eq:a_eqs_coup}-\eqref{eq:c_eqs_coup}, which describes the evolution of a 
triad of interacting helical modes. Using the notation \eqref{eq:notation}, the
general steady solutions were of the form $(0, \Up, 0;0,\Bp,0)$.   
In this section we now study the cases where either $\Up=0$ or $\Bp=0$, that is  
we analyse the stability of steady solutions of \eqref{eq:a_eqs_coup}-\eqref{eq:c_eqs_coup} 
of the form $(0, \Up, 0; 0, 0, 0)$ and $(0, 0, 0; 0, \Bp, 0)$. The former case may be
of particular interest due to its relation to dynamo action. 

\subsection{The kinematic dynamo}
For small magnetic fields the Lorentz force is small compared to inertial 
forces, and can be neglected in the momentum equation. This decouples the 
momentum equation from the induction equation and
defines the kinematic dynamo problem. In our setting, it corresponds 
to $|U_{s_p}|/|B_{s_p}|>>1$, and terms proportional to $|B_{s_p}|$ can 
be neglected as they are very small compared to terms proportional to 
$|U_{s_p}|$. 

Alternatively, one could also consider the steady solution $B_{s_p}=0$ 
while $U_{s_p} \neq 0$, which would correspond to a stability analysis 
of a flow field at a particular length scale subject to small 
perturbations of the magnetic and velocity fields, where the magnetic field
perturbation may be viewed as the magnetic seed field to be amplified by dynamo action. 
In this setting we observe from 
\eqref{eq:a_eqs_coup} that the term corresponding to the Lorentz force 
disappears 
while in \eqref{eq:c_second_der} terms involving $B_{s_p}$ disappear, thus the system simplifies to
\begin{align}
\partial_t^2 u_{s_k}&= |g_{kpq}|^2 (s_pp-s_q q)(s_kk-s_pp) \, |U_{s_p}|^2 u_{s_k} \ , \\
\partial_t^2 b_{s_k}&= -|g_{kpq}|^2 s_k k \, s_q q \, |U_{s_p}|^2 b_{s_k} \ . 
\label{eq:k-dyn} 
\end{align}
As the only contribution to the evolution of the magnetic field now comes from the 
velocity field, we associate the remaining terms in \eqref{eq:c_eqs_coup}
with dynamo action. From \eqref{eq:k-dyn} we observe that this system has exponential solutions 
leading to magnetic field growth if $s_k \neq s_q$, regardless of wavenumber ordering. 
So for energy transfer from $U_{s_p}$ into $\bk$ (and $\bq$)
 to become possible, the magnetic modes at wavenumbers $k$ and $q$ should be of 
opposite helicity. 

For small $k$, nonlocal interactions with $k << p\simeq q$ provide most transfer 
into $\bk$. This is because the eigenvalue determining the growth of the 
exponential solution of \eqref{eq:k-dyn} is larger for $q >> k$ than for $q \simeq k$, 
thus the perturbations should grow faster in the former than in the latter case. 
Hence, according to the instability assumption, $\Up$ loses energy in favour of $\bk$
mainly due to nonlocal interactions if $\bk$ describes the largest scales of the system. 

\subsubsection{The $\alpha$-effect}
One well-known example of a large-scale dynamo is the $\alpha$-effect of mean-field
electrodynamics (see e.~g.~\cite{Moffatt78}), 
where $\alpha$ is a coefficent in the mean-field induction equation related
to the kinetic helicity of the flow. The $\alpha$-effect leads to a generation of large 
and small-scale magnetic helicities of opposite sign \citep{Brandenburg01,Brandenburg03}. 
A positively helical velocity field generates magnetic field perturbations 
leading to the large-scale component of the magnetic field becoming negatively 
helical. By conservation of magnetic helicity, the small-scale 
component of the magnetic field then has to become positively helical 
(and more so if the initial magnetic field was positively helical). That is, 
the small-scale magnetic and kinetic helicities are of the same sign. 

It is plausible that the type of interaction $(0,U_{s_p},0;0,B_{s_p}=0,0)$
for $k<p,q$ with $s_k \neq s_p=s_q$ can be associated with the $\alpha$-effect. 
First, nonzero small-scale kinetic helicity (we have $s_p=s_q$) is present. 
Second, the magnetic field growth at the large scales is described by 
\eqref{eq:k-dyn}, where magnetic fluctuations at $k$ and $q$ of opposite 
helicities are necessary to obtain an unstable solution. That is, the large-scale
magnetic field has opposite helicity to the small scale one, reminiscent 
of the $\alpha$-dynamo. We also note that this combination of helicities 
produces a transfer of kinetic energy from small to large scales \citep{Waleffe92}.  
Thus this type of interaction feeds 
into the magnetic and velocity fields on scales larger than the characteristic scale 
$L=1/p$ of the velocity field. The magnetic field mode which is 
amplified by this process has helicity opposite to the velocity field at $p$, 
which conforms to expectations in terms of the $\alpha$-effect.


\subsection{Excitation of a flow by the Lorentz force}
For the other special solution $(0,0,0;0,B_{s_p} \neq 0,0)$ the system 
\eqref{eq:a_second_der} - \eqref{eq:c_second_der} simplifies to
\begin{equation}
\partial_t^2 u_{s_k}= |g_{kpq}|^2 (s_pp-s_q q)s_qq \, |B_{s_p}|^2 u_{s_k} \ , 
\label{eq:k-highPm_u} 
\end{equation}
\begin{equation}
\partial_t^2 b_{s_k}= -|g_{kpq}|^2 s_k k (s_kk-s_pp) \, |B_{s_p}|^2 b_{s_k} \ , 
\label{eq:k-highPm_b} 
\end{equation}
and we note that the inertial term in \eqref{eq:a_eqs_coup} and the `dynamo' term in \eqref{eq:c_eqs_coup} 
are now absent and the system of coupled ODEs has split into two decoupled ODEs. 
This case may perhaps be associated with the generation of turbulence 
caused by the action of the Lorentz force on the fluid (i.e.~energy conversion from $\Bp$ to $\uk$ or $\uq$) 
and interscale transfer of magnetic energy from $\Bp$ into $\bk$ or $\bq$. 
Exponentially growing solutions of \eqref{eq:k-highPm_b} only occur if $s_p=s_k$ and $k<p$, leading 
to a reverse transfer of magnetic energy.
Exponentially growing solutions of \eqref{eq:k-highPm_u} 
occur for $p>q$ and $s_p = s_q$ leading to forward and
reverse transfers corresponding to $k>p$ and $k<p$, respectively. 
Interestingly, energy transfer only becomes possible if the magnetic field is helical and 
the helicity of the velocity field mode does not affect the analysis. 

\section{Conclusions}
\label{sec:discussion}
The four main results of the present work are: First, 
unlike in non-conducting fluids \citep{Waleffe92}, the stability analysis shows that 
in MHD turbulence energy can be transferred
away from the smallest scales in a triad interaction. Second, the stability analysis
reveals mechanisms for reverse energy transfer 
for nonhelical magnetic fields, in which case it does not need to be 
driven by the inverse transfer of magnetic helicity. Third, forward energy 
transfers are more quenched in regions of high cross-helicity than
reverse energy transfers. Fourth, we expect significant cancellations to occur 
between the contributions to forward and reverse transfers, as on several occasions 
they occur with opposite signs in the same equation.
Our theoretical analysis was conducted within the framework of the instability assumption, 
and it is crucial to discuss the results within the wider context of MHD turbulence research.    

Interscale energy transfers between the two different vector fields as well as within the same 
fields have been studied by several groups for freely decaying \citep{Debliquy05,Brandenburg15} 
and stationary \citep{Brandenburg01,Alexakis05a,Carati06,Cho10} 
MHD turbulence as well as for the kinematic dynamo regime 
\citep{Mininni05a} and for magnetic helicity transfer \citep{Alexakis06}, 
using shell-filtered transfer terms calculated from DNSs or from a helical shell model \citep{Stepanov15}.
 In the stationary case, it was found that transfers between 
the same fields are mainly local while transfers between different fields were nonlocal, and
transfers from the injection scale to the largest scales in the system were observed.
In the decaying case, energy transfers were generally found to be mainly local. However, 
transfers between different fields were more nonlocal than transfers between the same 
fields. Furthermore, large cancellations occurred between the contributions to forward and reverse 
transfers \citep{Debliquy05}. The analysis presented here also predicts cancellations between these 
contributions to occur, thus being consistent with the aforementioned numerical results.  

In terms of locality and nonlocality of energy (and helicity) transfer, we found that
nonlocal interactions contribute to forward transfer only for 
interactions of helical modes with unlike helicity and mainly if the kinetic energy 
exceeds the magnetic energy. 
Interestingly, for inverse transfers we find less constraints on nonlocal interactions. In
particular for magnetically dominated systems nonlocal interaction between modes of like 
helicity contribute to  reverse energy transfer.  
In view of the cancellations that occur between forward and reverse transfers, 
the inverse cascade may thus have a significant 
nonlocal component which is not cancelled by forward transfers within
the same triad interaction.

A numerical study of large-scale magnetic field generation in helically forced
isotropic MHD turbulence was carried out by \citet{Brandenburg01}.
It was found that the injection of energy from the velocity field into the 
magnetic field occurs directly from the forcing scale into the largest 
resolved scale, implying that this is a nonlocal process.
Due to the non-locality of the observed increase in spectral power of the 
magnetic field at the lowest resolved wavenumber $k=1$ and the excellent 
agreement of numerical results with an $\alpha$-dynamo model, the transfer of 
energy into the $k=1$ mode is explained by the $\alpha$-effect 
rather than an inverse cascade, and it is shown 
to occur after saturation of the small-scale dynamo. Our results in sec.~\ref{sec:dynamo} 
suggest that one type of helical mode interaction may be mapped to 
the $\alpha$ effect, and we established that large-scale dynamo action is more active 
in the nonlocal limit than certain other types of interactions.       

One of the main results of the present work is the possibility of inverse energy transfer
for nonhelical magnetic fields. Such inverse transfer has recently been found in high resolution
DNSs of slightly compressible \citep{Brandenburg15} and relativistic \citep{Zrake14} MHD turbulence. 
An analysis of interscale transfers showed that this 
inverse transfer was mainly due to energy transfer away from the medium scale 
(see Supplemental Material of \cite{Brandenburg15}, last figure), while energy transfer away from 
the smallest scales also occurred. The analytic approach put forward in the present paper 
also shows that energy is transferred away from the medium and small scales for interactions
of modes with unlike helicities, thus being qualitatively consistent with these numerical results.
However, since no numerical work decomposing the MHD equations into helical contributions as suggested
by \citet{Biferale12} and \citet{Biferale13} has been carried out so far, we are not in a position to 
claim numerical confirmation of our results. 

Having discussed numerical results, we now turn to measurements
of energy transfer in the solar wind. Unlike in our own analysis and in the numerical 
results discussed so far except for \citep{Cho10}, a background magnetic field 
is present in the solar wind. 
Recent measurements at 1 AU \citep{Stawarz10} showed
negative Els\"{a}sser fluxes in regions of high cross-helicity, giving possible evidence
of inverse energy transfer in these regions, which cannot be explained by selective
decay as cross-helicity cascades forwards \citep{Frisch75}. Our results may be helpful in explaining
this phenomenon as one of the results we obtained was a quenching of forward
energy transfer in regions of high cross-helicity, leaving more inverse transfer to perhaps
dominate the dynamics in these regions. 

In subsequent work \citep{Coburn14} concerns were raised on the implications of the effect 
of expansion in the solar wind especially in regions of high $H_c$. Expansion effects had been neglected 
in the previous analysis. The authors restrict their measurements to regions where the relative 
cross helicity $\sigma_c$ is not too large, that is $0 \leqslant |\sigma_c| \leqslant 0.5$ and 
measure positive energy fluxes on average, while the instantaneous flux shows large variations 
including negative values.  
It is shown that the broad distribution of the measured instantaneous 
fluxes are related to intermittency of the energy cascade in terms of 
the variability of the energy flux \citep{Politano95,Karimabadi13} and not caused by
experimental uncertainty. 
The various possibilities of energy transfer in forward and 
reverse directions determined in the present work are consistent with these measurements, as they also 
would result in broader tails of the probability distribution of the energy flux, 
even if on average energy transfer proceeds in the forward direction.
As for the concerns about the validity of the negative fluxes measured by \citet{Stawarz10}, 
our results do suggest that the measured inverse fluxes may be a 
genuine effect due to quenching of forward energy transfers if $H_c$ is large.

Since most of this discussion is based on statements of plausibility rather than certainty, 
more work clearly has to be carried out before a decisive result on energy transfer in MHD 
turbulence can be achieved, and we hope that our analysis constitues a step forwards in this direction.    
As suggested by \citet{Biferale12,Biferale13}, one could study energy and helicity transfers
numerically by projecting out helical modes of a particular sign, similar to work done by \citet{Biferale12,Biferale13}
and \citet{Sahoo15} 
in hydrodynamic turbulence. However, numerical verification of reverse spectral transfer due to the 
particular nonhelical interactions found in the present work may be difficult to obtain in that
framework, and a particular DNS study concentrating on inverse transfer for
nonhelical magnetic fields using the full MHD equations subject to
small-scale forcing may be needed in order to provide further insight.      
An analysis of Fourier-filtered transfer terms from DNSs of highly unbalanced MHD turbulence
compared to balanced MHD turbulence could be carried out in order to verify (or otherwise) 
the proposed quenching of forward transfers by high values of $H_c$, especially as 
it is not possible to quantify this effect from theoretical analysis only.  
On the analytical side, the present work may be extended to include the effects of a 
background magnetic field and of compressive fluctuations, which would be included
in the decomposition of the velocity field as modes parallel to the wavevector $\vec{k}$.
Asides from providing an advance in MHD turbulence research on a fundamental level, 
this work may contribute to the further theoretical
understanding of various physical processes involving inverse
cascade and dynamo action. This includes the evolution and origin of 
cosmological and galactic magnetic fields as well as solar physics and 
the dynamics of laboratory plasmas and turbulence in liquid metal flows.  

\section*{Acknowlegdements}
A.~B. acknowledges support from the UK Science and Technology Facilities Council, 
M.~F.~L. and M.~E.~M. are funded by the UK Engineering and Physical Sciences
Research Council (EP/K503034/1 and EP/M506515/1) and 
J.~J. was supported by a University of Edinburgh Physics and Astronomy
Summer Student Scheme Fellowship.

\appendix

\section{The dependence of $Q$ on the cross-helicity}
\label{app:crosshel}

In section \ref{sec:steady_state}, the parameter $Q$ was defined as $Q=\alpha \delta-\beta\gamma$, where
$\alpha$, $\beta$, $\gamma$ and $\delta$ were the entries of the matrix in \eqref{eq:matrix_ode}. Using 
the expressions for these terms given in \eqref{eq:entries}, we obtain  
\beq
Q=|g_{kpq}|^4 s_kk s_qq (s_k k -s_p p)(s_q q -s_p p)\left (|U_{s_p}|^4+|B_{s_p}|^4-2\mbox{Re}([U_{s_p}^*B_{s_p}]^2) \right) \ .
\label{eqapp:Q}
\eeq 
However, in \eqref{eq:Q}, instead of the term $\mbox{Re}([\Up^*\Bp]^2)$ a term involving $H_c(p)$ appeared.  

In general, the helical coefficients $\Up$ and $\Bp$ are related by a complex number $M= m+in$
such that $\Bp=M\Up$. Expressions for $m$ and $n$ can be found
by decomposing the two fields into their real and imaginary parts.
Let $\Up=U_1+iU_2$ and $\Bp=B_1+iB_2$. Then
\begin{align}
m=&\frac{1}{|\Up|^2}(U_1B_1+U_2B_2) \text{ and}\\
n=&\frac{1}{|\Up|^2}(U_1B_2-U_2B_1),
\end{align}
and we note the constraint $n^2=|\Bp|^2/|\Up|^2-m^2$ which follows from the definition of $M$.
Decomposing the cross-helicity in the same way results in $H_c(p)=|\Up|^2 m$.
Now we can relate $\mbox{Re}([\Up^*\Bp]^2)$ to the cross-helicity
by rewriting it in terms of the components of $\Up$ and $\Bp$:
\begin{align}
\mbox{Re}([\Up^*\Bp]^2)=&(U_1B_1+U_2B_2)^2-(U_1B_2-U_2B_1)^2\\
=&|\Up|^4m^2-|\Up|^4\left(\frac{|\Bp|^2}{|\Up|^2}-m^2\right)\\
=&2H_c(p)^2-|\Up|^2|\Bp|^2 \ ,
\end{align}
and we obtain \eqref{eq:Q} by
substitution of this expression for $\mbox{Re}([\Up^*\Bp]^2)$ into \eqref{eqapp:Q}

Since the maximum and minimum values of $|H_c(p)|$ are
$|\Up||\Bp|$ and $0$ respectively, it is useful to define the relative cross-helicity
$\rho=H_c(p)/(|\Up||\Bp|)$, which takes values between -1 and 1.
We obtain
\beq
\mbox{Re}([\Up^*\Bp]^2)=|\Up|^2|\Bp|^2 (2\rho^2-1) \ ,
\eeq
which is bounded by $-|\Up|^2|\Bp|^2$ and $|\Up|^2|\Bp|^2$,
where the first value is the case of vanishing cross-helicity
and the latter occurs when there is maximal cross-helicity.
This implies that the term $(|U_{s_p}|^4+|B_{s_p}|^4+2|\Up|^2|\Bp|^2-4H_c(p)^2)$ 
in \eqref{eq:Q} cannot be negative. 

\section{$x^2-Q>0$ for $s_k = s_q \neq s_p$}
\label{app:Q_calc}
In section \ref{sec:no_trans}, the result was dependent on whether $x^2-Q$ is positive 
or negative. Recall that the helicity combination in question was $s_k = s_q \neq s_p$.
Therefore $x^2-Q$ becomes
\begin{align}
\label{eq:appQ1}
x^2-Q =& \frac{1}{4}  [|U_{s_p}|^4 (kq+(k+p)(q+p))^2 + |B_{s_p}|^4 (k(k+p)+q(q+p))^2 \nonumber \\ 
      &+ 2 |U_{s_p}|^2 |B_{s_p}|^2 [kq+(k+p)(q+p)][k(k+p)+q(q+p)]  ] \nonumber \\
      & -[kq (k +p)(q +p)](|U_{s_p}|^4+|B_{s_p}|^4-2\mbox{Re}([U_{s_p}^*B_{s_p}]^2)) \nonumber \\
      =& \frac{1}{4}  [|U_{s_p}|^4 (kq-(k+p)(q+p))^2 + |B_{s_p}|^4 (k(k+p)-q(q+p))^2 \nonumber \\ 
      &+ 2 |U_{s_p}|^2 |B_{s_p}|^2 [kq+(k+p)(q+p)][k(k+p)+q(q+p)]  ] \nonumber \\
      & +2\mbox{Re}([U_{s_p}^*B_{s_p}]^2)[kq (k +p)(q +p)] \ .
\end{align}
If $\mbox{Re}([U_{s_p}^*B_{s_p}]^2)>0$, then this implies $x^2-Q>0$. If $\mbox{Re}([U_{s_p}^*B_{s_p}]^2)<0$,
then some more steps are required to show that $x^2-Q >0$. 
In general, we have $|\mbox{Re}([U_{s_p}^*B_{s_p}]^2)|\leqslant|U_{s_p}|^2 |B_{s_p}|^2$
(see appendix \ref{app:crosshel}), hence assume 
$\mbox{Re}([U_{s_p}^*B_{s_p}]^2)=-|U_{s_p}|^2 |B_{s_p}|^2$ as this would be the most 
negative value this term can take. It corresponds to zero cross-helicity at $\vec{p}$.
Equation \eqref{eq:appQ1} is now an inequality and reads
\begin{align} 
x^2-Q \geqslant & \frac{1}{4} \left [|U_{s_p}|^4 (kq-(k+p)(q+p))^2 + |B_{s_p}|^4 (k(k+p)-q(q+p))^2 \right ] \nonumber \\ 
      &+ \frac{1}{2} |U_{s_p}|^2 |B_{s_p}|^2 [kq+(k+p)(q+p)][k(k+p)+q(q+p)]   \nonumber \\
      & -2 |U_{s_p}|^2 |B_{s_p}|^2[kq (k +p)(q +p)] \ .
\end{align}
The result $x^2-Q>0$ follows immediately if we can show 
\beq
\label{eq:ineq_pq}
\frac{[kq+(k+p)(q+p)][k(k+p)+q(q+p)]}{kq(k+p)(q+p)} \geqslant 4 \ .
\eeq
Defining $v\equiv p/k$ and $w\equiv q/k$, this inequality can also be written as
\beq
\label{eq:ineq_vw}
\frac{1}{v+w} + \frac{w}{1+v} + \frac{1+v + w(v+w)}{w} \geqslant 4 \ ,
\eeq
where $v$ and $w$ must satisfy $1\leqslant v \leqslant w < 1+v$ due to the 
triad geometry. This implies that the first and second terms on the 
left-hand side (LHS) of the inequality are smaller than unity, and the third term is the 
largest of the three. In order to prove the inequality, we therefore concentrate 
on the third term and derive a lower bound for it. 
We proceed by defining a family of curves $F_v(w)$ by 
\beq
F_v(w)\equiv \frac{1+v + w(v+w)}{w} \ , 
\eeq  
and aim to find their minima depending on the parameter $v$. The triad geometry 
will then give a smallest allowed value of $v$ and thus provide the smallest possible 
value of the third term on the LHS of \eqref{eq:ineq_vw}.
Differentiating $F_v(w)$ with respect to $w$ yields
\beq
\frac{d}{dw}F_v(w)=\frac{w(2w+v)-(1+v+w[v+w])}{w^2} =\frac{w^2-(1+v)}{w^2}  \ ,
\eeq  
setting this expression equal to zero results in $w=\pm\sqrt{1+v}$, and it is easily checked 
that $(d^2/dw^2)F_v(w) > 0$. By definition $w>0$, hence $w=\sqrt{1+v}$, and we obtain
\beq
F_v(\sqrt{1+v})=\frac{1+v + \sqrt{1+v}(v+\sqrt{1+v})}{\sqrt{1+v}}=2\sqrt{1+v}+v \ . 
\eeq  
This expression is minimal for $v=1$ and in this case equal to 
$F_1(\sqrt{2})=2\sqrt{2}+1 \simeq 3.828$, which is the lower bound for 
the largest term on the LHS of \eqref{eq:ineq_vw}. Substituting the 
corresponding values $v=1$ and $w=\sqrt{2}$ into the remaining terms on the  
LHS of \eqref{eq:ineq_vw} we obtain
\beq
\frac{1}{1+\sqrt{2}} + \sqrt{2} + 2\sqrt{2}+1 \simeq 1.828 + 3.828 > 4 \ ,
\eeq
thus the inequality \eqref{eq:ineq_pq} is satisfied and $x^2-Q >0$. 

\section{Graphical determination of contraints on stability}
\label{app:graphs}
As explained in the main body of the text, the term $Q$ given in \eqref{eq:Q} 
determines the stability of the system \eqref{eq:matrix_ode} if $x<0$. As such, 
a solution is unstable if $Q<0$ or if $x^2-Q < 0$, where the latter case is the more 
difficult to determine, as the sign of $x^2-Q$ depends on the shape of the wavenumber 
triad, the cross-helicity and the ratio $|U_{s_p}|/|B_{s_p}|$. 
Given the multitude of possibilities that can emerge for this, the simplest way of determining 
the constraints on the stability of a solution of \eqref{eq:matrix_ode} is using a 
graphical method. For each combination of helicities
we plot $x^2 - Q$ for several set values of $|U_{s_p}|/|B_{s_p}|$ and $H_c(p)$ 
in order to show in which parameter range instabilities are more likely to 
occur. 

The dependence of $x^2-Q$ on the triad $k,p,q$ can be reduced to a dependence 
on the triad's {\em shape} by rescaling each wavenumber similar to the 
procedure in appendix \ref{app:Q_calc}, which enables us to use 
two-dimensional plots and the triad geometry to obtain the necessary information. 
Figures \ref{fig:case1}-\ref{fig:case4} show the function $x^2(v,w)-Q(v,w)$ for the different 
cases shown in tables \ref{tbl:split}-\ref{tbl:forward}, where $v$ and $w$ correspond to 
the smallest and largest wavenumber in a given triad, rescaled by the middle one such 
that the triad geometry enforces the contraint $0 < v \leqslant 1 \leqslant w < 1+v$,
hence each wavenumber pair $(v,w)$ describes a shape of triad.
Each subfigure corresponds to set values of $H_c(p)$ and $|U_{s_p}|/|B_{s_p}|$, 
while each point $(v,w)$ in a particular graph corresponds to a class of triad interactions
characterised by their shape. Regions in wavenumber space excluded from the analysis by
the constraints of the triad geometry are shaded in grey, positive values of $x^2-Q$ 
leading to stability are indicated in black and negative values of $x^2-Q$ leading 
to unstable solutions are marked white.
Across the main four figures, $H_c(p)$ increases towards
the bottom of the figure while $|U_{s_p}|/|B_{s_p}|$ increases from left to right, leading 
to the constraints summarised in tables \ref{tbl:split}-\ref{tbl:forward}.
Depending on the wavenumber ordering, the definitions of $v$ and $w$
are slightly different, and we explain the procedures for each case individually. 
\begin{itemize} 
\item $s_k \neq s_p = s_q$ and $k<p<q$ \\
In this case we rescale all wavenumbers by $p$, such that $v\equiv  k/p$ and $w\equiv q/p$. 
As can be seen in fig.~\ref{fig:case1}, for decreasing $|\Up|/|\Bp|$ and increasing $H_c(p)$ 
less and less unstable solutions occur and we obtain the constraints on split transfer shown in table \ref{tbl:split}.

\begin{figure}
  \centering
  \subfigure{\includegraphics[width=0.3\textwidth]{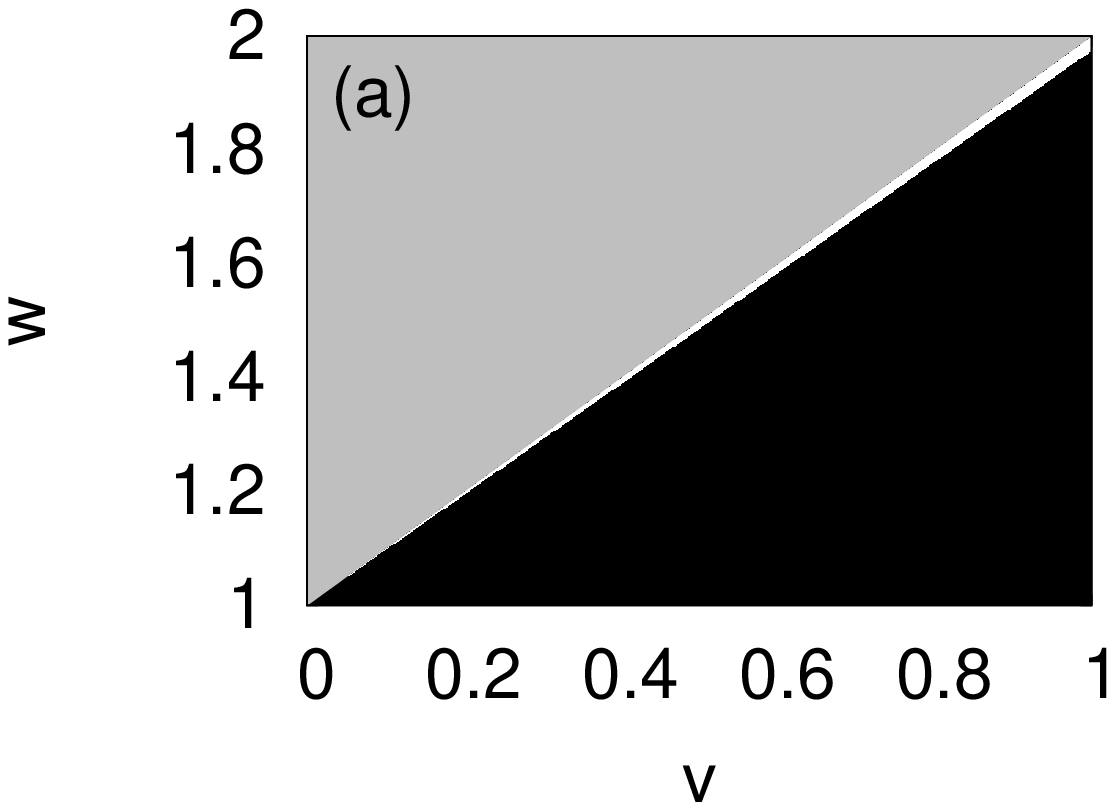}}
  \subfigure{\includegraphics[width=0.3\textwidth]{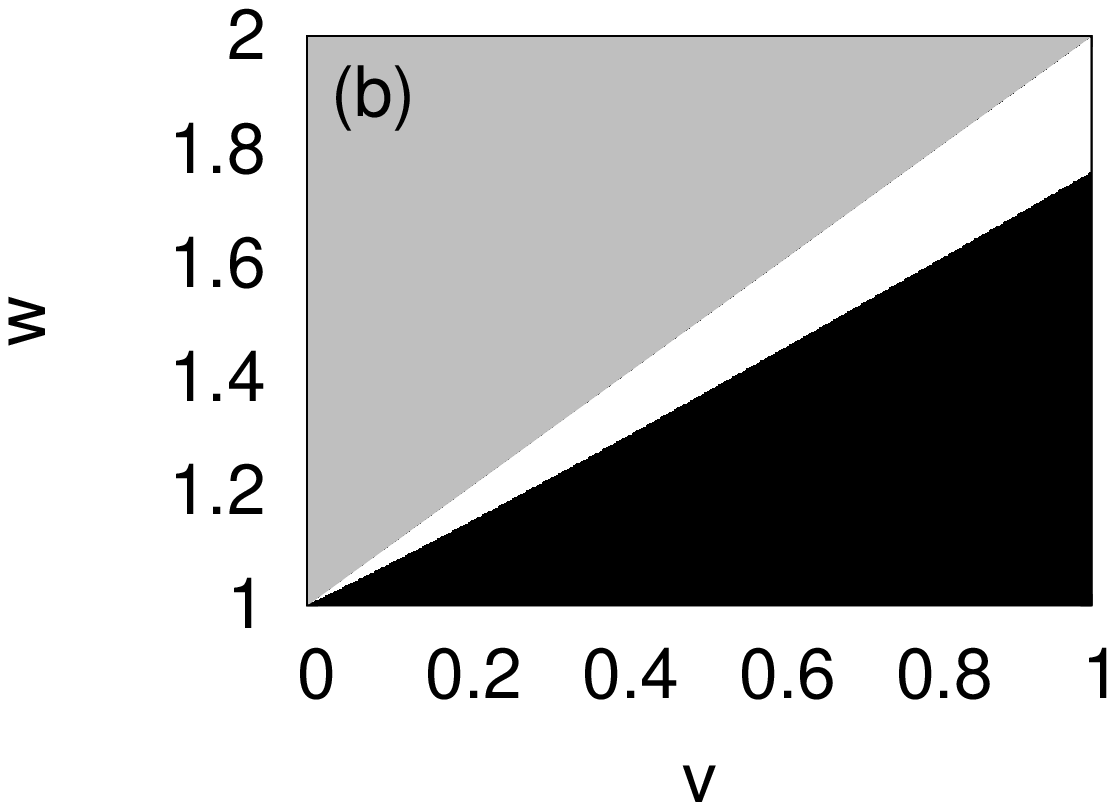}}
  \subfigure{\includegraphics[width=0.3\textwidth]{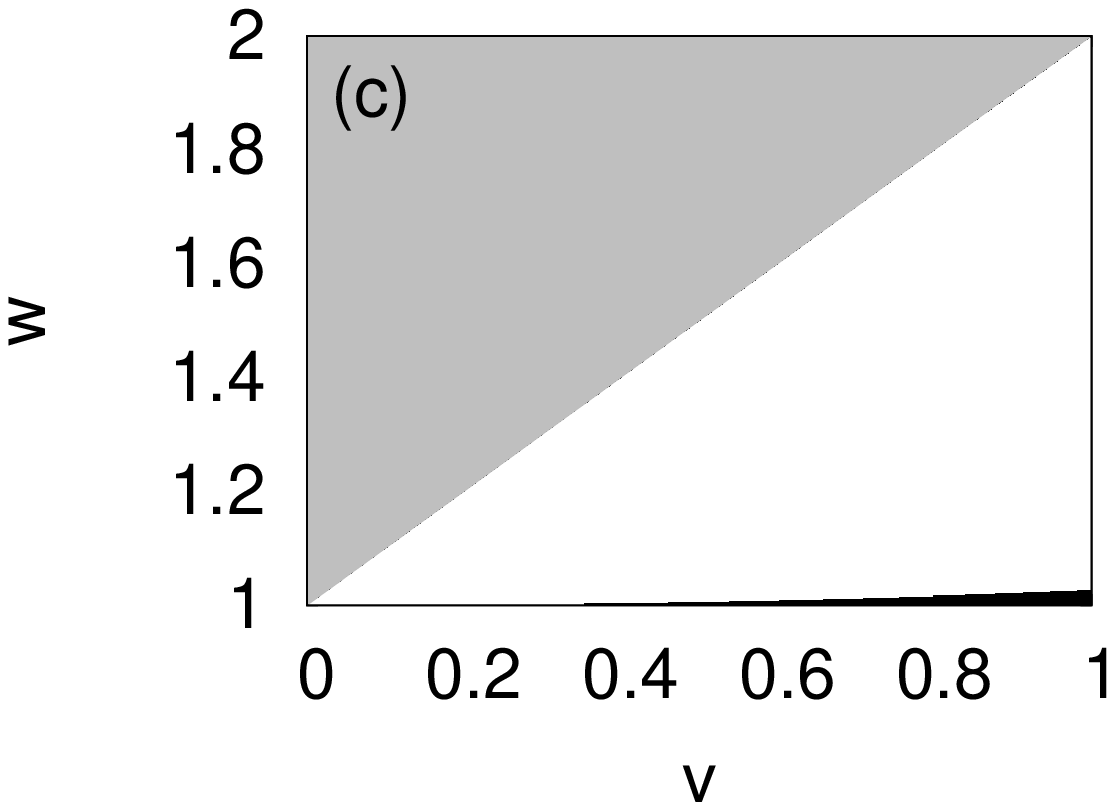}}
  \subfigure{\includegraphics[width=0.3\textwidth]{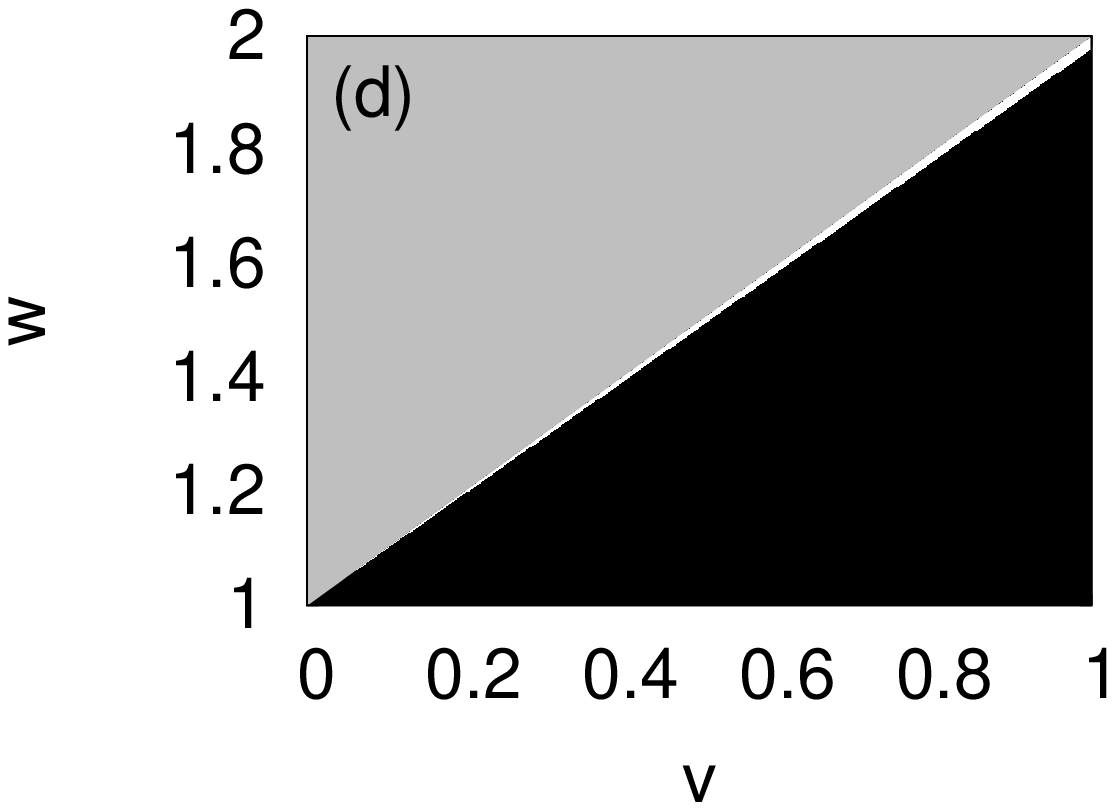}}
  \subfigure{\includegraphics[width=0.3\textwidth]{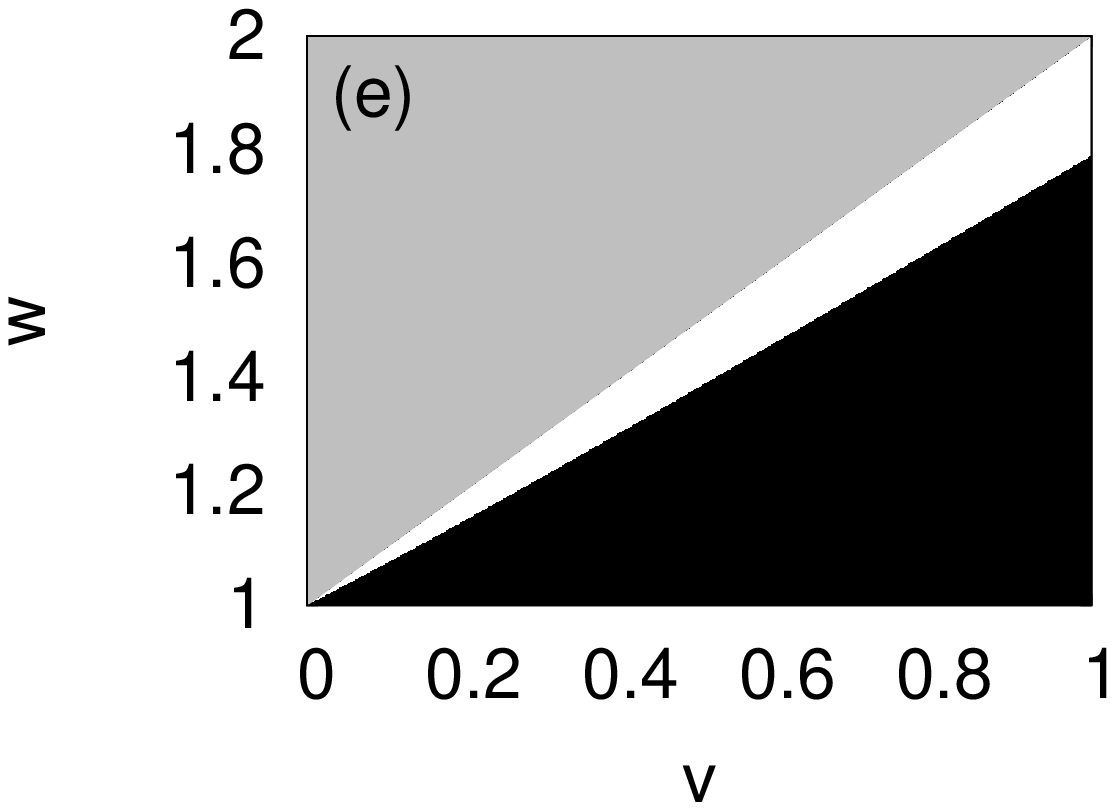}}
  \subfigure{\includegraphics[width=0.3\textwidth]{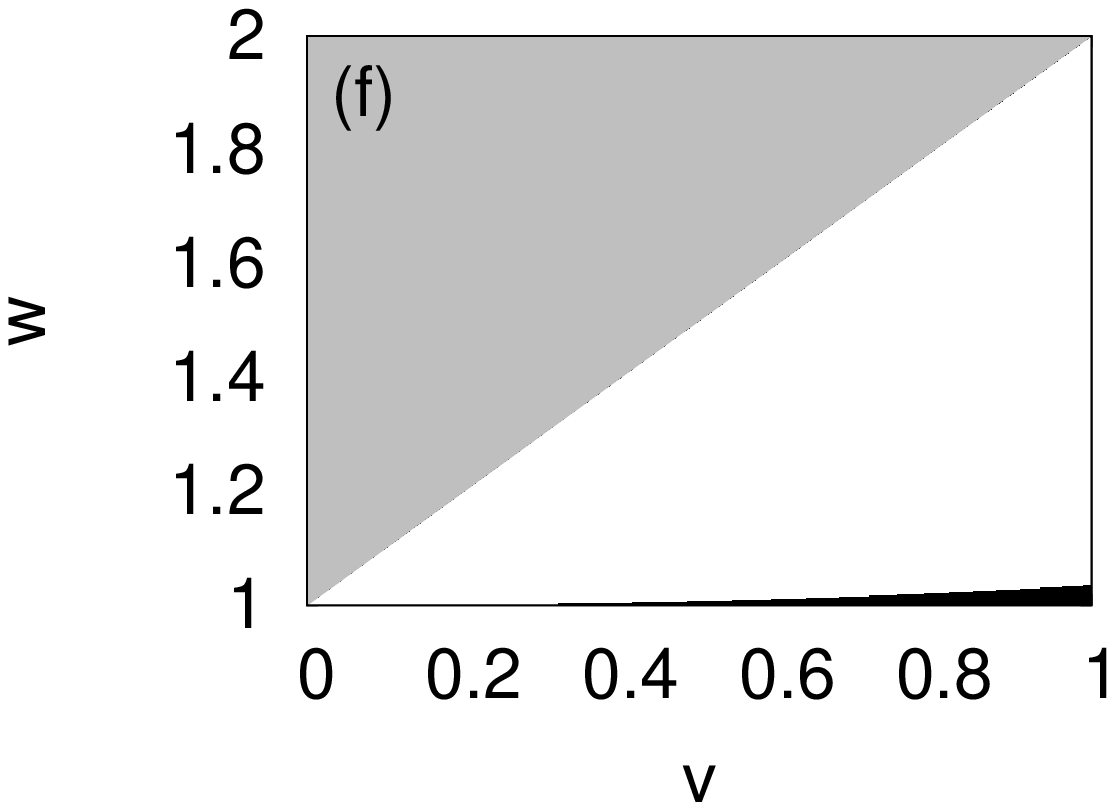}}
  \subfigure{\includegraphics[width=0.3\textwidth]{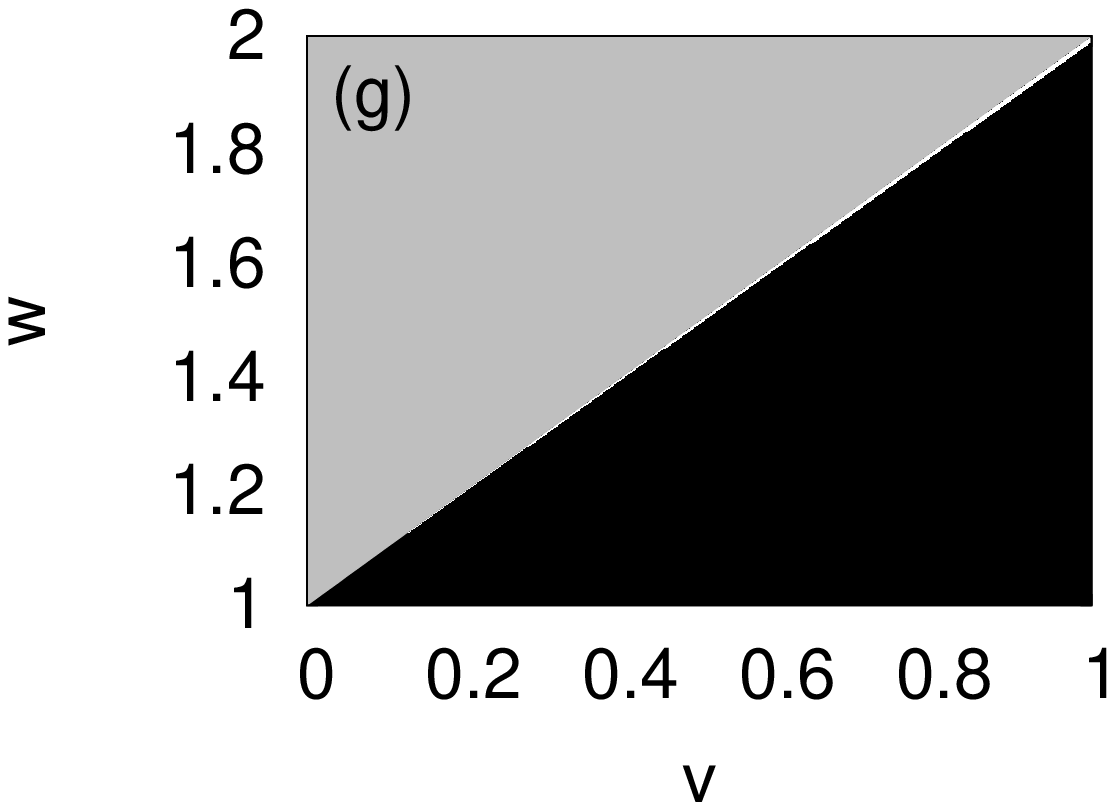}}
  \subfigure{\includegraphics[width=0.3\textwidth]{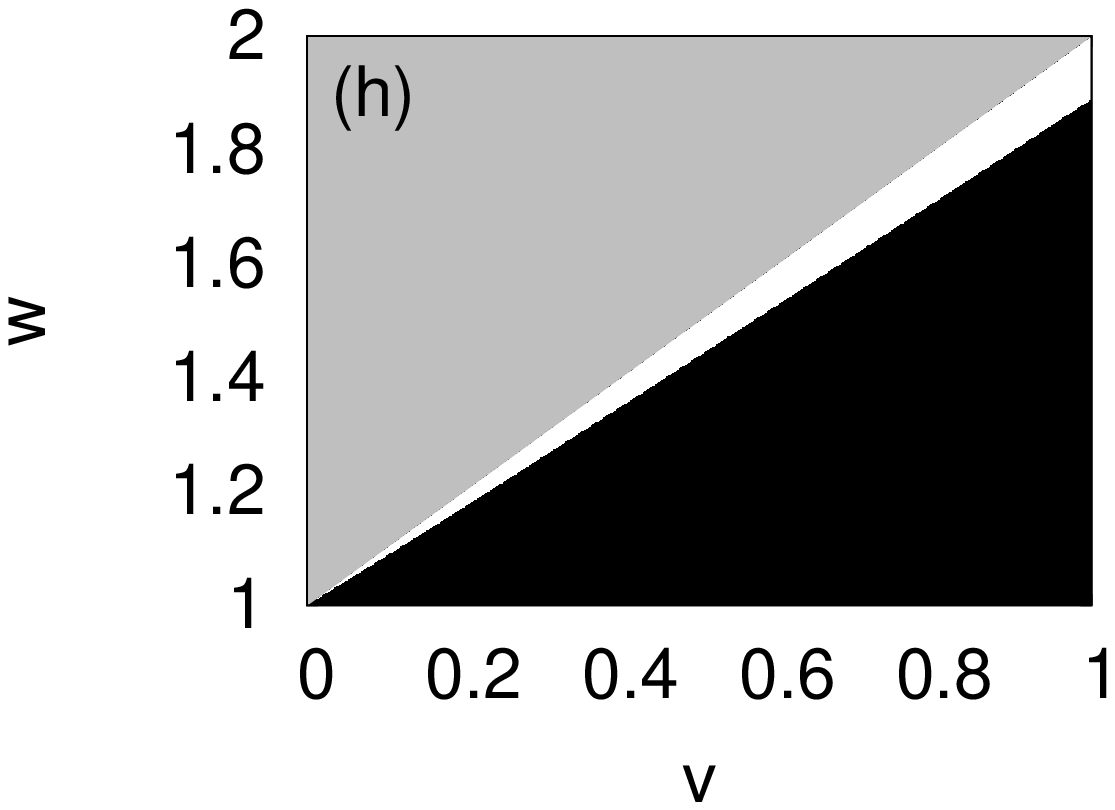}}
  \subfigure{\includegraphics[width=0.3\textwidth]{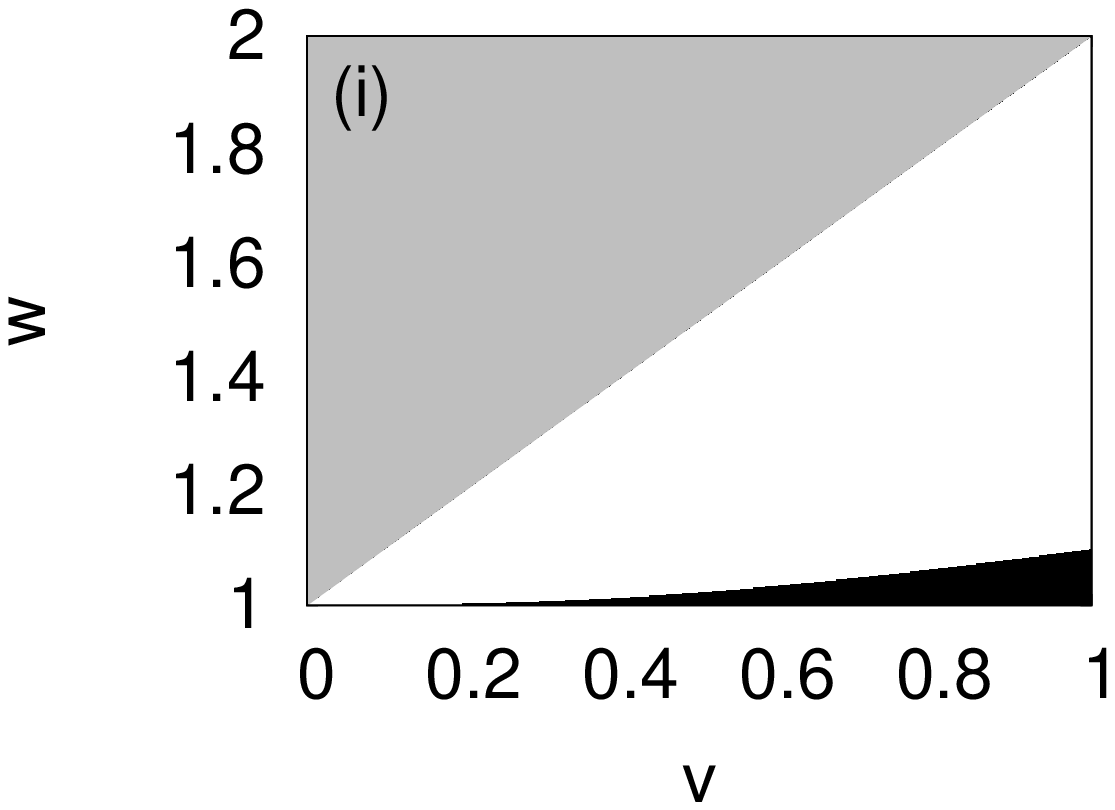}}
  \subfigure{\includegraphics[width=0.3\textwidth]{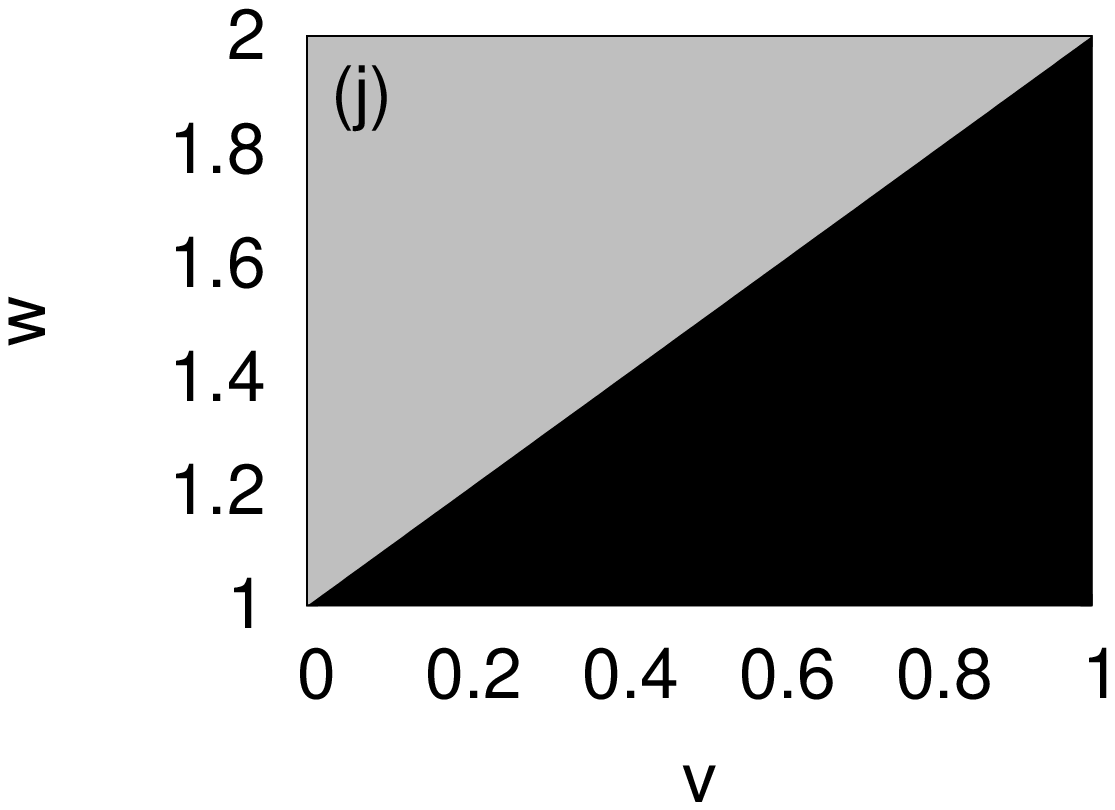}}
  \subfigure{\includegraphics[width=0.3\textwidth]{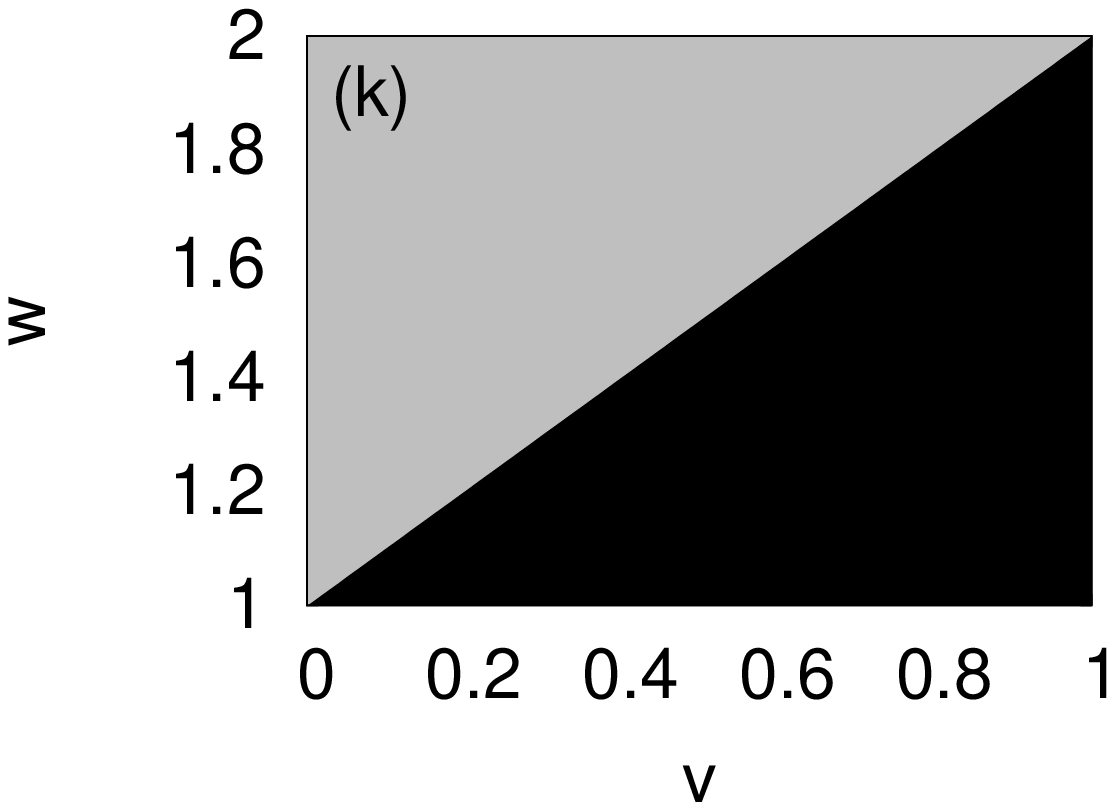}}
  \subfigure{\includegraphics[width=0.3\textwidth]{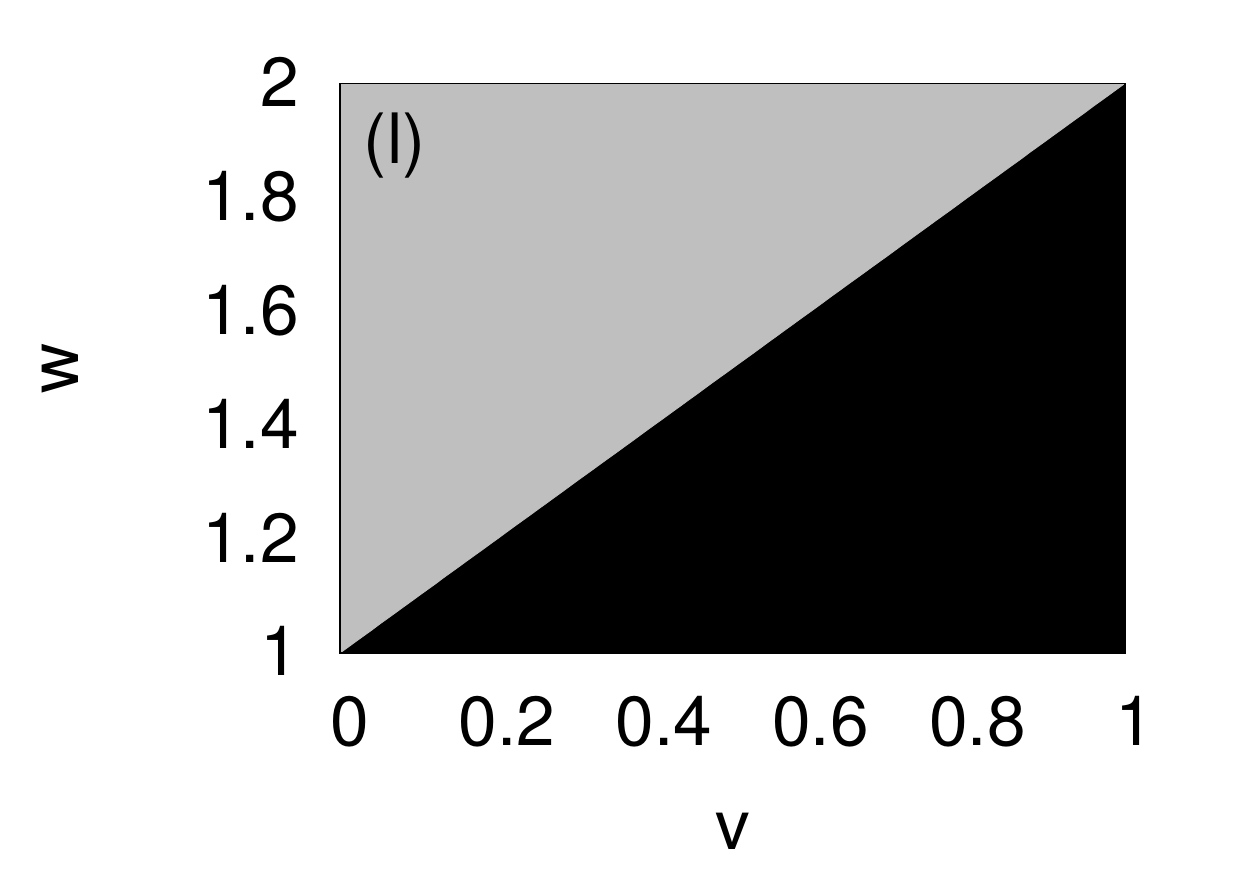}}
  \caption{Plots of $f(v,w)=x^2-Q$ for various values of $|\Up|/|\Bp|$ and cross-helicity
           for case 1 in appendix \ref{app:graphs} ($s_k\neq s_p=s_q, k<p<q$).
           The upper grey triangle is ruled out by the condition $w<1+v$ and unstable values are shown in white.
           The ratio $|\Up|/|\Bp|$ increases from left to right, with each column of subfigures taking the values
           0.01, 0.1 and 1 respectively, while each row takes the following values of relativ cross-helicity:
           $H_c(p)/(|\Up||\Bp|)=0, 0.5, 0.9$ and $1$.}
  \label{fig:case1}
\end{figure}

\item $s_k \neq s_p = s_q$ and $p<k<q$ \\ 
In this case we rescale all wavenumbers by $k$, such that $v\equiv  p/k$ and $w\equiv q/k$. 
As can be seen in 
fig.~\ref{fig:case2}, for decreasing $|\Up|/|\Bp|$ and increasing $H_c(p)$ less and less unstable solutions occur and
we obtain the constraints on forward transfer shown in table \ref{tbl:forward}.

\begin{figure}
  \centering
  \subfigure{\includegraphics[width=0.3\textwidth]{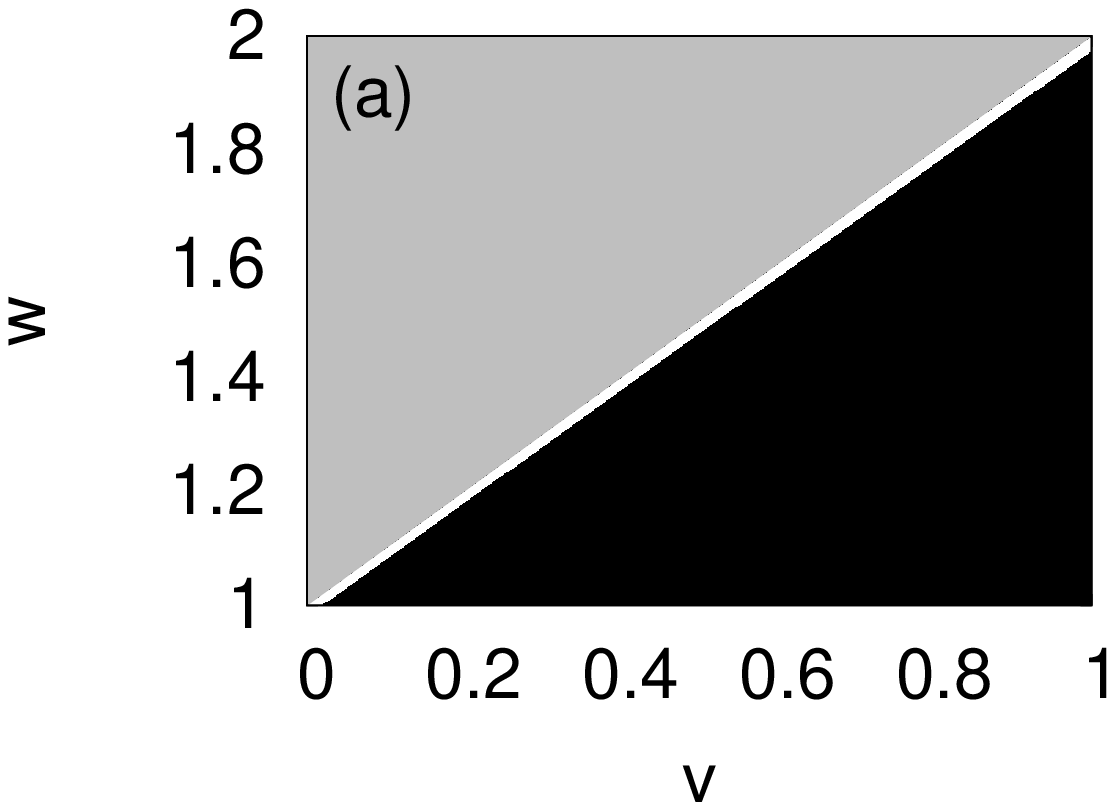}}
  \subfigure{\includegraphics[width=0.3\textwidth]{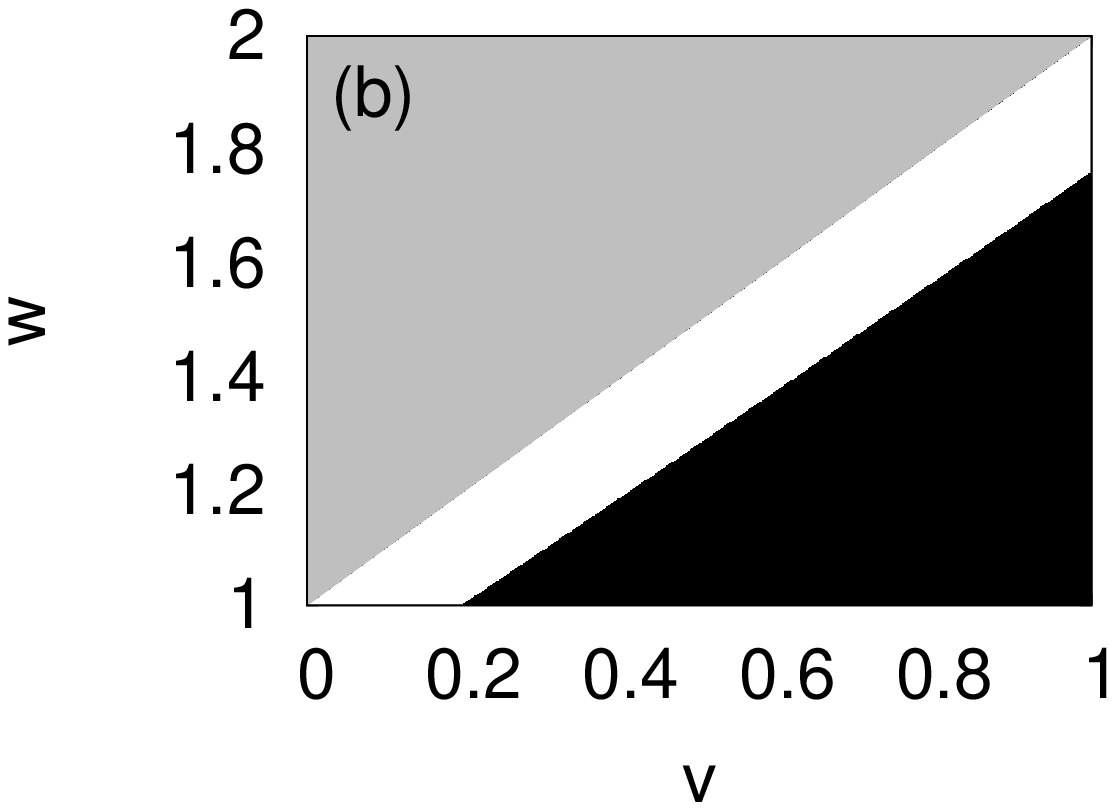}}
  \subfigure{\includegraphics[width=0.3\textwidth]{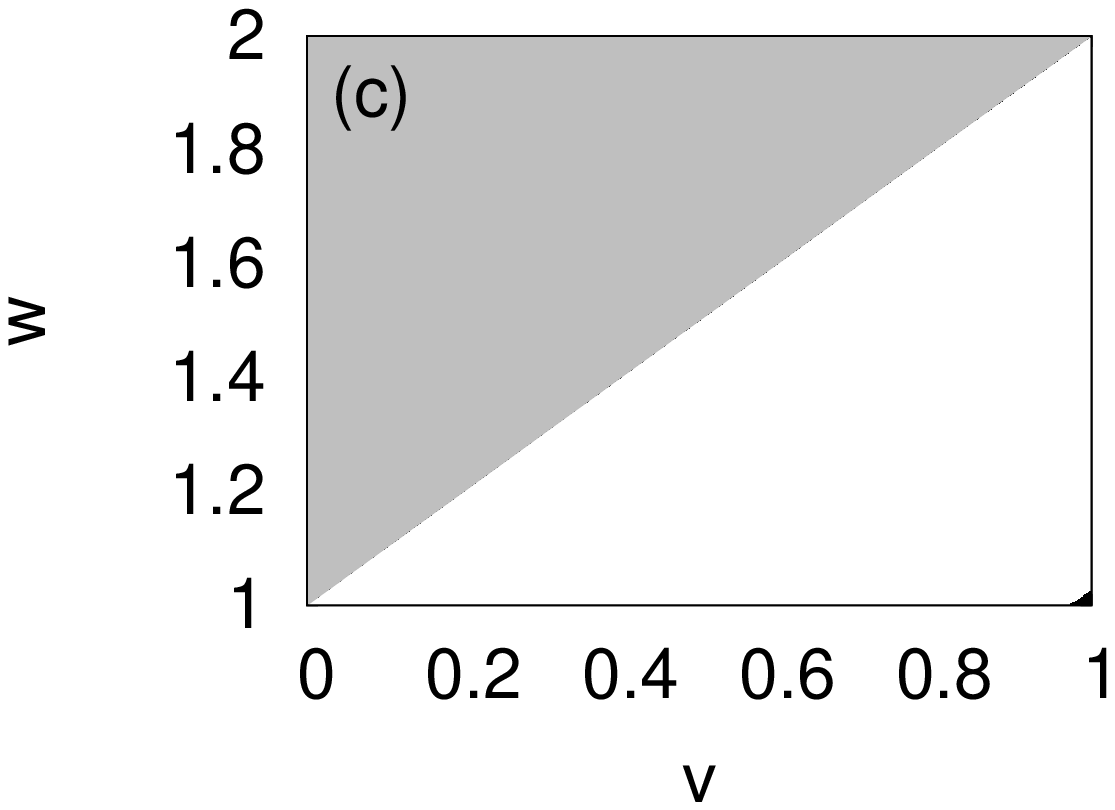}}
  \subfigure{\includegraphics[width=0.3\textwidth]{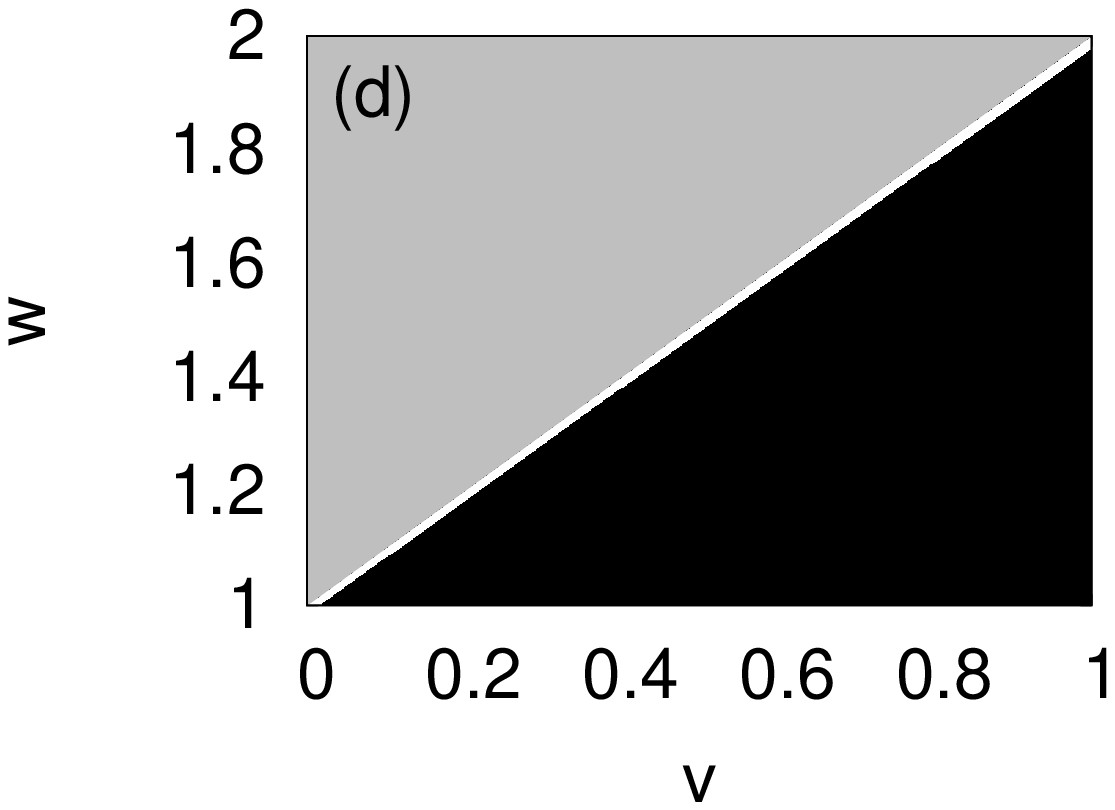}}
  \subfigure{\includegraphics[width=0.3\textwidth]{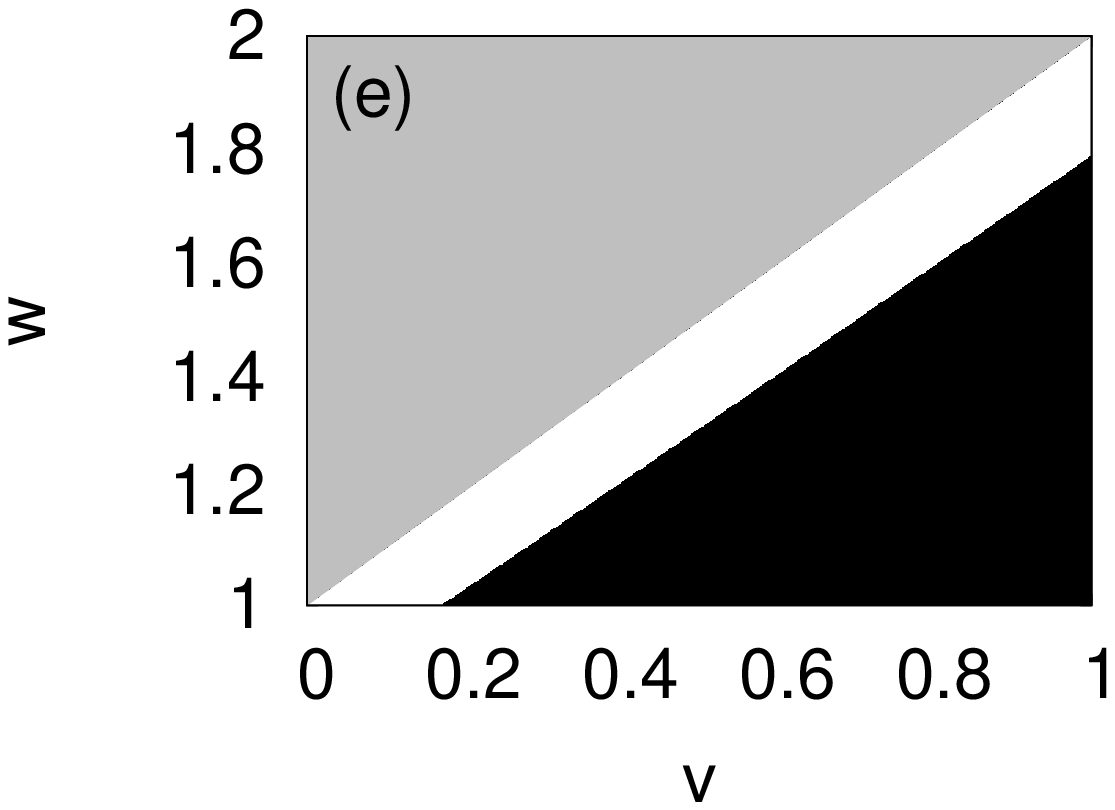}}
  \subfigure{\includegraphics[width=0.3\textwidth]{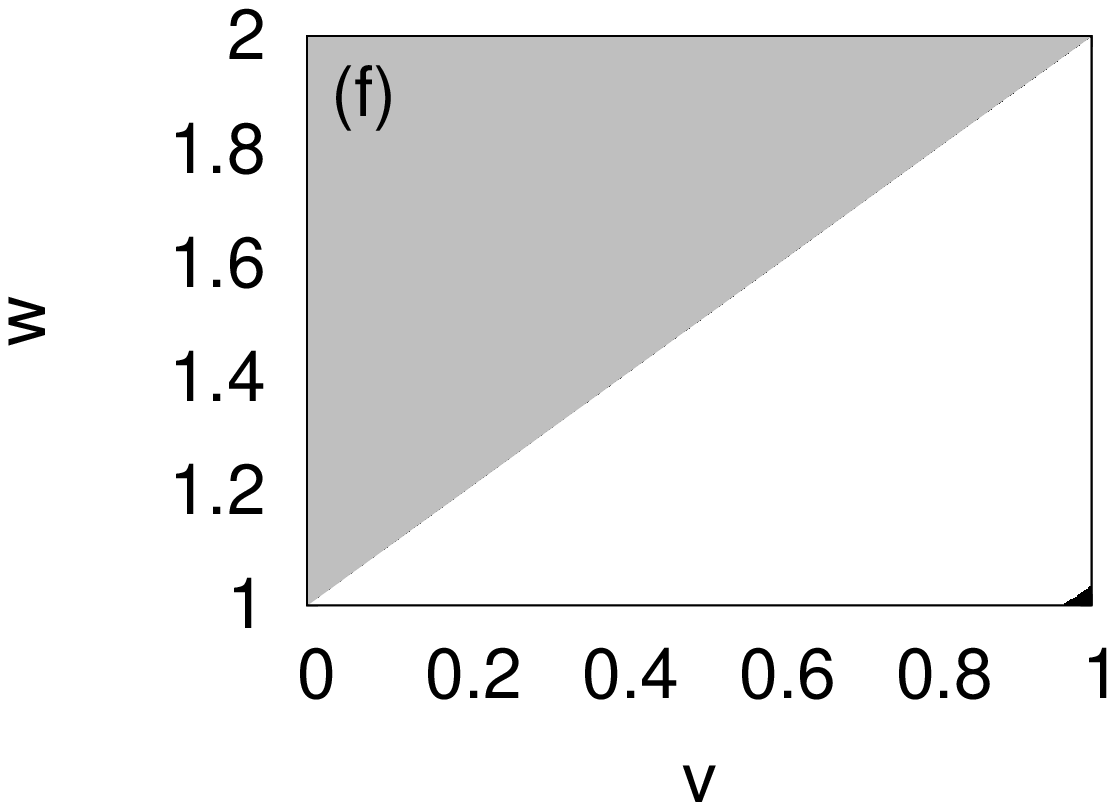}}
  \subfigure{\includegraphics[width=0.3\textwidth]{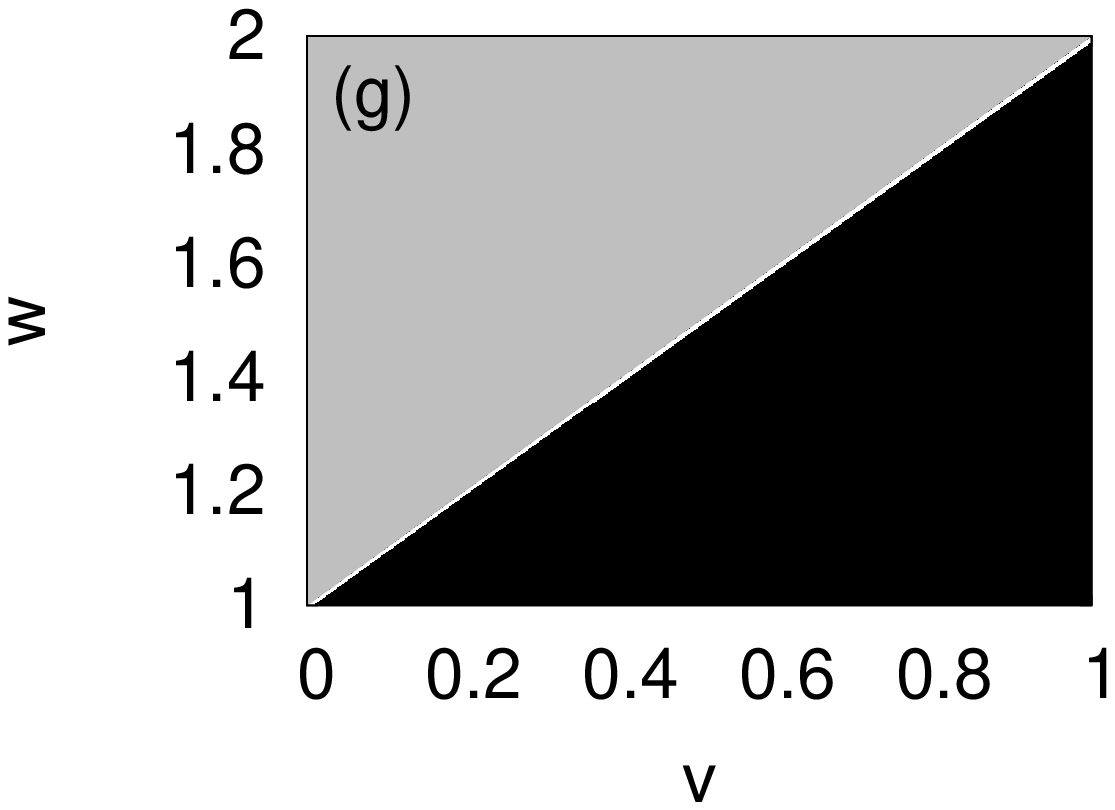}}
  \subfigure{\includegraphics[width=0.3\textwidth]{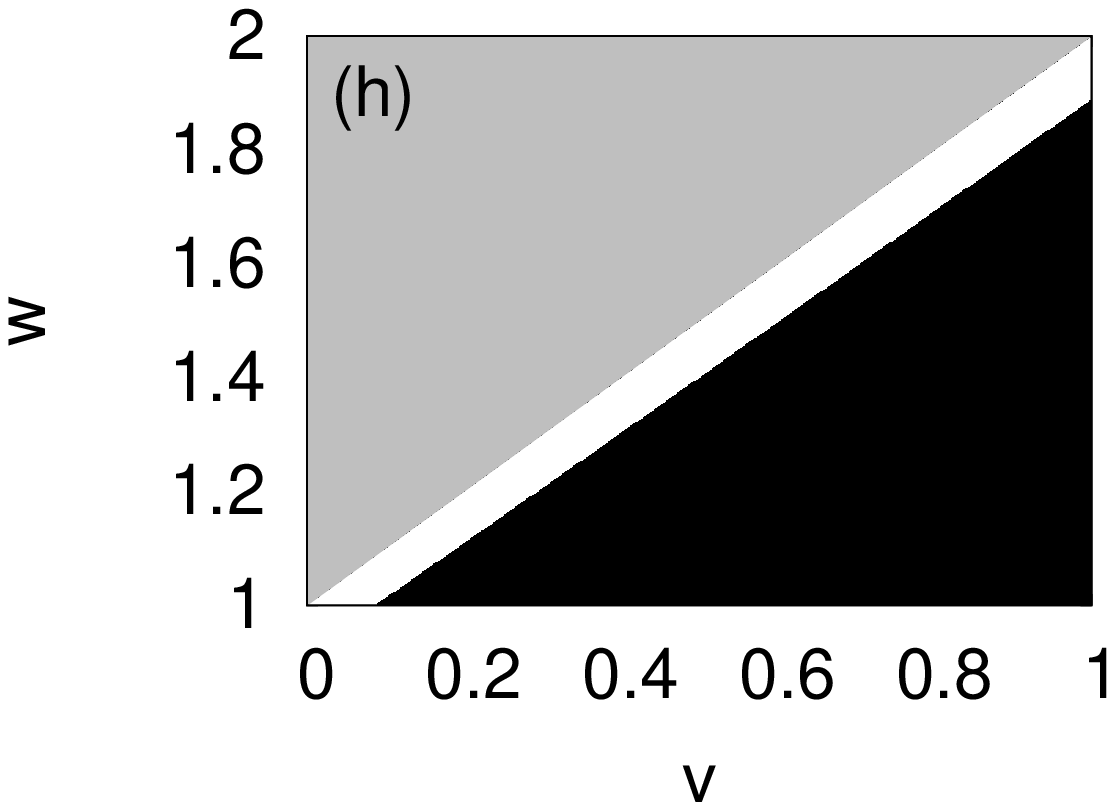}}
  \subfigure{\includegraphics[width=0.3\textwidth]{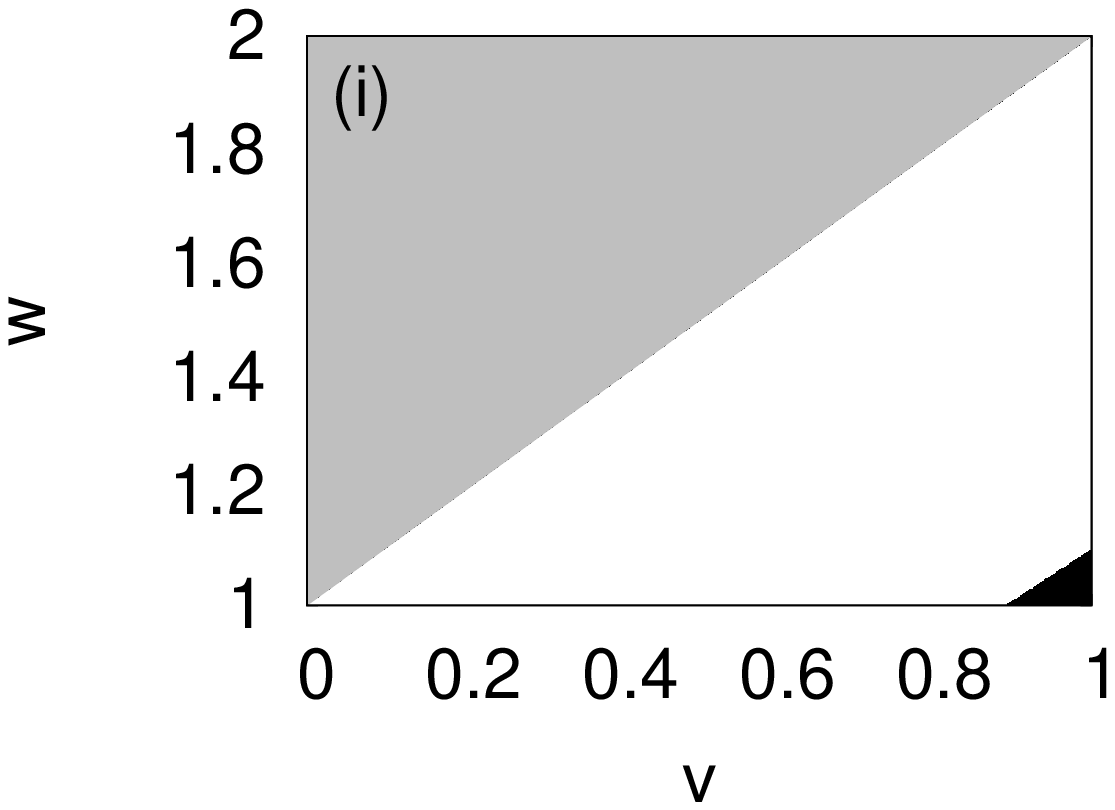}}
  \subfigure{\includegraphics[width=0.3\textwidth]{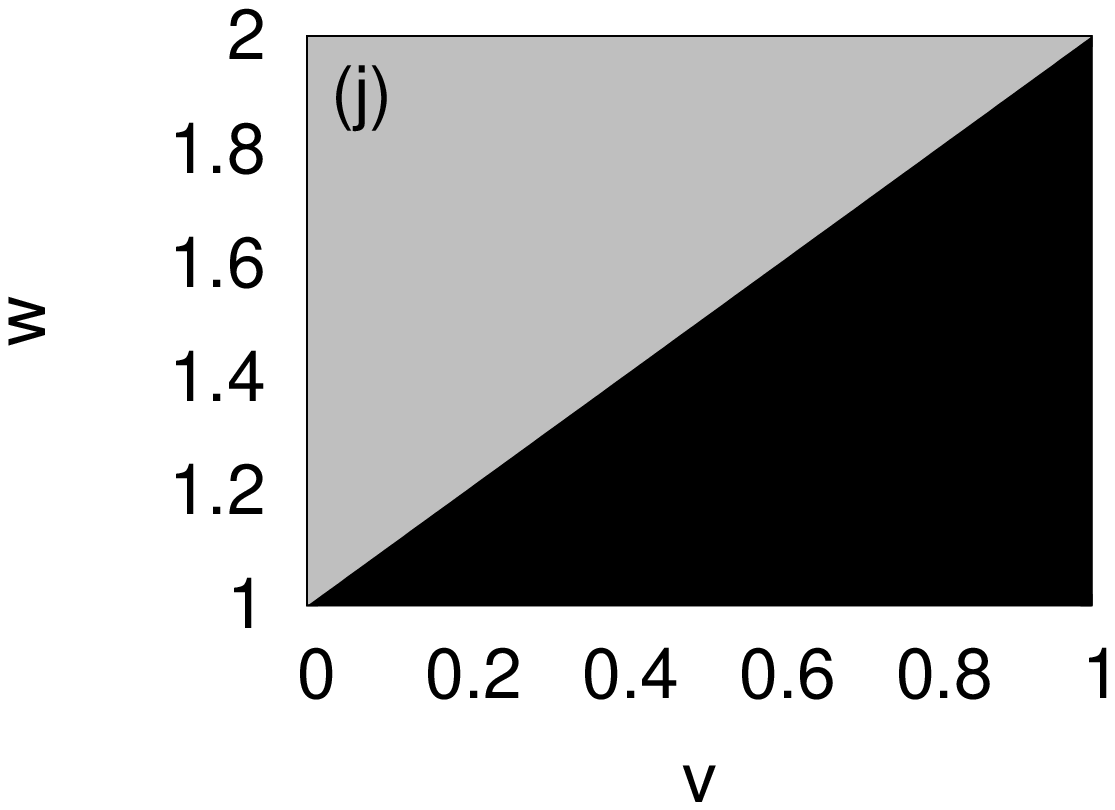}}
  \subfigure{\includegraphics[width=0.3\textwidth]{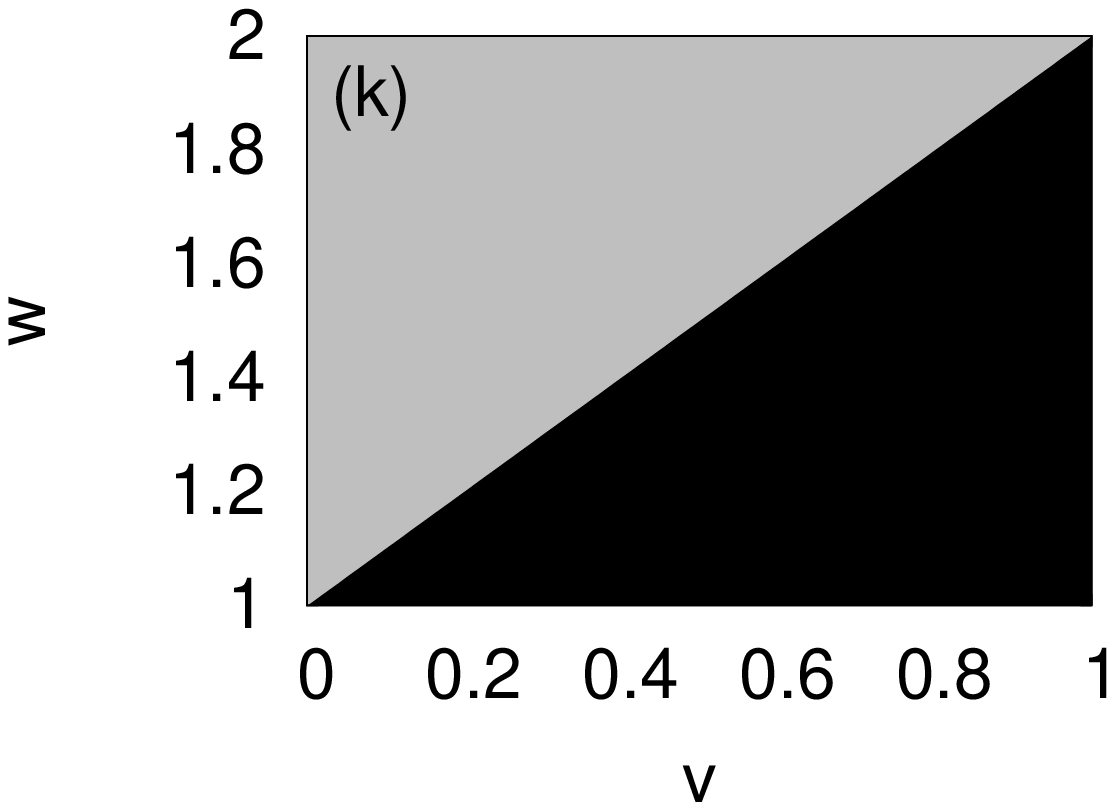}}
  \subfigure{\includegraphics[width=0.3\textwidth]{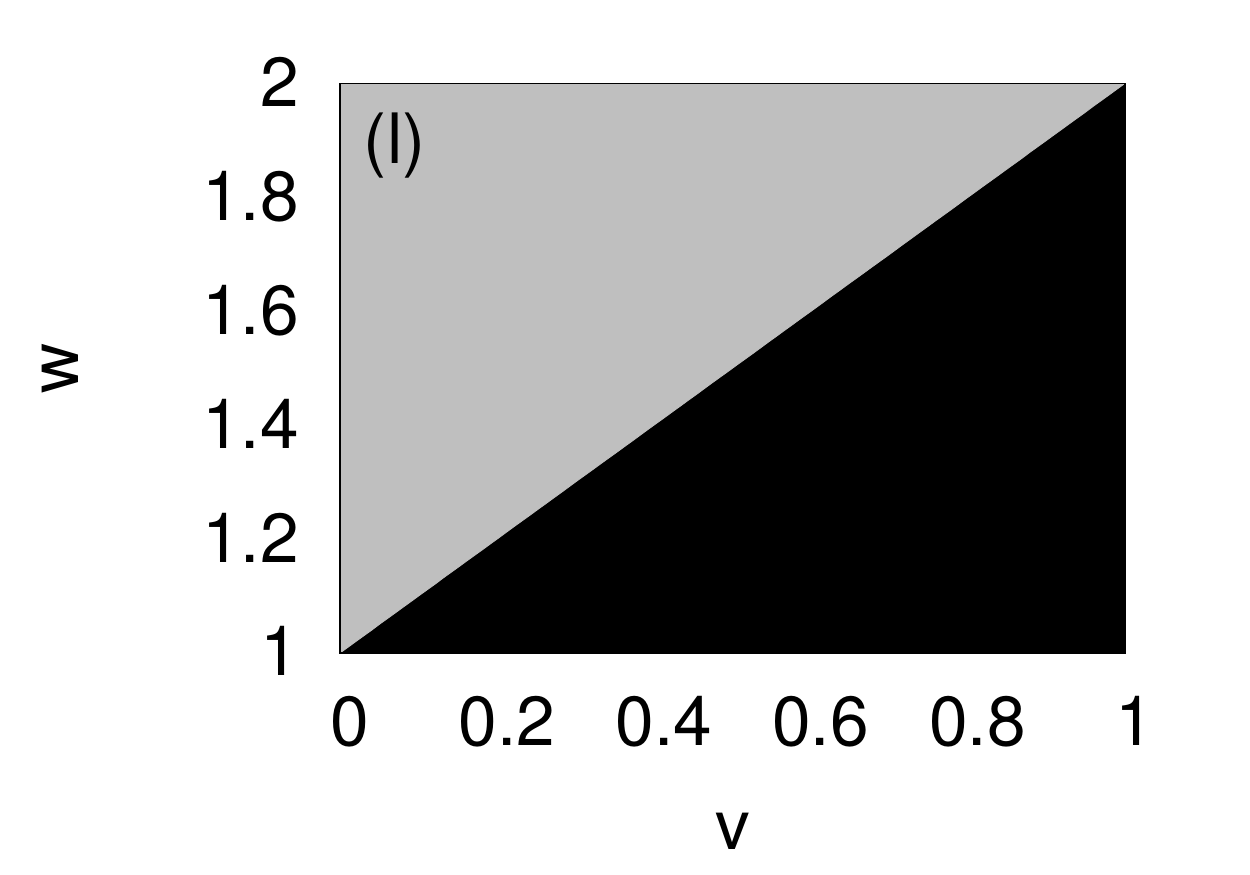}}
  \caption{Plots of $f(v,w)=x^2-Q$  for various values of $|\Up|/|\Bp|$ and cross-helicity
           for case 2 in appendix \ref{app:graphs} ($s_k\neq s_p=s_q, p<k<q$).
           The upper grey triangle is ruled out by the condition $w<1+v$ and unstable values are shown in white.
           The ratio $|\Up|/|\Bp|$ increases from left to right, with each column of subfigures taking the values
           0.01, 0.1 and 1 respectively, while each row takes the following values of relative cross-helicity:
           $H_c(p)/(|\Up||\Bp|)=0, 0.5, 0.9$ and $1$.}
  \label{fig:case2}
\end{figure}

\item $s_k = s_p \neq s_q$ and $p<k<q$ \\ 
In this case we rescale all wavenumbers by $k$, such that $v\equiv  p/k$ and $w\equiv q/k$. 
As can be seen in fig.~\ref{fig:case3}, for decreasing $|\Up|/|\Bp|$ and increasing $H_c(p)$ less and less unstable solutions occur and
we obtain the constraints on forward transfer shown in table \ref{tbl:forward}.
 
\begin{figure}
  \centering
  \subfigure{\includegraphics[width=0.3\textwidth]{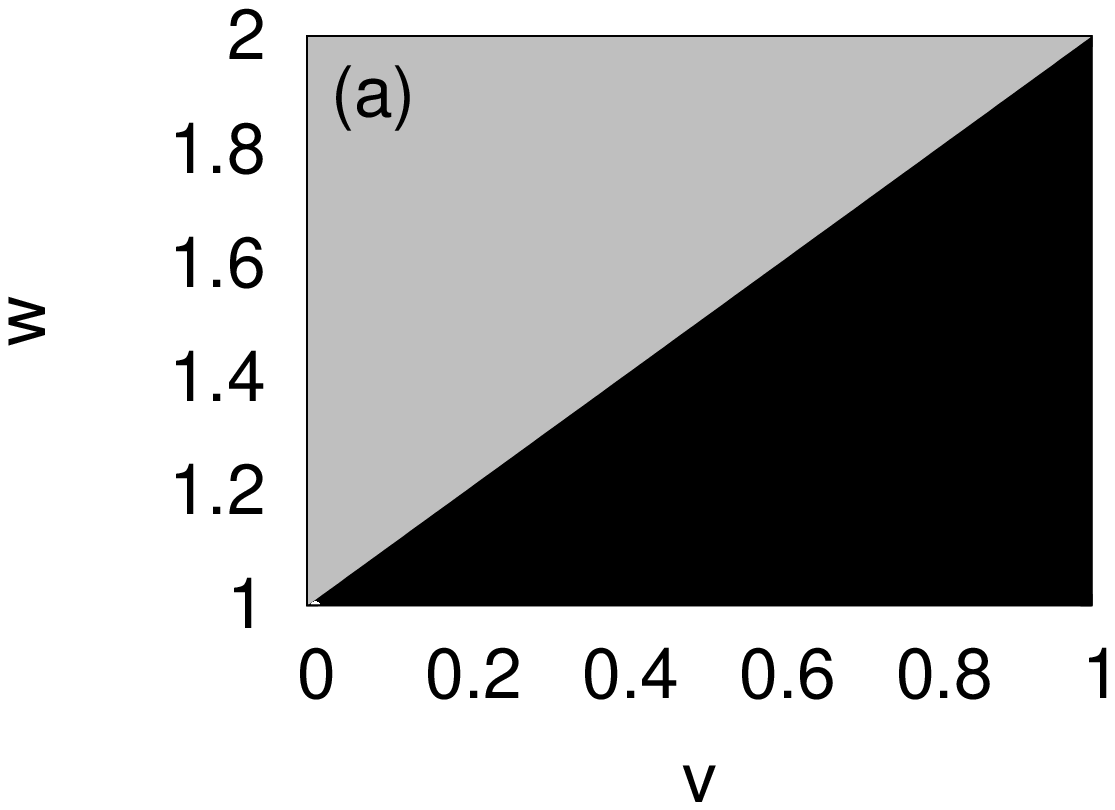}}
  \subfigure{\includegraphics[width=0.3\textwidth]{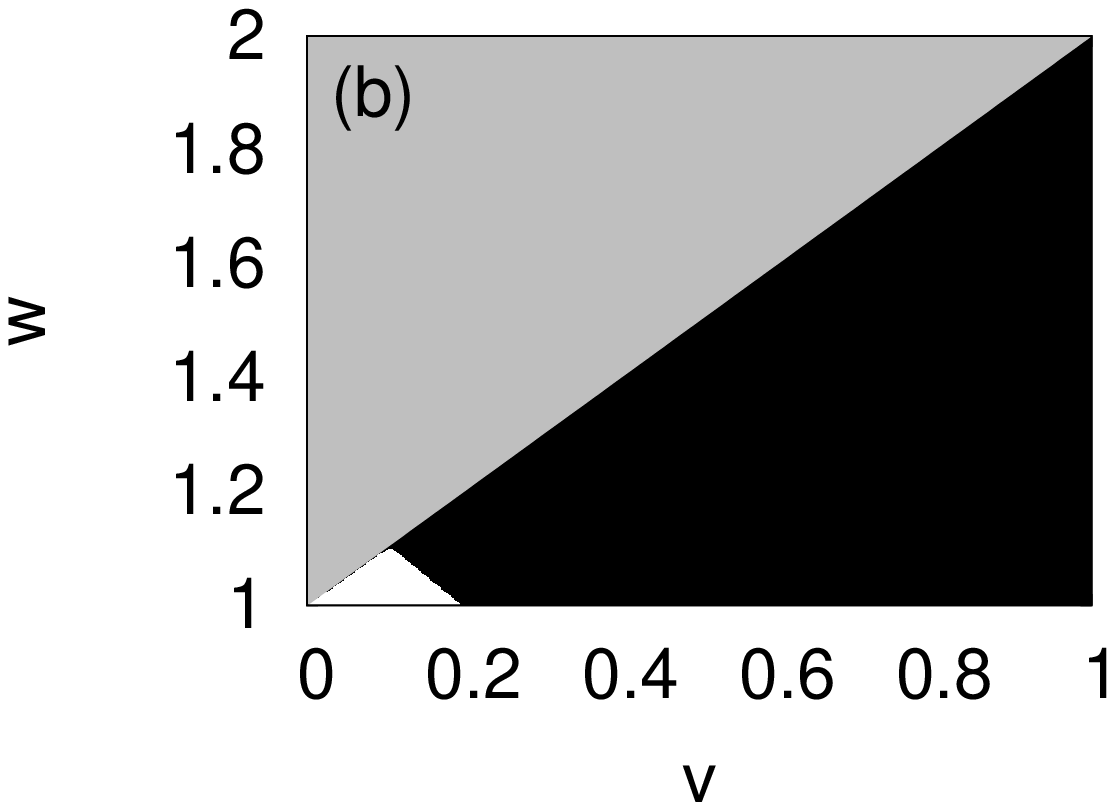}}
  \subfigure{\includegraphics[width=0.3\textwidth]{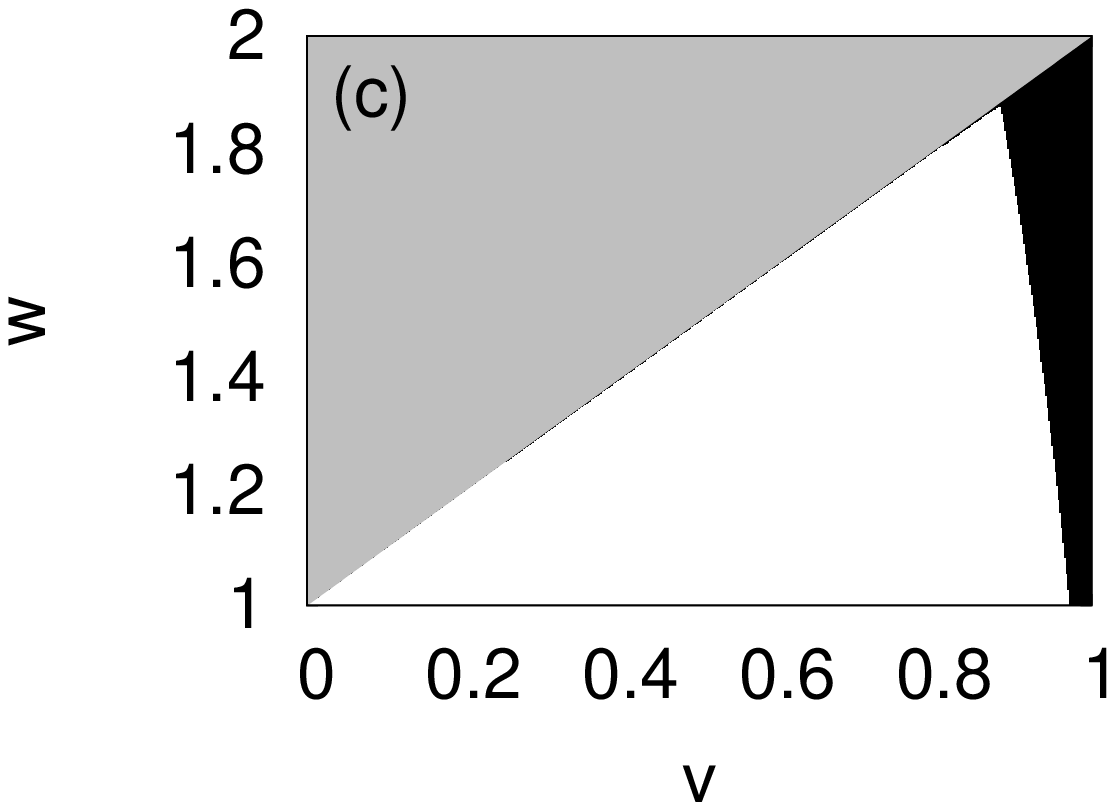}}
  \subfigure{\includegraphics[width=0.3\textwidth]{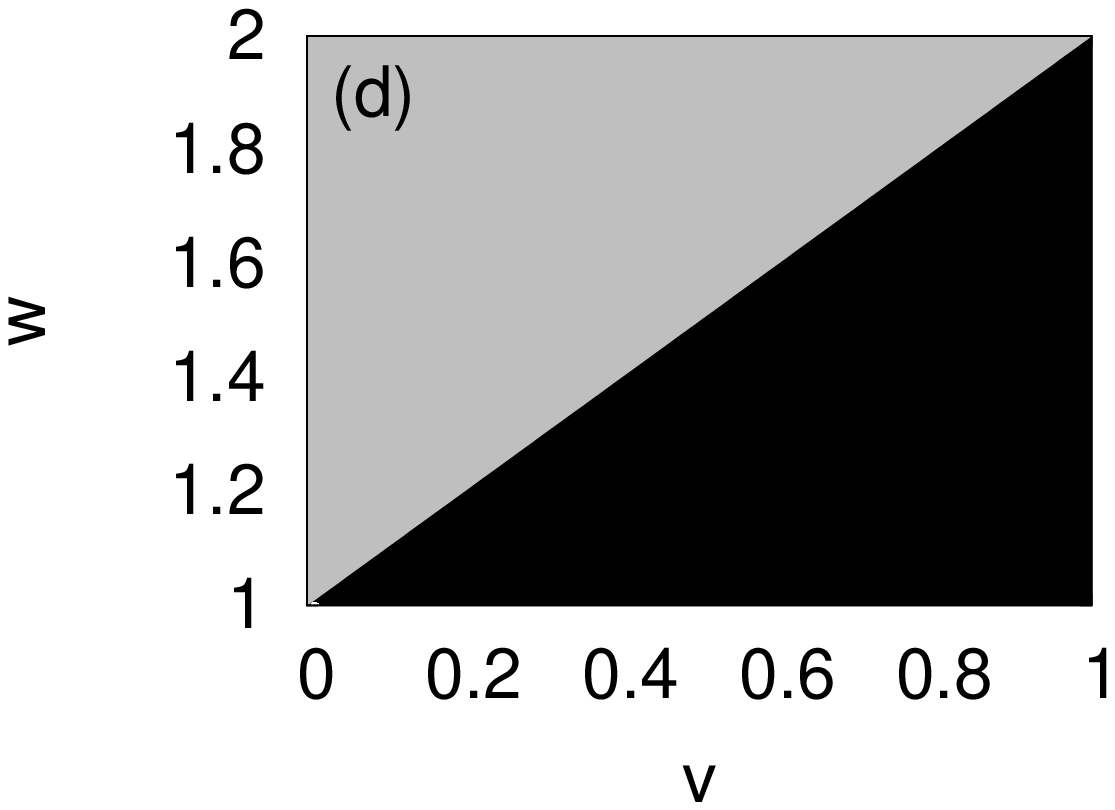}}
  \subfigure{\includegraphics[width=0.3\textwidth]{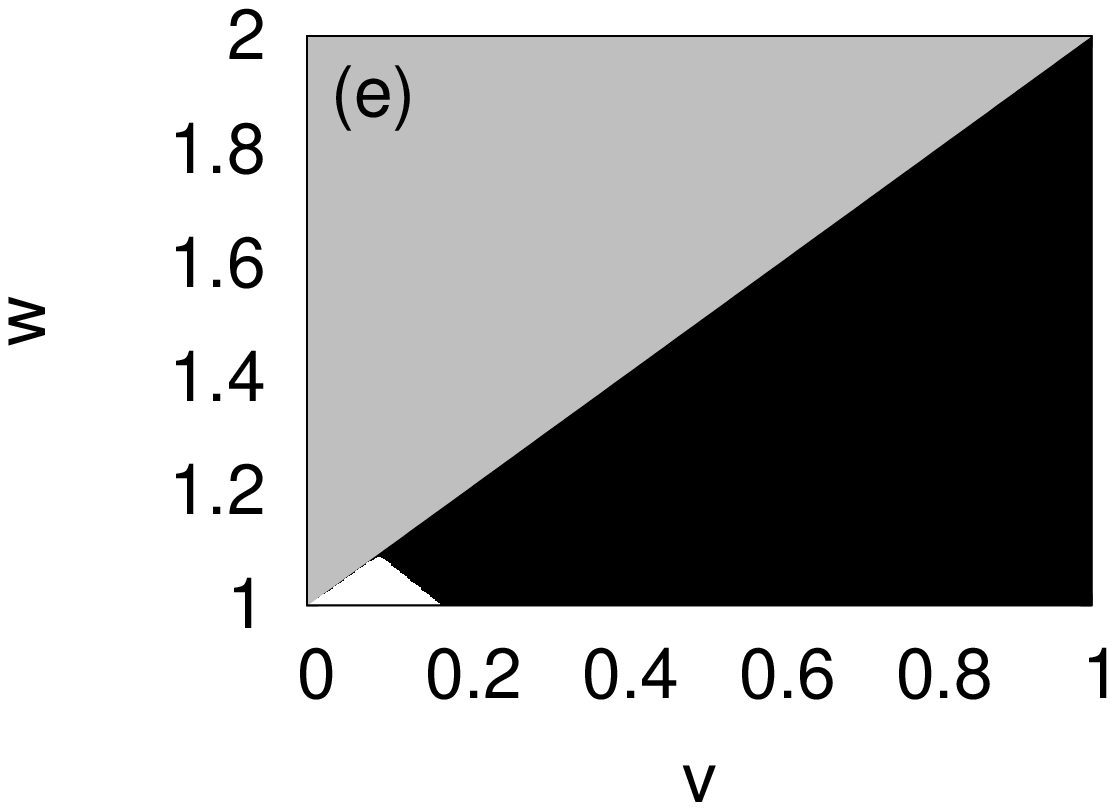}}
  \subfigure{\includegraphics[width=0.3\textwidth]{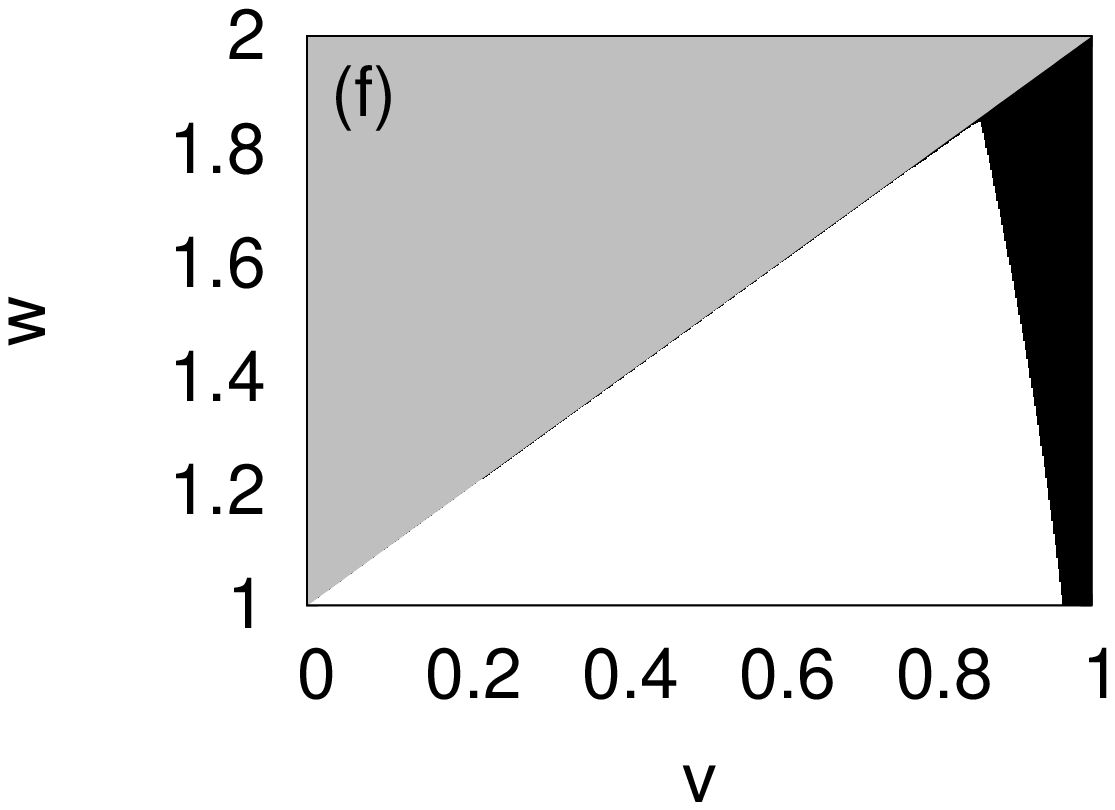}}
  \subfigure{\includegraphics[width=0.3\textwidth]{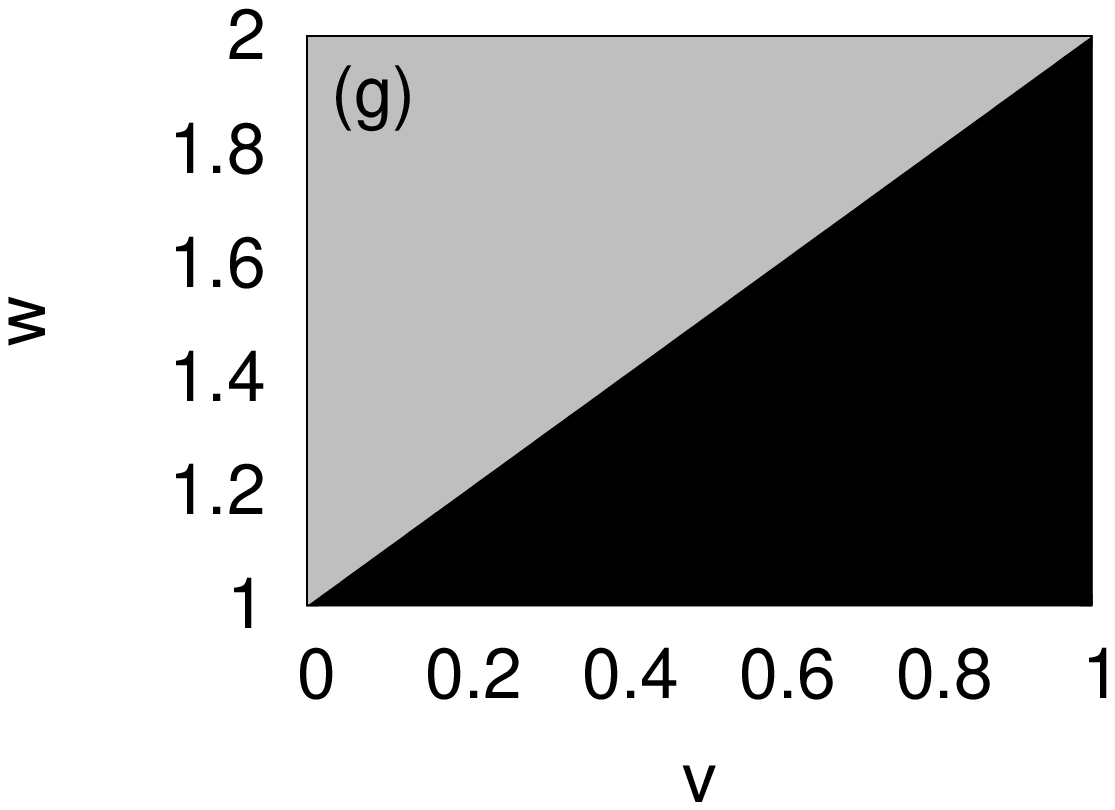}}
  \subfigure{\includegraphics[width=0.3\textwidth]{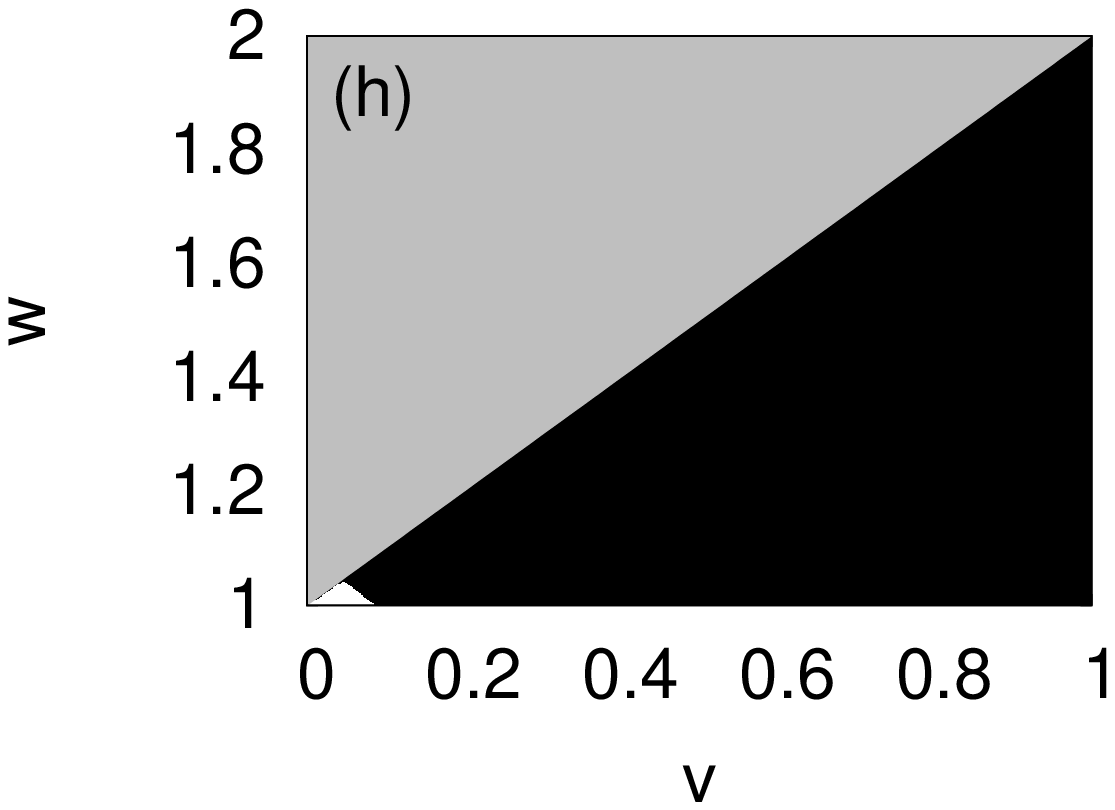}}
  \subfigure{\includegraphics[width=0.3\textwidth]{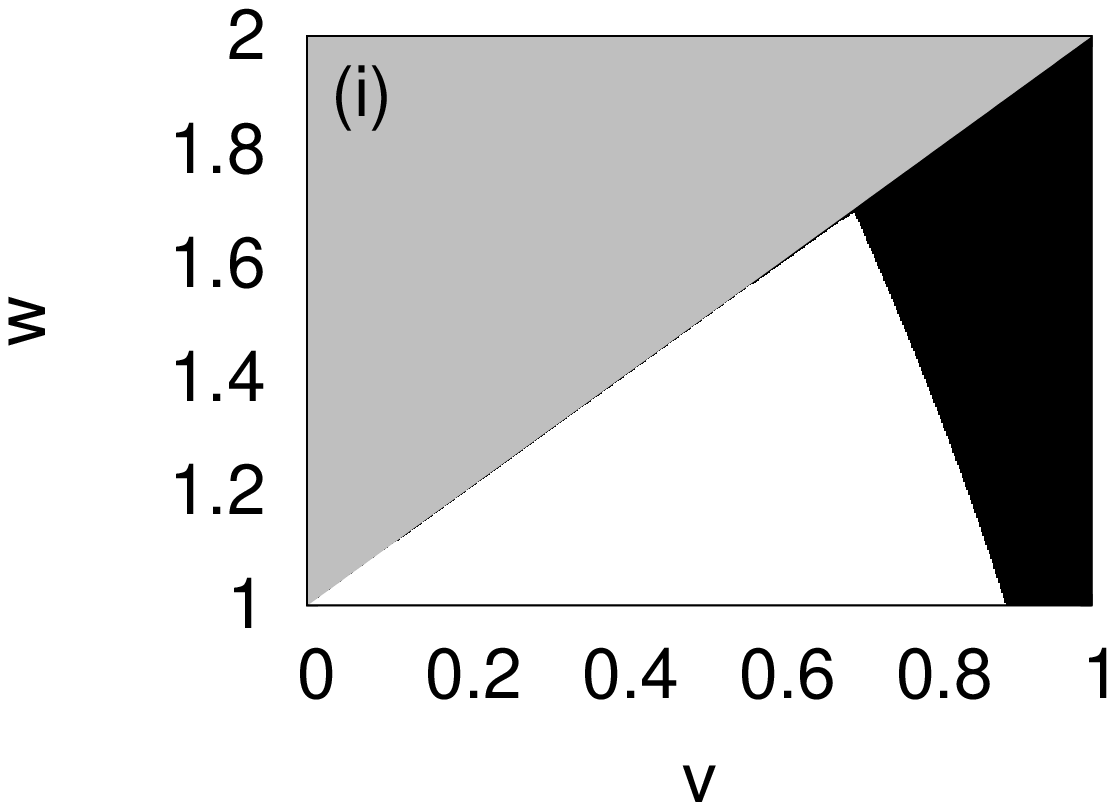}}
  \subfigure{\includegraphics[width=0.3\textwidth]{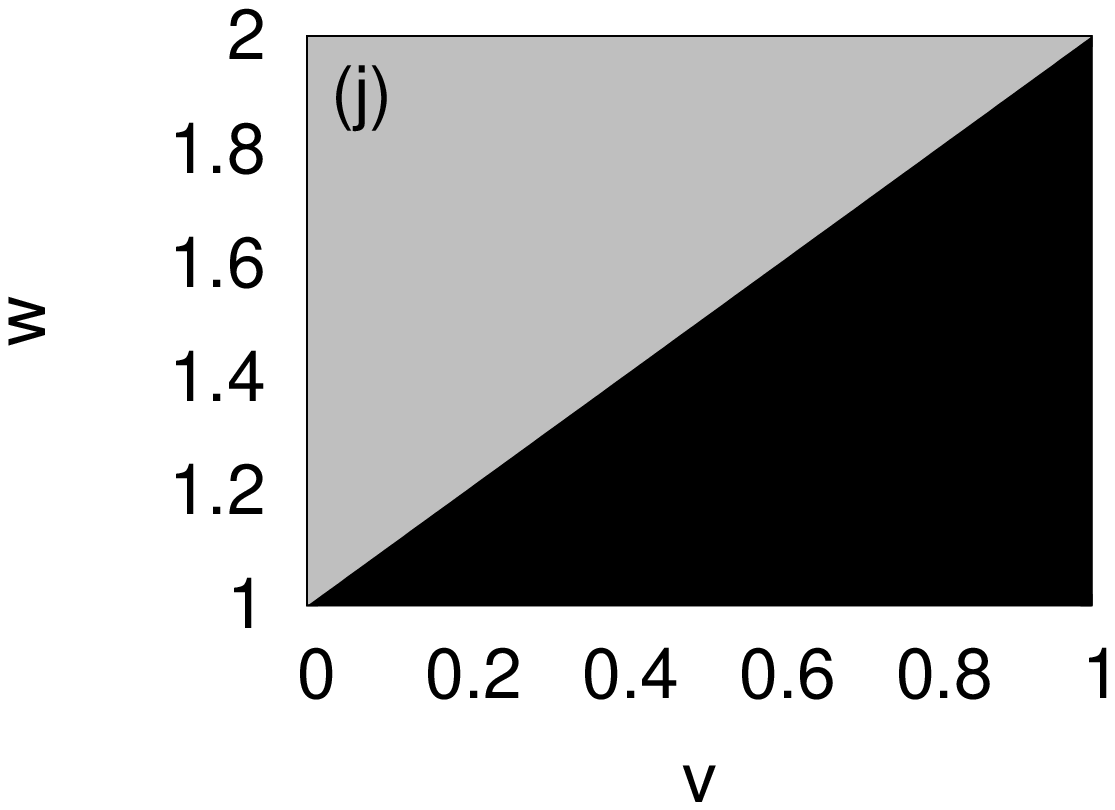}}
  \subfigure{\includegraphics[width=0.3\textwidth]{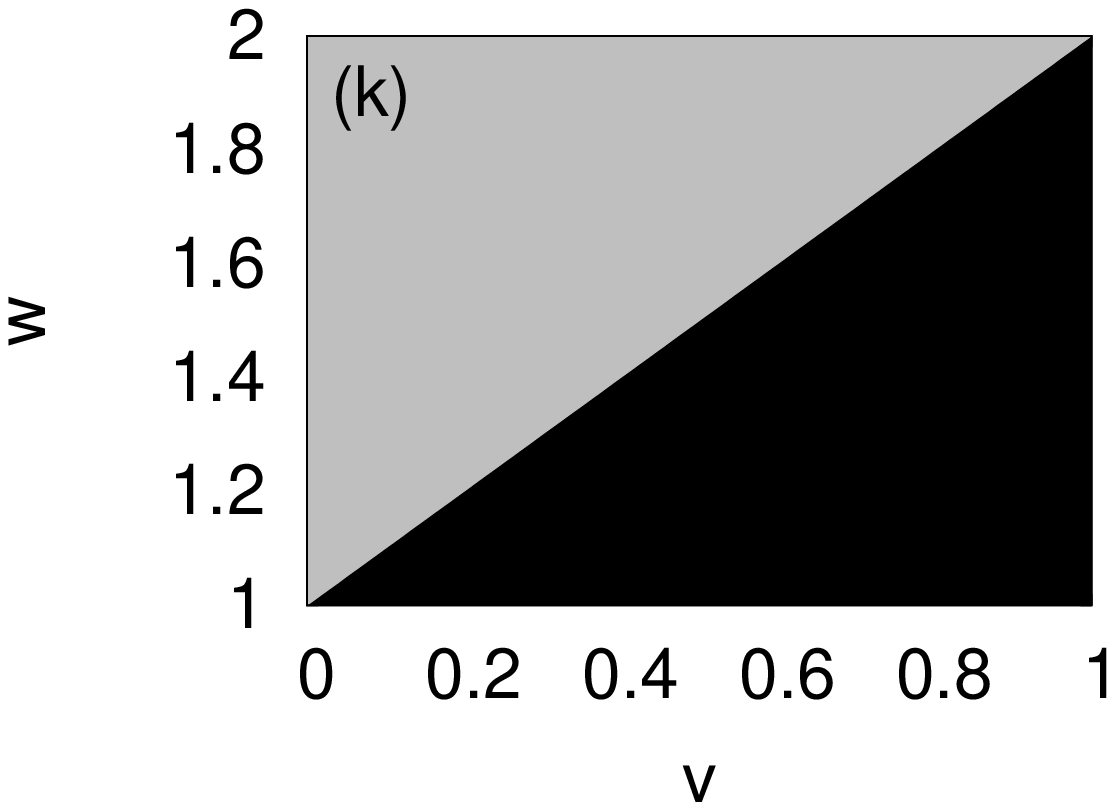}}
  \subfigure{\includegraphics[width=0.3\textwidth]{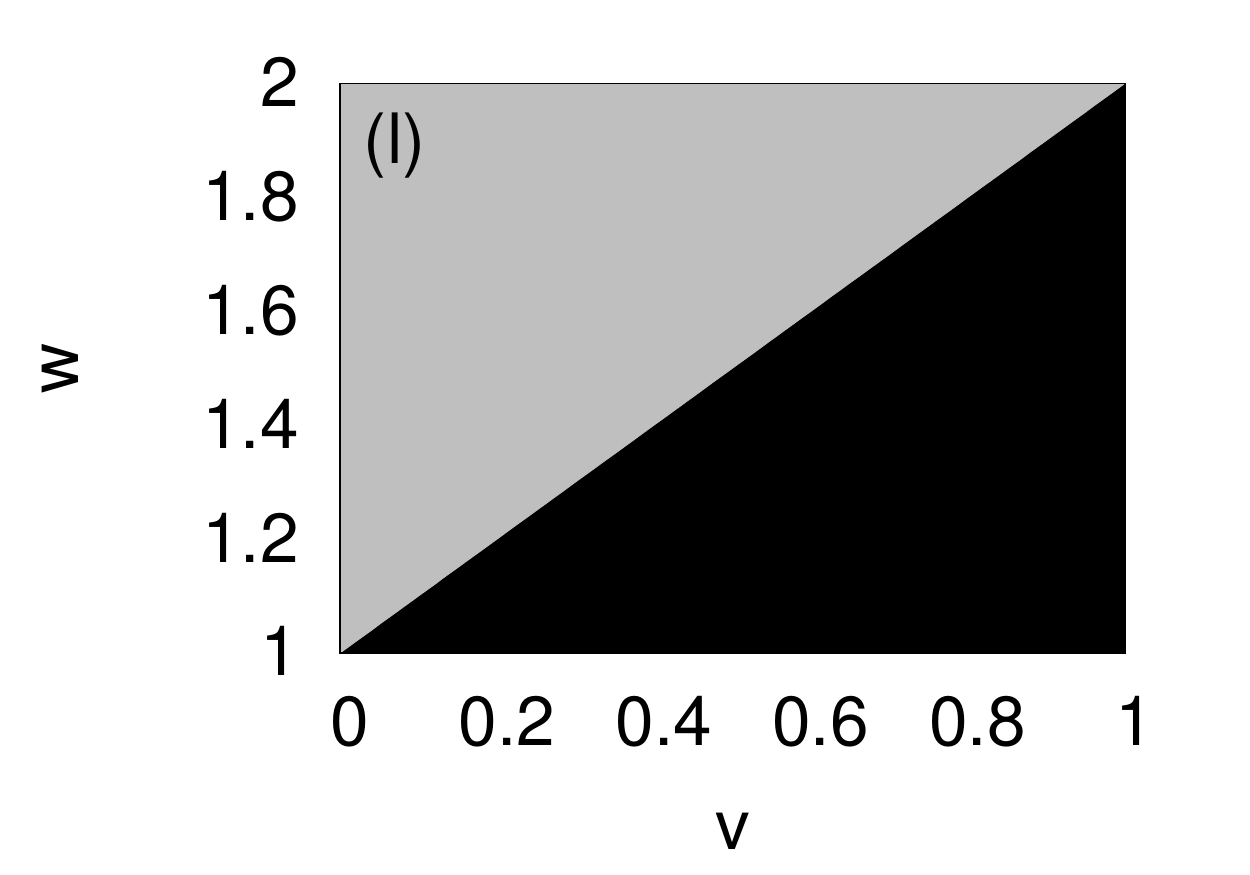}}
  \caption{Plots of $f(v,w)=x^2-Q$ for various values of $|\Up|/|\Bp|$ and cross-helicity
           for case 3 in appendix \ref{app:graphs} ($s_k= s_p \neq s_q, p<k<q$).
           The upper grey triangle is ruled out by the condition $w<1+v$ and unstable values are shown in white.
           The ratio $|\Up|/|\Bp|$ increases from left to right, with each column of subfigures taking the values
           0.01, 0.1 and 1 respectively, while each row takes the following values of relative cross-helicity:
           $H_c(p)/(|\Up||\Bp|)=0, 0.5, 0.9$ and $1$.}
  \label{fig:case3}
\end{figure}

\item $s_k = s_p = s_q$ and $k<q<p$ \\ 
In this case we rescale all wavenumbers by $q$, such that $v\equiv  k/q$ and $w\equiv p/q$. 
As can be seen in 
fig.~\ref{fig:case4}, now for {\em increasing} $|\Up|/|\Bp|$ and increasing $H_c(p)$ less and less triads lead to 
unstable solutions and we obtain the constraints shown in table \ref{tbl:reverse} on reverse transfer.
 
\begin{figure}
  \centering
  \subfigure{\includegraphics[width=0.3\textwidth]{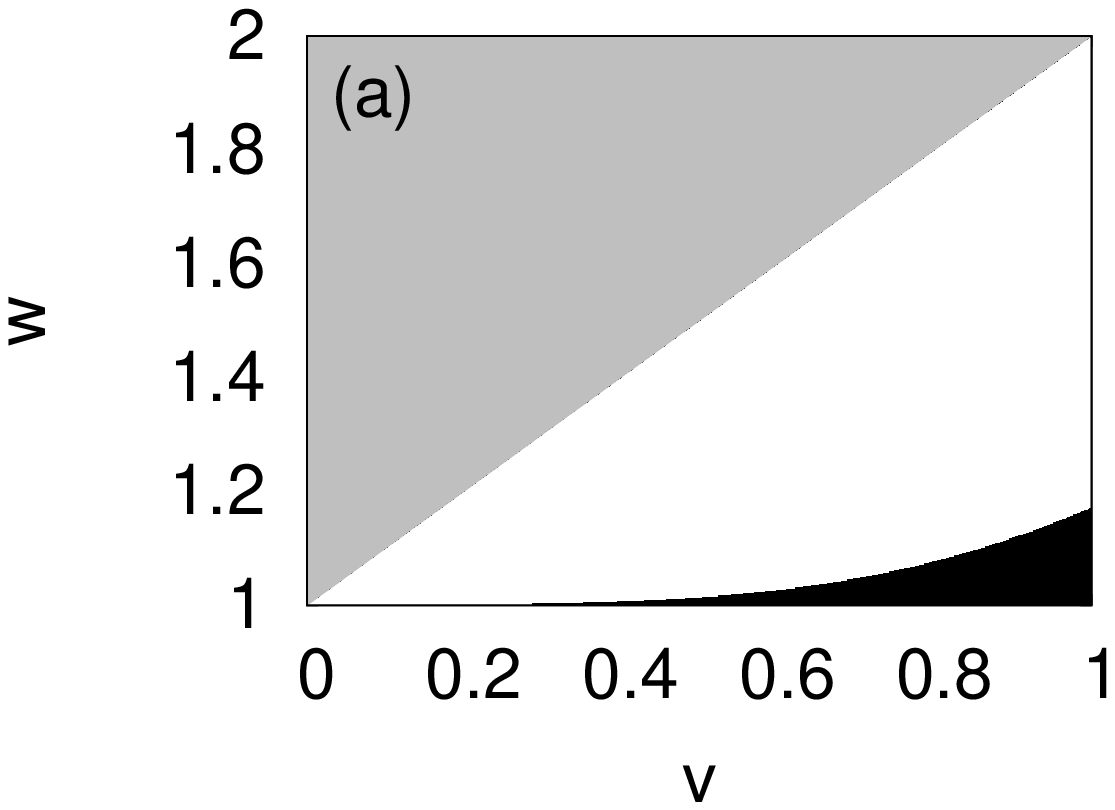}}
  \subfigure{\includegraphics[width=0.3\textwidth]{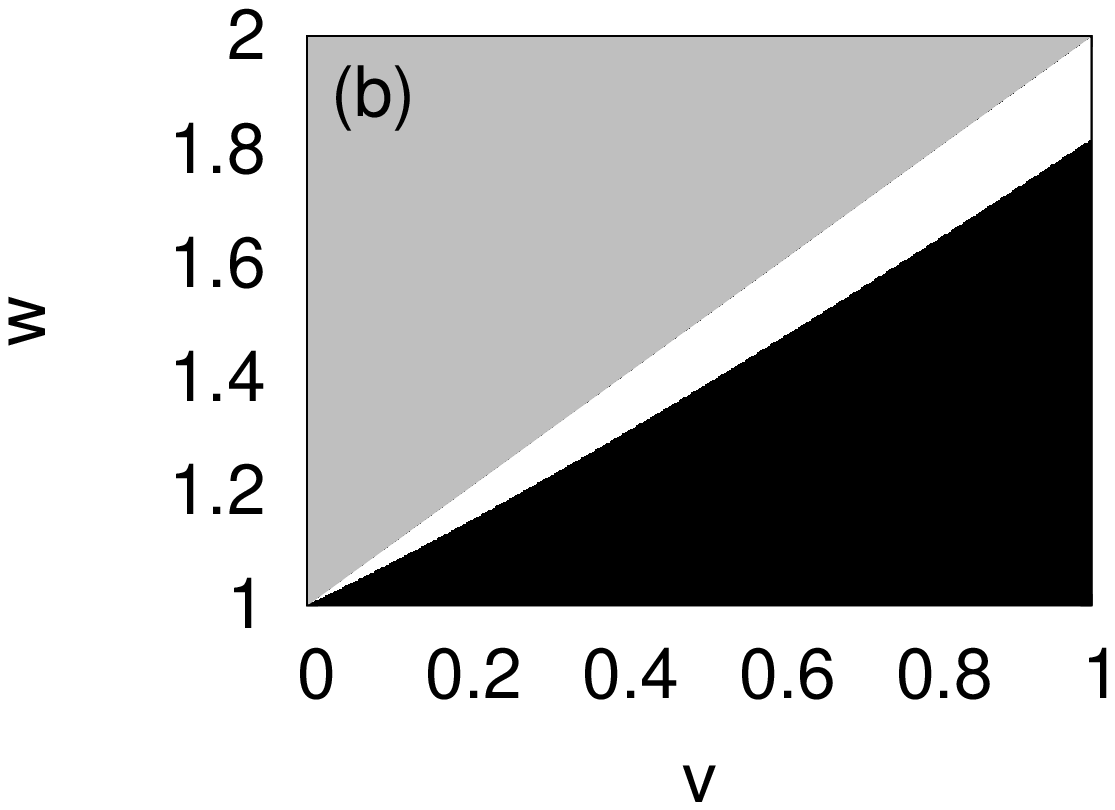}}
  \subfigure{\includegraphics[width=0.3\textwidth]{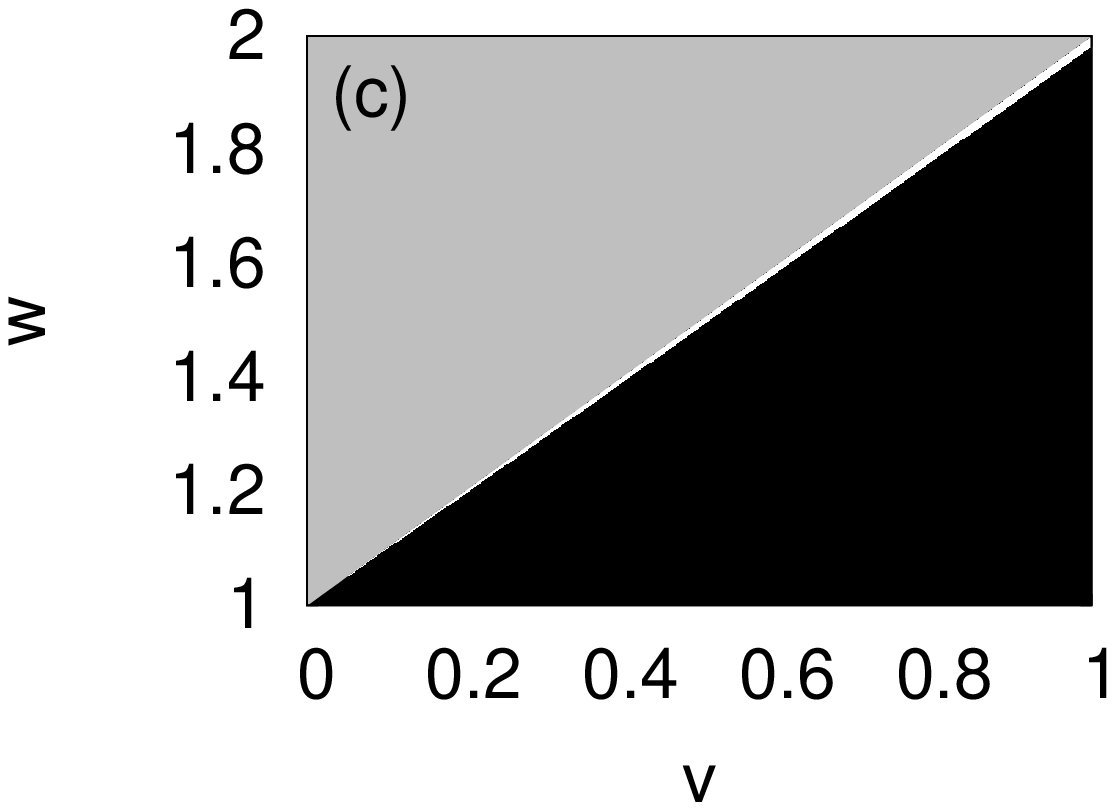}}
  \subfigure{\includegraphics[width=0.3\textwidth]{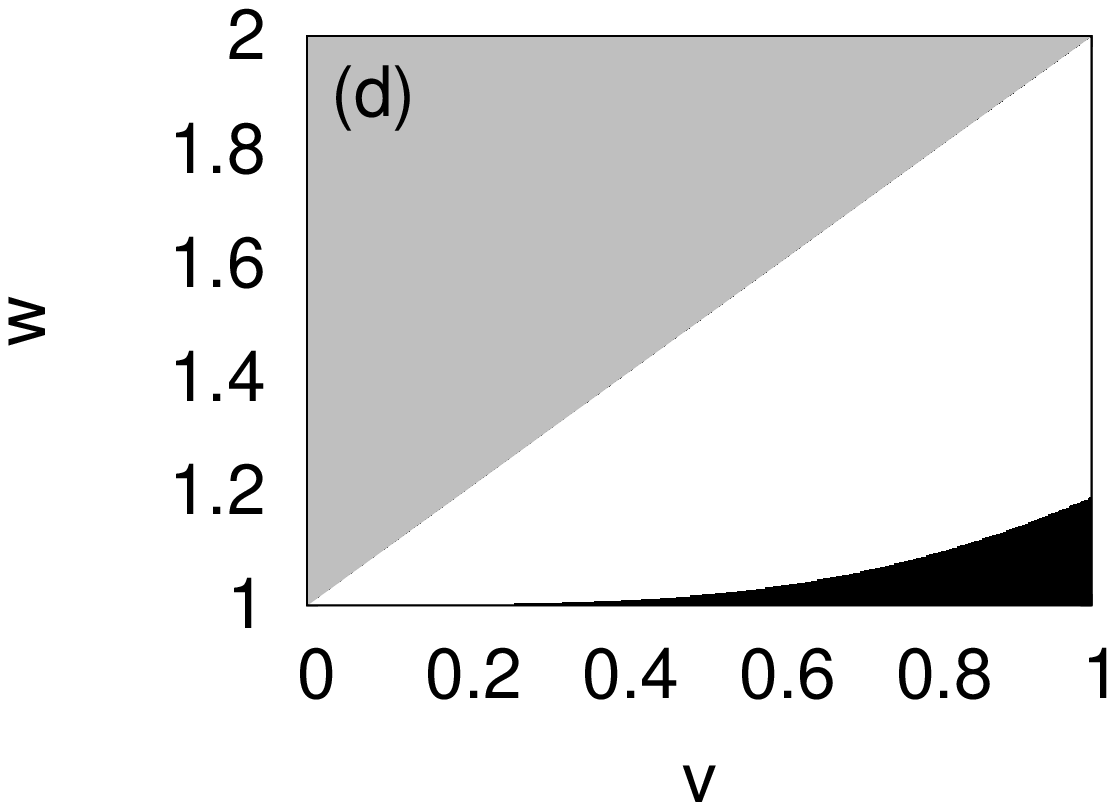}}
  \subfigure{\includegraphics[width=0.3\textwidth]{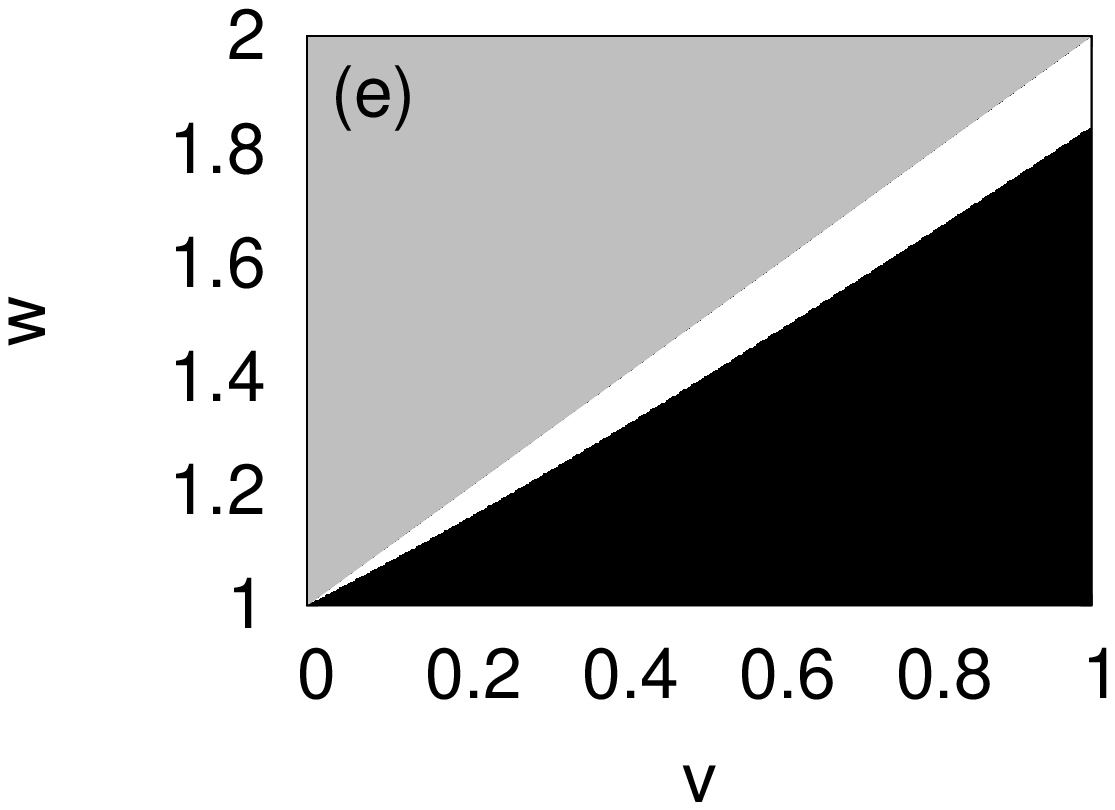}}
  \subfigure{\includegraphics[width=0.3\textwidth]{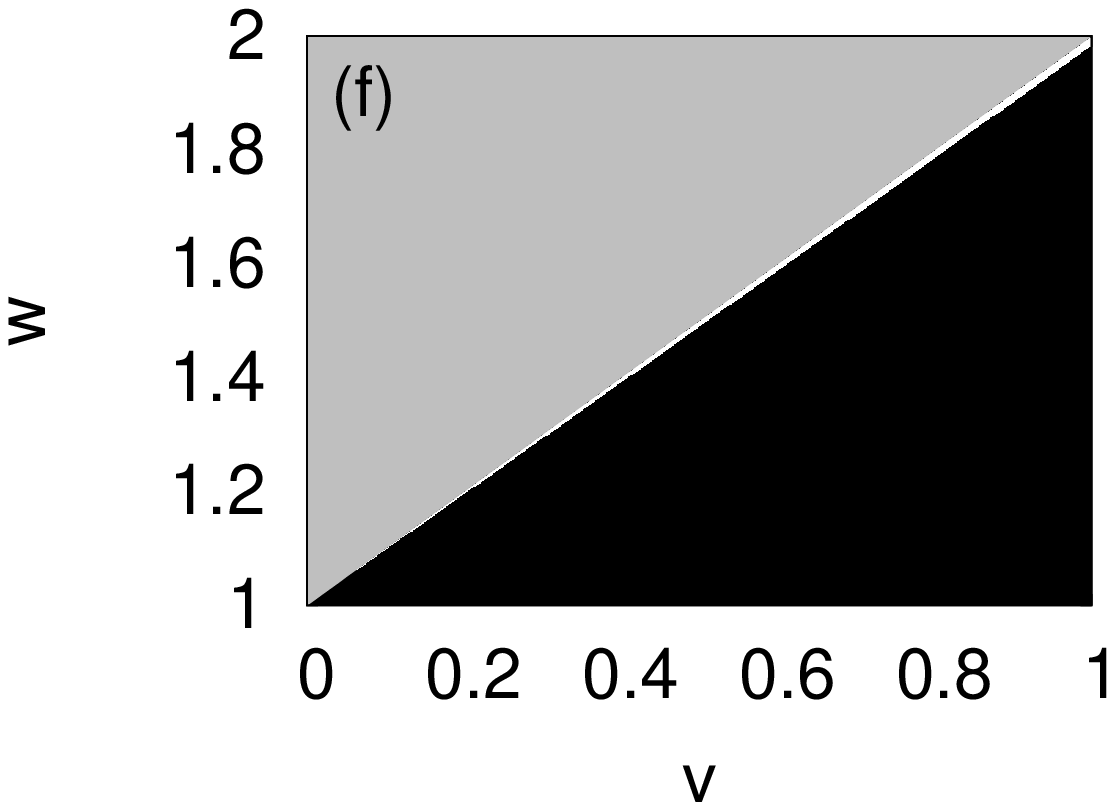}}
  \subfigure{\includegraphics[width=0.3\textwidth]{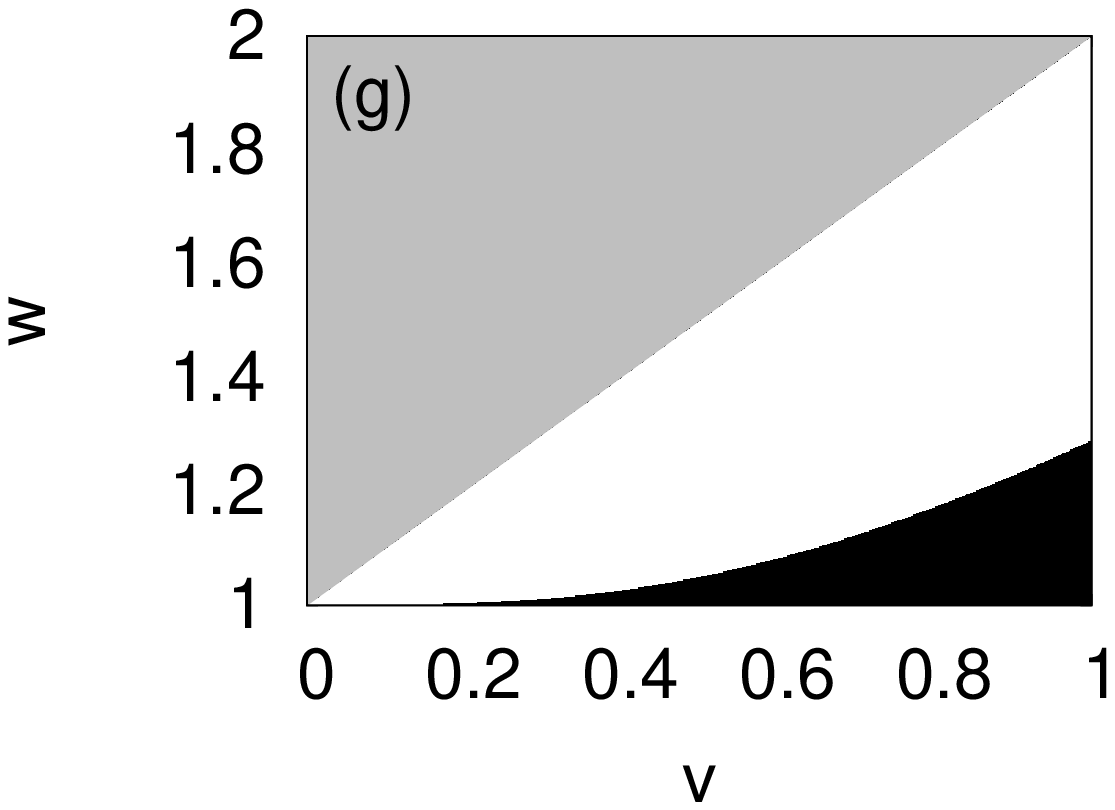}}
  \subfigure{\includegraphics[width=0.3\textwidth]{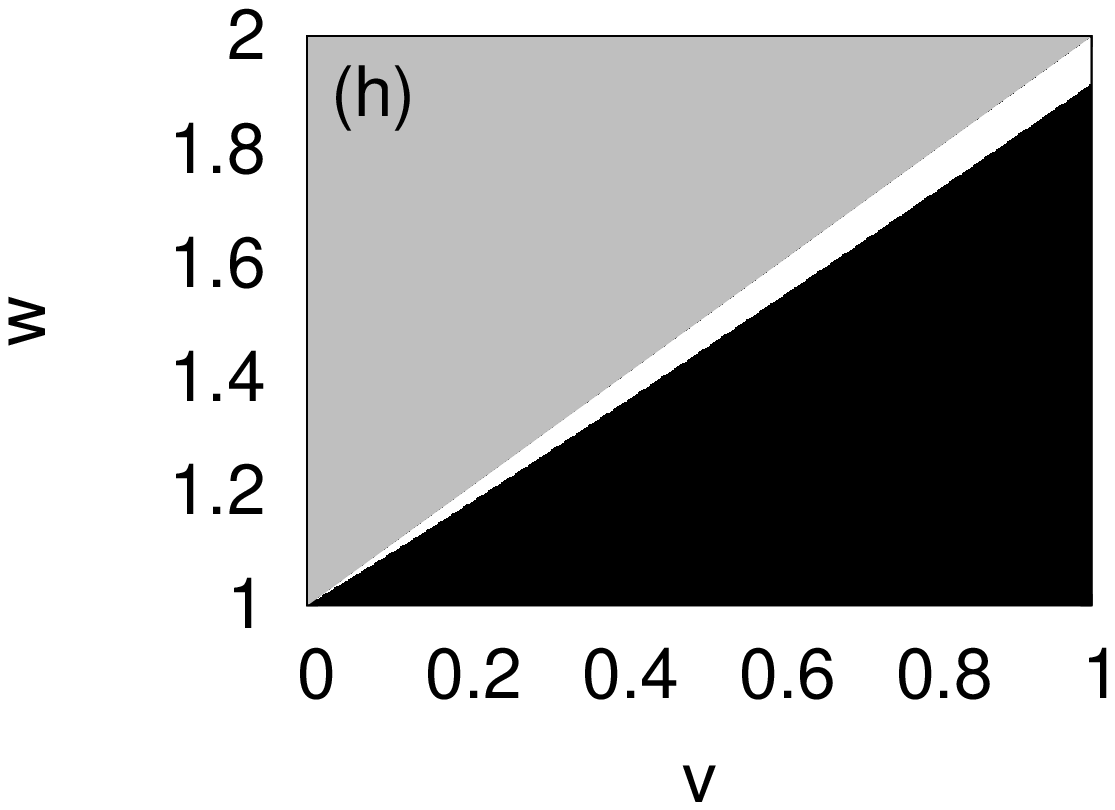}}
  \subfigure{\includegraphics[width=0.3\textwidth]{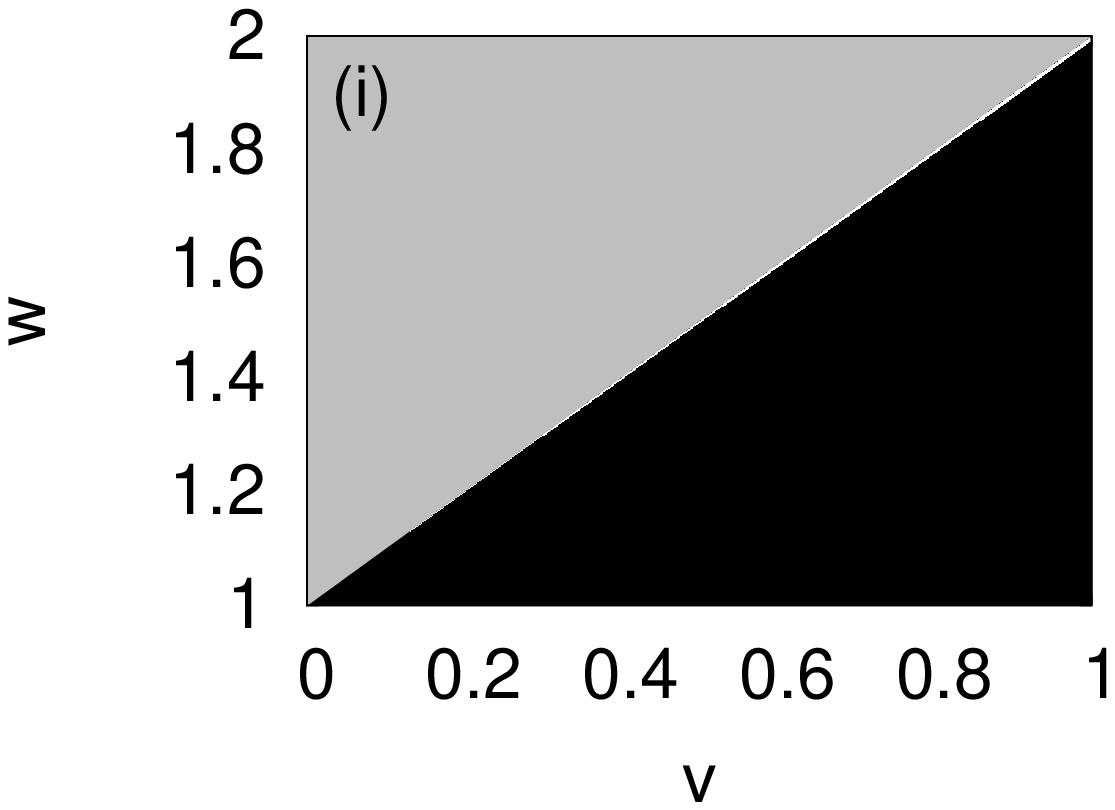}}
  \subfigure{\includegraphics[width=0.3\textwidth]{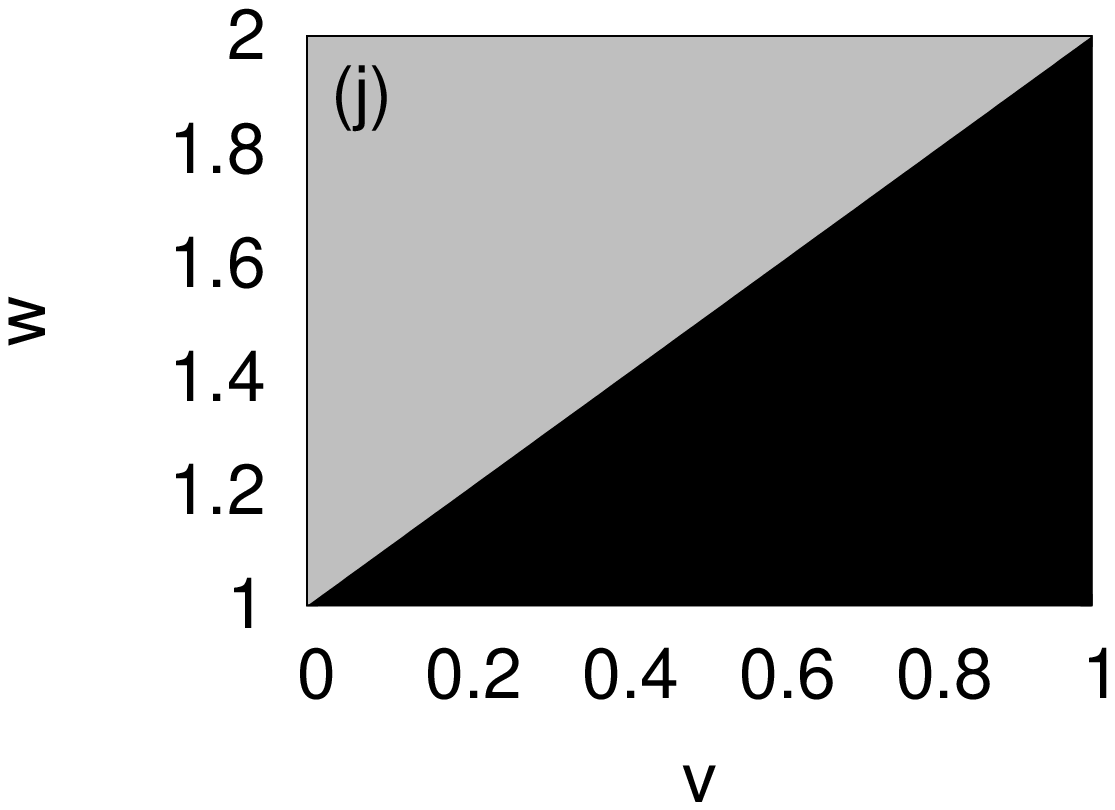}}
  \subfigure{\includegraphics[width=0.3\textwidth]{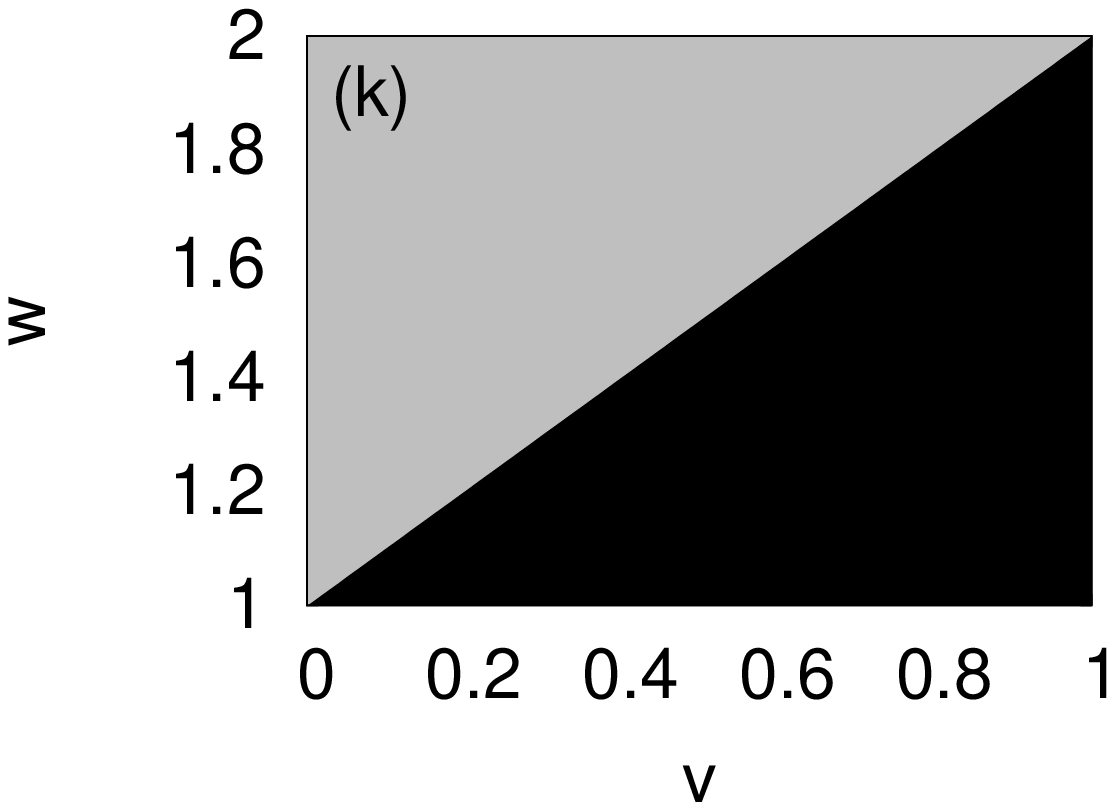}}
  \subfigure{\includegraphics[width=0.3\textwidth]{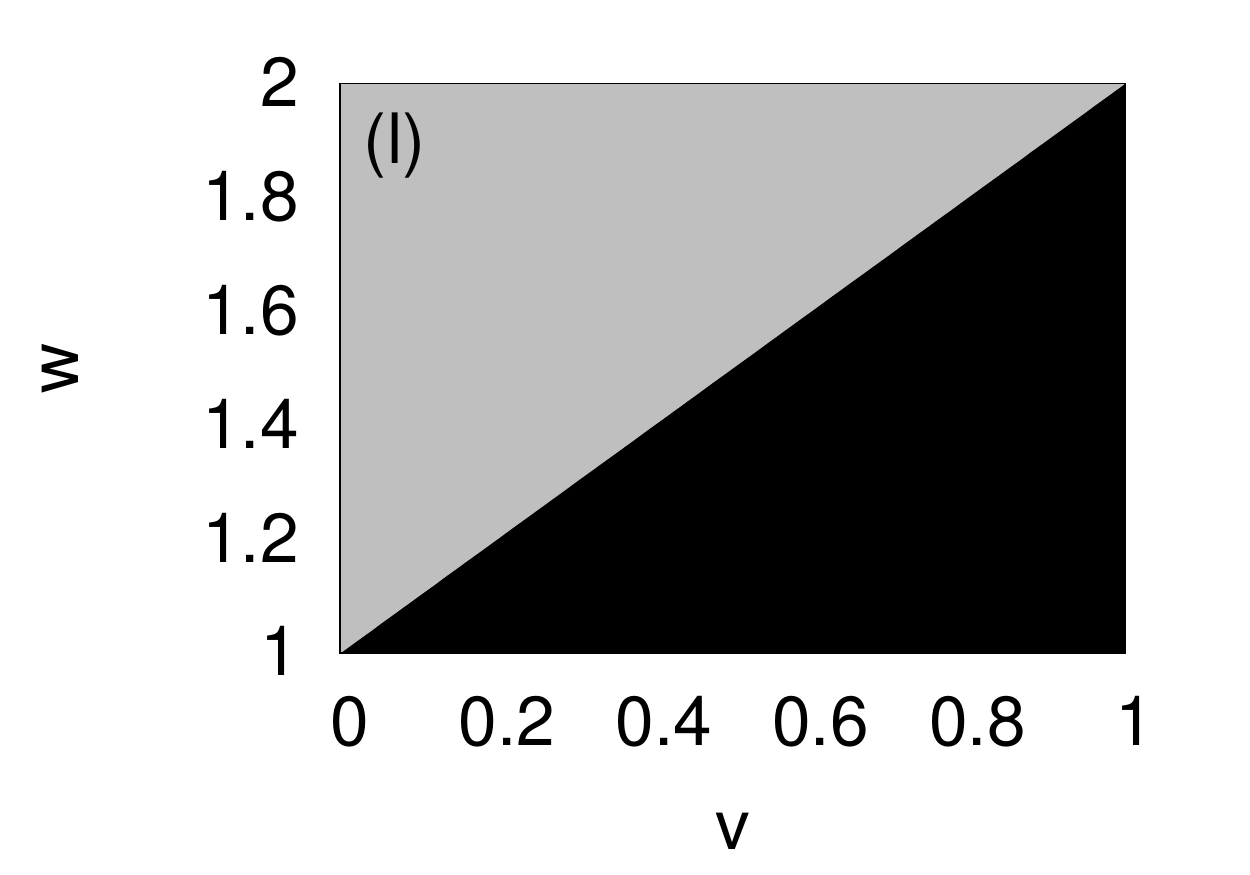}}
  \caption{Plots of $f(v,w)=x^2-Q$ for various values of $|\Up|/|\Bp|$ and cross-helicity
           for case 4 in appendix \ref{app:graphs} ($s_k= s_p=s_q, k<q<p$).
           The upper grey triangle is ruled out by the condition $w<1+v$ and unstable values are shown in white.
           The ratio $|\Up|/|\Bp|$ increases from left to right, with each column of subfigures taking the values
           1, 10 and 100 respectively, while each row takes the following values of relative cross-helicity:
           $H_c(p)/(|\Up||\Bp|)=0, 0.5, 0.9$ and $1$.}
  \label{fig:case4}
\end{figure}

\end{itemize}

\section{Similarity scaling}
\label{app:similarity}
In the respective inertial ranges 
\begin{equation} E_{kin}(\alpha k)/E_{kin}(k) = \alpha^{-n}, \ \  E_{mag}(\alpha k)/E_{mag}(k) = \alpha^{-m}\label{eq:E_scaling}\end{equation}
where $\alpha$ is a real number and $n >0$  and $m>0$ are the spectral indices
of the kinetic and magnetic energy spectra, respectively. 
From
\begin{equation} E_{kin}(k) dk = \frac{1}{2}\int_{|\vec{k}|=k} \langle |u_+(\vec{k})|^2 + |u_-(\vec{k})|^2 \rangle d\vec{k} \ ,
\end{equation}
\begin{equation} E_{mag}(k) dk = \frac{1}{2}\int_{|\vec{k}|=k} \langle |b_+(\vec{k})|^2 + |b_-(\vec{k})|^2 \rangle d\vec{k} \ ,
\end{equation}
we then find a scaling of the helical coefficients 
\begin{equation} u_s(\alpha \vec{k})=\alpha^{-(5+n)/2} u_s(\vec{k}), \ \ 
b_s(\alpha \vec{k})=\alpha^{(-5+m)/2} b_s(\vec{k}) \ .
\end{equation}
From \eqref{eq:E_scaling} the scaling of $T_{HD}^{(i)}(k,p,q)$, $T_{LF}^{(i)}(k,p,q)$ and $T_{mag}^{(i)}(k,p,q)$ is then given by
\begin{align} 
&\frac{T^{(i)}_{HD}(\alpha k,\alpha p,\alpha q)}{T^{(i)}_{HD}(k,p,q)} = \alpha^{-(1+3n)/2}=\alpha^{-\beta}  \ , \\
&\frac{T^{(i)}_{LF}(\alpha k,\alpha p,\alpha q)}{T^{(i)}_{LF}(k,p,q)} = \alpha^{-(1+n+2m)/2}=\alpha^{-\beta'}  \ , \\
&\frac{T^{(i)}_{mag}(\alpha k,\alpha p,\alpha q)}{T^{(i)}_{mag}(k,p,q)} = \alpha^{-(1+n+2m)/2}=\alpha^{-\beta'}  \ , 
\end{align}
In hydrodynamics $n=5/3$, while in MHD there are
different predictions for the spectral exponent, either $m=3/2$ (Iroshnikov-Kraichnan) or
$m=5/3$ (Kolmogorov). Note that $n=m=5/3$ implies $\beta'=\beta = 3$ while $n=5/3$ and $m=3/2$ implies $\beta'=2+5/6$. 
In both cases $\beta-2>0$.

\bibliographystyle{jfm}
\bibliography{helmodes,wdm_sos}

\end{document}